\documentclass[aps,reprint,nofootinbib,nobibnotes,notitlepage,superscriptaddress,onecolumn,prd,preprintnumbers,
 amsmath,amssymb
]{revtex4-2}

\usepackage[caption=false]{subfig}
\usepackage{braket}
\usepackage{float}
\usepackage{lipsum}
\usepackage{graphicx}
\usepackage{dcolumn}
\usepackage{bm}
\usepackage{natbib}
\usepackage{hyperref}
\hypersetup{
	colorlinks = true,
    linkcolor = Red,
    urlcolor  = Red,
    citecolor = Red
}
\usepackage[skip=4pt plus1pt, indent=10pt]{parskip}
\usepackage{microtype}

\usepackage{verbatim}
\usepackage[amssymb]{SIunits}
\usepackage{tabularx}
\usepackage[dvipsnames]{xcolor}
\usepackage{wasysym}
\usepackage{multirow}
\usepackage{diagbox}
\usepackage[normalem]{ulem}

\usepackage{tikz}
\usepackage[compat=1.1.0]{tikz-feynman}

\renewcommand{\arraystretch}{1.3}
\def\be{\begin{equation}}
\def\ee{\end{equation}}

\usepackage{color}
\definecolor{darkgreen}{RGB}{0,120,0}
\definecolor{darkgreen}{RGB}{0,120,0}

\newcommand{\Mpch}{h^{-1}\mathrm{Mpc}}
\newcommand{\hMpc}{h\,\mathrm{Mpc}^{-1}}

\newcommand{\delD}[1]{(2\pi)^3\delta_\mathrm{D}\left({#1}\right)}

\newcommand{\av}[1]{\left\langle{#1}\right\rangle} 

\newcommand{\vk}{\vec k}
\newcommand{\hk}{\hat{\vec k}}

\newcommand{\vd}{\vec d}

\newcommand{\vz}{\vec z}

\newcommand{\vx}{\vec x}
\newcommand{\hx}{\hat{\vec x}}
\newcommand{\vy}{\vec y}
\newcommand{\vw}{\vec w}

\newcommand{\C}{\mathsf{C}}
\newcommand{\Si}{\mathsf{S}^{-1}}
\newcommand{\Sid}{\mathsf{S}^{-\dagger}}
\newcommand{\F}{\mathcal{F}}

\newcommand{\G}{\mathcal{G}}

\newcommand{\N}{\mathsf{N}}
\newcommand{\Ci}{\mathsf{C}^{-1}}

\newcommand{\hy}{\hat{\vec y}}

\newcommand{\hz}{\hat{\vec z}}

\renewcommand{\vr}{\vec r}

\renewcommand{\P}{\mathcal{P}}

\renewcommand{\P}{\mathsf{P}}
\newcommand*{\polybin}{\selectfont\textsc{PolyBin3D}\xspace}
\newcommand{\hun}{\,\mathrm{km}\,\mathrm{s}^{-1}\mathrm{Mpc}^{-1}}
\newcommand{\ld}{\Lambda{\rm CDM}}
\newcommand{\kmin}{k_{\rm min}}
\newcommand{\kmax}{k_{\rm max}}
\newcommand{\eff}{{\rm eff}}
\newcommand{\kmsMpc}{{\rm km\,s^{-1}\,Mpc^{-1}}}
\newcommand{\Mpc}{\text{Mpc}}
\newcommand{\eV}{{\,\rm eV}}
\newcommand{\Pshot}{P_{\rm shot}}
\newcommand{\Bshot}{B_{\rm shot}}
\newcommand{\Ashot}{A_{\rm shot}}

\DeclareSymbolFont{toneletters}{T1}{\familydefault}{m}{it}
\SetSymbolFont{toneletters}{bold}{T1}{\familydefault}{bx}{it}

\DeclareMathSymbol\edth{\mathord}{toneletters}{"F0}

\usepackage{adjustbox}
\usepackage{dashbox}

\def\beq{\begin{eqnarray}}
\def\eeq{\end{eqnarray}}

\let\vec\mathbf

\usepackage{empheq}

\def\mnras{\rm{MNRAS}}

\def\aap{\rm{A\&A}}

\def\apj{\rm{ApJ}}                 

\def\pasj{\rm{PASJ}}

\defcitealias{desi2}{Paper 2}
\defcitealias{desi3}{Paper 3}

\usepackage{xspace}

\definecolor{darkgreen}{RGB}{0,120,0}

\begin{document}
\preprint{MIT-CTP/5890}

\title{\texorpdfstring{{\Large Reanalyzing DESI DR1:}\\
1. $\Lambda$CDM Constraints from the Power Spectrum \& Bispectrum}{Reanalyzing DESI DR1: LCDM Constraints from the Power Spectrum \& Bispectrum}}


\author{Anton~Chudaykin}
\email{anton.chudaykin@unige.ch}
\affiliation{D\'epartement de Physique Th\'eorique and Center for Astroparticle Physics,\\
Universit\'e de Gen\`eve, 24 quai Ernest  Ansermet, 1211 Gen\`eve 4, Switzerland}
\author{Mikhail M.~Ivanov}
\email{ivanov99@mit.edu}
\affiliation{Center for Theoretical Physics -- a Leinweber Institute, Massachusetts Institute of Technology, 
Cambridge, MA 02139, USA} 
 \affiliation{The NSF AI Institute for Artificial Intelligence and Fundamental Interactions, Cambridge, MA 02139, USA}
\author{Oliver~H.\,E.~Philcox}
\email{ohep2@cantab.ac.uk}
\affiliation{Simons Society of Fellows, Simons Foundation, New York, NY 10010, USA}
\affiliation{Center for Theoretical Physics, Columbia University, New York, NY 10027, USA}
\affiliation{Leinweber Institute for Theoretical Physics at Stanford, 382 Via Pueblo, Stanford, CA 94305, USA}
\affiliation{Kavli Institute for Particle Astrophysics and Cosmology, 382 Via Pueblo, Stanford, CA 94305, USA}

\begin{abstract} 
    \noindent 
    We present the first independent re-analysis of the galaxy clustering data from DESI Data Release 1, utilizing an effective field theory full-shape model. We analyze the power spectra and bispectra of the public catalogs using a custom-built pipeline based on window-deconvolved quasi-optimal estimators, accounting for a number of systematic effects. Compared to the official collaboration analysis, we add the galaxy power spectrum hexadecapole and the bispectrum monopole, and also introduce a novel stochastic estimator for fiber collisions, which facilitates robust bispectrum analyses. As a first application, we perform a full-shape analysis of the DESI power spectra and bispectra in the context of the standard cosmological model, $\Lambda$CDM. Using external priors on the physical baryon density and the primordial power spectrum tilt, we constrain the matter density fraction to $\Omega_m=0.284\pm 0.011$, the Hubble constant to $H_0=70.7\pm 1.1\,\hun$, and the mass fluctuation amplitude to $\sigma_8=0.811\pm 0.030$. The bispectrum sharpens constraints on $\sigma_8$ and $\Omega_m$ by $\approx 10$\% and shifts $\Omega_m$ by $\approx 1\sigma$ towards the \textit{Planck} $\Lambda$CDM value. Combining our full-shape likelihood with the official DESI DR2 BAO measurements, cosmological parameters shift further towards the \textit{Planck} values, with $\Omega_m=0.296\pm 0.007$, $H_0=68.8\pm 0.6\,\hun$, $\sigma_8=0.818\pm 0.029$ (with tighter constraints obtained in joint analyses). Similar results are obtained in a joint analysis with DR1 BAO, accounting for the cross-covariance. Finally, the bispectrum data improves measurements of quadratic bias parameters, which are consistent with predictions from halo occupation distribution models. Our work highlights the importance of higher-order statistics and sets the stage for upcoming full-shape analyses of non-minimal cosmological models.
\end{abstract}

\maketitle

\section{Introduction}

\setlength{\parskip}{2pt plus1pt}

\noindent Since the 1970s, three-dimensional galaxy redshift surveys have played a pivotal role in our development of the standard cosmological model. Early experiments including the Center for Astrophysics (CfA) redshift surveys revealed a cosmic web of clusters, filaments, and voids hypothesized to arise from the gravitational growth of initial density perturbations \citep{deLapparent1986,Geller1989,1980lssu.book.....P,1978ApJ...221....1D}. Moreover, the statistical analysis of these measurements provided early hints for a Universe with $\Omega_m<1$ \citep[e.g.,][]{Davis:1985rj}, a result which was later confirmed by supernovae observations \citep{Riess1998,Perlmutter1999}. Whilst originally seen as a tracer of the matter-radiation-equality scale ($\Gamma\equiv\Omega_mh$),\footnote{This has seen a recent revival of interest in the context of sound-horizon-free measurements of the Hubble scale \citep[e.g.,][]{Philcox:2020xbv,Philcox:2022sgj,Zaborowski:2024wpo,DESI:2025euz}.} it was later demonstrated that galaxy surveys could provide a geometric probe of the Universe, utilizing the baryon acoustic oscillation (BAO) feature imprinted by recombination-era physical processes \citep{Eisenstein:1997ik,Eisenstein:2006nk,Cole:2005sx,Eisenstein2005}.

With the advent of large-scale projects such as the Sloan Digital Sky Survey (SDSS), the detection significance of the BAO feature has continued to increase, which has led to precision constraints being placed on a various models of the early- and late-Universe, mostly in concert with the cosmic microwave background (CMB) anisotropies measured by \textit{Planck} and other experiments \citep[e.g.,][]{Percival2009,Reid2012,Padmanabhan2012,BOSS:2016wmc,Aghanim:2018eyx,BOSS:2016psr}. Of course, the galaxy dataset contains much more information than just the oscillations. This has been previously exploited to yield constraints on the growth parameter $f\sigma_8(z)$, extensions to general relativity, and the bias parameters describing galaxy formation, amongst other things \citep[e.g.,][]{Gil-Marin:2014sta,Guo:2014iga,Obuljen:2020ypy,Slepian:2015hca,Gil-Marin:2016wya,Gil-Marin:2014pva,Gil-Marin:2014baa,eBOSS:2020fvk,Guo:2014iga,Beutler:2013yhm,Beutler:2014yhv,BOSS:2012vpn}. An important recent development has been the introduction of robust perturbative models for the galaxy distribution, in particular those based on the effective field theory of large-scale structure (EFT) \citep[e.g.,][]{Ivanov:2019pdj,DAmico:2019fhj,Chen:2021wdi,Vlah:2015zda,Pajer:2013jj,Carrasco:2012cv,Mercolli:2013bsa,Porto:2013qua,Vlah:2015sea,Senatore:2014via,Senatore:2014vja,Senatore:2014eva,Angulo:2015eqa,Senatore:2017hyk,Baumann:2010tm,Assassi:2014fva,Assassi:2015jqa,Chen:2020zjt,Blas:2015qsi,Blas:2016sfa,McDonald:2006mx,McDonald:2009dh,Agarwal:2020lov,Rubira:2023vzw,Rubin:2023ovl,MoradinezhadDizgah:2020whw,Desjacques:2018pfv,Desjacques:2016bnm,Mirbabayi:2014zca}, see~\cite{Ivanov:2022mrd} for a recent review). Facilitated by fast numerical codes (including \textsc{class-pt}, \textsc{pybird}, \textsc{velocileptors}, \textsc{folps} and \textsc{class one-loop} \citep{classpt,DAmico:2020kxu,Chen:2020fxs,Noriega:2022nhf,Moretti:2023drg,Linde:2024uzr}), this EFT-based `full-shape' approach (also known as `direct modeling'), unlocks the potential of galaxy surveys be allowing them to be analyzed in the same manner as the CMB. This has led to a wide variety of constraints on dark energy, neutrinos, dark matter interactions, novel particles, inflationary physics, and beyond \citep[e.g.,][]{Ibanez:2024uua,DAmico:2022osl,DAmico:2022gki,Ivanov:2024hgq,Ivanov:2023qzb,Cabass:2024wob,Zhang:2021yna,Chen:2022jzq,Cabass:2022ymb,Chen:2021wdi,Philcox:2021kcw,Holm:2023laa,Wadekar:2020hax,Colas:2019ret,Holm:2023laa,Ivanov:2019hqk,Chen:2024vuf,He:2023oke,He:2023dbn,Xu:2021rwg,Chudaykin:2022nru,Chudaykin:2020ghx,Zaborowski:2024wpo,Ishak:2024jhs,DESI:2025ejh,DESI:2025euz,DESI:2024jis,DESI:2024hhd}.

Currently, we are in the era of stage-four spectroscopic experiments, including the Dark Energy Spectroscopic Instrument (DESI; \citep{Aghamousa:2016zmz}), \textit{Euclid} \citep{Laureijs:2011gra}, the Vera C. Rubin observatory (Rubin; \citep{LSST:2008ijt}), and the Nancy Grace Roman space telescope (Roman; \citep{Spergel:2015sza}). These promise to revolutionize the spectroscopic field, providing redshifts for over a hundred million galaxies and quasars across a broad range of redshifts. After just a single year of operation, DESI already delivered the most precise constraints on the BAO and growth parameters to date \citep{DESI:2024uvr,DESI:2024lzq,DESI:2024jis,DESI:2024mwx,DESI:2024hhd}, with significantly tighter constraints expected soon (and present already for the BAO \citep{DESI:2025zgx,DESI:2025zpo}). Besides placing tight constraints on $\Lambda$CDM parameters (through EFT modeling \citep[e.g.,][]{Maus:2024dzi}), this has yielded tantalizing evidence for non-standard cosmological expansion histories \citep{DESI:2024hhd,DESI:2025zgx}.

Recently, DESI announced the public release of their year-one data in Data Release 1 (DR1; \citep{DESI:2025fxa}). This includes the processed galaxy catalogs for four types of galaxy populations (bright galaxies, luminous red galaxies, emission line galaxies, and quasars), accompanied by systematic weights and collections of random points, which can be used to model the survey geometry. Turning this data into constraints on cosmology is not a simple task however. In particular, the clustering statistics (\textit{i.e.}\ the power spectra or two-point functions) were not included in the public release, and neither were the covariance matrices required to form the experimental likelihood \citep[e.g.,][]{Forero-Sanchez:2024bjh}. Whilst the former can be generated using public codes, this is non-trivial, particularly for statistics beyond the power spectrum. Furthermore, the data release did not include simulations, which can be used both in consistency checks of the pipeline and to form covariance matrices.

\begin{figure}[!t]
	\centering
	\includegraphics[width=0.6\textwidth]{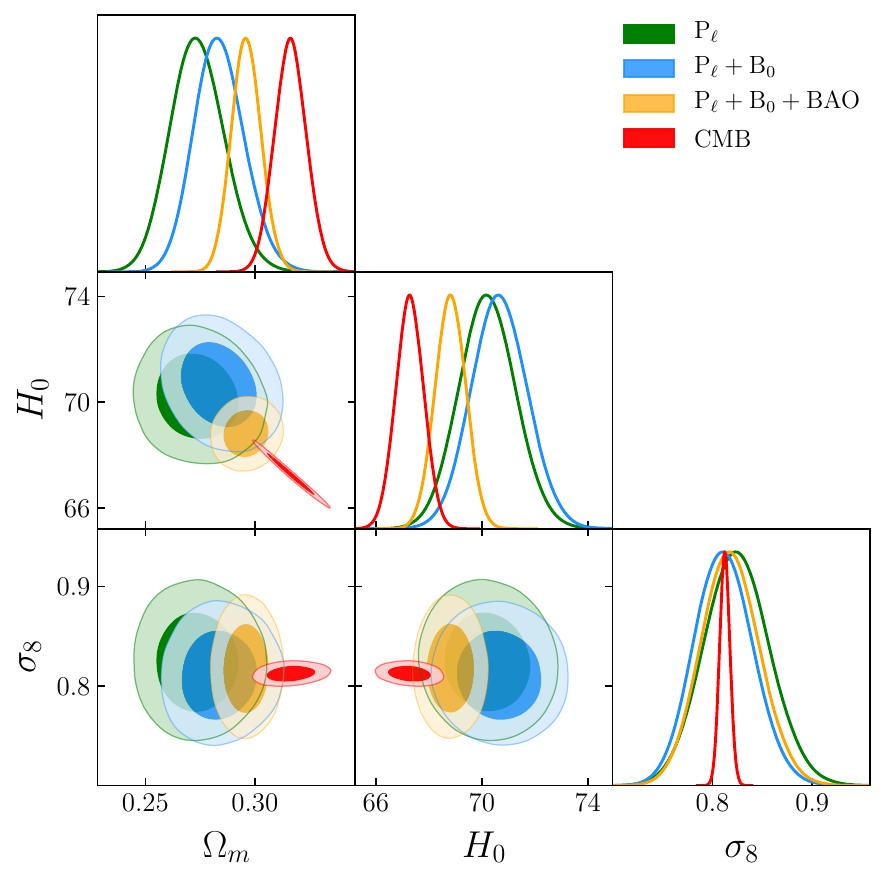}
	\caption{\textbf{Summary of Results}: Projected constraints on the $\Omega_m$, $H_0$ and $\sigma_8$ parameters in the $\ld$ model, using the following data combinations: the DESI DR1 redshift-space power spectrum $P_\ell$ (green), adding the DESI DR1 bispectrum monopole $B_0$ (blue), and including the DESI DR2 BAO measurements (orange). We include a BBN prior on $\omega_b$ and a wide prior on $n_s$ in all cases. The power spectrum and bispectrum measurements are new to this work; for the BAO, we take the latest results from DESI DR2 \citep{DESI:2025zgx}, assuming negligible cross-correlation with the DR1 dataset (as validated in Appendix \ref{app:cross}). Results obtained using DR1 BAO (including the cross-correlation) are presented in Appendix \ref{app:joint-bao}. For comparison we also show the \textit{Planck} 2018 CMB constraints, including lensing reconstruction from both the \textit{Planck} \textsc{npipe} PR4 maps and ACT DR6 (red). The DESI contours, which are the main result of this work, show good agreement with the CMB, particularly when additional information such as the bispectrum and BAO are included. The numerical constraints are listed in Tab.\,\ref{tab:main}.
    }
    \label{fig:main}
\end{figure}

In this series, we present an independent reanalysis of the DESI DR1 full-shape data. Starting from the galaxy catalogs, we compute summary statistics via quasi-optimal estimators using the \polybin code, analogous to the pseudo-$C_\ell$ statistics used in the CMB community \citep{Hivon2002,Alonso:2018jzx}. These extend the standard estimators \citep[e.g.,][]{Hand:2017irw,Scoccimarro:2015bla}, allowing for approximate `de-windowing' of the data, as well as efficient computation of the residual window functions and the masked Gaussian covariance matrices \citep{Philcox:2024rqr,Philcox:2020vbm,Philcox:2021ukg}. The fiducial DESI analysis focussed on the galaxy two-point function \citep{DESI:2024jis}; here, we include both the power spectrum multipoles and (large-scale) bispectrum monopole in our baseline analysis (see also \citep{NovellMasot:2025fju}), and account for a wide variety of systematic effects including stochasticity, integral constraints and wide-angle-effects in all relevant statistics. 
In addition, we use a more general theory model in our analysis compared to \citep{DESI:2024jis}, which does not fix to zero any EFT parameters (such as the cubic tidal bias
$b_{\Gamma_3}$ or the scale-dependent shot-noise amplitude). To avoid the bias induced by fiber-collisions \citep[e.g.,][]{Bianchi:2024fnl,Pinon:2024wzd}, we additionally introduce a novel stochastic scheme for mitigating fiber collisions in the bispectrum. Combining this dataset with robust theory models derived from the EFT, we place tight constraints on $\Lambda$CDM both from the full-shape data alone and in concert with CMB measurements and BAO (from both DR1 and DR2). Future work will include discussion of non-minimal cosmological models,
additional statistics, and beyond. The main results of this work are summarized in Fig.\,\ref{fig:main}\,\&\,Tab.\,\ref{tab:main}, comparing $\ld$ constraints from the full-shape power spectrum and bispectrum to those incorporating BAO and CMB information. 

\vskip 6pt
The remainder of this work is as follows. Our main cosmological results are summarized in Fig.~\ref{fig:main} and in Tab.~\ref{tab:main}.
In \S\ref{sec: data} we discuss the datasets underlying this work, including the DESI DR1 galaxy catalogs. \S\ref{sec: pspec}\,\&\,\ref{sec: bspec} present a detailed overview of the power spectrum and bispectrum, including discussion of the underlying estimators, the treatment of various systematic effects, and the measured summary statistics. Our theory model is outlined in \S\ref{sec: theory} before we discuss data consistency tests in \S\ref{sec: consistency}. The main $\Lambda$CDM constraints are presented in \S\ref{sec: lcdm-results}, before we conclude in \S\ref{sec: conclusions}. Additional discussion of fiber collision modeling is given in Appendix \ref{app: pk-fc} with the full set of parameter constraints provided in Appendix \ref{app:tables}. Finally, Appendix \ref{app:cross} discusses the cross-covariance between DESI DR1 full-shape and DESI DR2 BAO, whilst Appendix \ref{app:joint-bao} presents a joint analysis using DR1 BAO.

\begin{table}[!t]
    \centering
  \begin{tabular}{|c|ccc|} \hline
    \textbf{Dataset} 
    & $\Omega_m$ 
    & $H_0$ 
    & $\sigma_8$
    \\
    \hline
    $P_\ell(k)$ 
    & $\quad 0.274_{-0.013}^{+0.012}\quad$ 
    & $\quad 70.22_{-1.06}^{+1.06}\quad$ 
    & $\quad 0.825_{-0.033}^{+0.033}\quad$ 
    \\
    $P_\ell(k)+B_0(k)$ 
    & $0.284_{-0.012}^{+0.010}$ 
    & $70.67_{-1.05}^{+1.05}$ 
    & $0.811_{-0.031}^{+0.028}$ 
    \\
    $P_\ell(k)+B_0(k)+{\rm BAO}$ 
    & $0.296_{-0.007}^{+0.007}$ 
    & $68.82_{-0.58}^{+0.58}$ 
    & $0.818_{-0.029}^{+0.029}$ 
    \\
    $\quad \rm CMB\quad$ 
    & $0.316_{-0.007}^{+0.007}$ %
    & $67.28_{-0.53}^{+0.53}$ %
    & $0.812_{-0.005}^{+0.005}$ %
    \\
  \hline
    \end{tabular}
    \caption{\textbf{Summary of Results}: Mean and 68\% confidence intervals on $\ld$ cosmological parameters from the main analyses of this work (matching Fig.\,\ref{fig:main}). Here, $P_\ell(k)$ and $B_0(k)$ represent the DESI DR1 power spectrum multipoles and bispectrum monopole measured using quasi-optimal estimators (\S\ref{sec: pspec}\,\&\,\ref{sec: bspec}) The BAO and CMB datasets are external to this work (from DESI DR2 and \textit{Planck} respectively) and are described in \S\ref{sec:data_extra}. A full summary of these analyses can be found in Tab.\,\ref{tab:main2}. Here and throughout, we quote $H_0$ posteriors are given in $\hun$ units.}
\label{tab:main}
\end{table}

Throughout this work we assume a \textit{Planck} 2018 fiducial cosmology with $\{h = 0.6736, \omega_b = 0.02237, \omega_{\rm cdm} = 0.1200, N_{\rm ur} = 2.0328, \omega_\nu = 0.00064420\}$ assuming a single massive neutrino \citep{Aghanim:2018eyx}. We additionally notate Fourier-space integrals via $\int_{\vk} \equiv (2\pi)^{-3}\int d\vk$.

\section{Data and methodology}\label{sec: data}
\subsection{DESI DR1}
\noindent In this work, we utilize the full galaxy survey dataset from the first public data release (DR1) of DESI \citep{DESI:2025fxa}. This comprises four samples, themselves split in redshift: the Bright Galaxy Survey (BGS), the Luminous Red Galaxies (LRG), the Emission Line Galaxies (ELG), and the $z<2.1$ quasars (QSO) \citep{DESI:2024aax}.\footnote{Formally, these correspond to the \texttt{BGS\_BRIGHT-21.5}, \texttt{LRG}, \texttt{ELG\_LOPnotqso} and \texttt{QSO} clustering samples available in version $1.5$ of the public DESI DR1 catalog.} We do not include any Lyman-$\alpha$ forest data, and we further omit the low-redshift tail of the ELG sample ($z<1.1$) due to its comparatively large angular systematics \citep{DESI:2024jis,Rosado-Marin:2024xte}. The key parameters describing each sample are given in Tab.\,\ref{tab: desi-chunks}. 

\begin{table}[]
    \centering
    \begin{tabular}{|l|c|c|c|c|c|c|}
    \hline
     & Redshift & $z_{\rm eff}$ & $\bar{n}$ [$h^{3}\mathrm{Mpc}^{-3}$] & $P_{\rm FKP}$ [$h^{-3}\mathrm{Mpc}^{3}$] & $A_{\rm NGC}$ [$\mathrm{deg}^2$] & $A_{\rm SGC}$ [$\mathrm{deg}^2$]\\\hline
     \textbf{BGS} & $[0.1,0.4]$ & $0.2953$ & $1.9\times 10^{-4}$ & $7\times 10^3$ & $5108$ & $2072$\\
      \textbf{LRG1} & $[0.4,0.6]$ & $0.5096$ & $2.1\times 10^{-4}$ & $1\times 10^4$ & 3465 & 1770\\
      \textbf{LRG2} & $[0.6,0.8]$ & $0.7059$ & $2.1\times 10^{-4}$ & $1\times 10^4$ & 3465& 1770\\
      \textbf{LRG3} & $[0.8,1.1]$ & $0.9184$ & $1.1\times 10^{-4}$ & $1\times 10^4$ & 3465& 1770\\
      \textbf{ELG2} & $[1.1,1.6]$ & $1.3170$ & $1.0\times 10^{-4}$ & $4\times 10^3$ & 2865& 1270\\
      \textbf{QSO} & $[0.8,2.1]$ & $1.4901$ & $2.2\times 10^{-5}$ & $6\times 10^3$ & 4541& 2536\\
    \hline
    \end{tabular}
    \caption{\textbf{Sample Definitions}: Description of the six galaxy catalogs used in this work, which match those of \citep{DESI:2024aax,DESI:2024jis}. For each, we include the redshift range, effective redshift, mean number density and Feldman-Kaiser-Peacock (FKP) weight, $P_{\rm FKP}$. Following \citep{DESI:2024jis}, we omit the ELG1 sample due to systematic contamination. The effective redshifts are computed according to $z_{\rm eff} = \int dV\,z\,n^2(z) /\int dV n^2(z)$ for weighted number density $n(z)$. The mean density, $\bar{n}$, is computed approximately from the inverse power spectrum shot-noise. The $A_{\rm NGC, SGC}$ columns indicate the effective area of the Northern and Southern parts of the catalog.}\label{tab: desi-chunks}
\end{table}

For each sample, the public catalog contains both galaxies and randoms, with the latter (which have much greater density than the galaxies) emulating the survey selection function without clustering. Moreover, each is split into the Northern and Southern galactic cap (hereafter NGC and SGC). Where relevant, we will combine the two datasets weighting by the effective areas, \textit{i.e.}\ $X_{\rm tot} = (A_{\rm NGC}X_{\rm NGC}+A_{\rm SGC}X_{\rm SGC})/(A_{\rm NGC}+A_{\rm SGC})$ for some statistic $X$. 

Given the raw samples, we first convert angles and redshifts into comoving positions using our fiducial (\textit{Planck} 2018) cosmology. These catalogs define a continuous data field, $d$ and a background density $n$:
\beq\label{eq: raw-data}
	d(\vx) = \sum_{i=1}^{N_g}w_{g,i}\delta_{\rm D}(\vx-\vx_{g,i}) - \alpha\sum_{i=1}^{N_r}w_{r,i}\delta_{\rm D}(\vx-\vx_{r,i}), \qquad n(\vx) = \alpha\sum_{i=1}^{N_r}w_{r,i}\delta(\vx-\vx_{r,i}),
\eeq
where $\{w_{g,i},\vx_{g,i}\},\{w_{r,i},\vx_{r,i}\}$ are the weights and positions of the galaxies and randoms respectively and $\alpha = \sum_{i=1}^{N_g} w_{g,i}/\sum_{i=1}^{N_r}w_{r,i} \ll 1$.\footnote{The precise number of randoms depends on the sample of interest, but is always $\gg 10\times$ that of the number of galaxies.} In practice, we paint the discrete particles to a grid using a triangle-shaped-cell interpolation scheme using \textsc{nbodykit} \citep{Hand:2017pqn}, then deconvolve the window function in Fourier-space. The grid is chosen by demanding that the Nyquist frequency is at least $0.4\hMpc$ for power spectra and $0.2\hMpc$ for bispectra, with no less than $1000\Mpch$ padding in each dimension.

Following \citep{DESI:2024aax}, we define the weights as
\beq
	w = w_{\rm comp}w_{\rm zfail}w_{\rm imsys}w_{\rm FKP}, \qquad w_{\rm FKP}(\vx) = \left[1+\bar{n}(\vx)P_{\rm FKP}\right]^{-1}
\eeq
which correspond to completeness, redshift-failure, imaging systematics, and optimality weights, respectively. The latter are derived assuming the FKP form \citep{Feldman:1993ky}, and assume a (sample-dependent) constant fiducial power spectrum as specified in Tab.\,\ref{tab: desi-chunks} (taken from \citep{DESI:2024aax}), as well as an expected background galaxy density $\bar{n}$. Whilst these weights include a correction for angular systematics, such as galactic dust extinction, they do not attempt to account for pairwise `fiber-collision' systematics, which will be discussed below. For the bispectrum, we include an extra weight proportional to $n(z)^{-1/3}$, where $n(z)$ is the weighted radial galaxy density. This ensures that both statistics have the same effective redshift, as discussed in \S\ref{sec: theory}.

\subsection{Additional Datasets}
\label{sec:data_extra}

\paragraph{BAO}
\noindent We add BAO measurements from DESI DR2~\cite{DESI:2025zgx,DESI:2025zpo}, based on post-reconstruction clustering measurements in configuration space.
This dataset accumulates 14 million galaxy and quasars spectroscopic redshift observations, along with 820,000 Lyman-$\alpha$ forest spectra and their cross-correlation with 1.2 million quasar positions. 
They are divided into various redshift bins given by the different combinations of DESI tracers (similar to Tab.\,\ref{tab: desi-chunks}): the BGS in the redshift range $0.1<z<0.4$ ($z_\eff=0.295$), two LRG samples in $0.4<z<0.6$ ($z_\eff=0.51$) and $0.6<z<0.8$ ($z_\eff=0.706$), a combined LRG + ELG tracer in $0.8<z<1.1$ ($z_\eff=0.934$), an ELG sample in $1.1<z<1.6$ ($z_\eff=1.321$), a quasar sample covering $0.8<z<2.1$ ($z_\eff=1.481$), as well as Lyman-$\alpha$ forest auto-correlation and cross-correlation with the QSOs, both spanning a higher redshift range $1.8<z<4.2$ (and not included in the DR1 full-shape analysis).
For BGS the angle-average distance ratio $D_V/r_{\rm drag}$ is reported, while for the rest of the BAO DR2 measurements a combination of correlated measurements of $D_H/r_{\rm drag}$ and $D_M/r_{\rm drag}$ is used. 
In the remainder of this work, we will denote the DESI BAO DR2 measurements as ``BAO''. 

To combine the full-shape analysis of the DR1 pre-reconstruction statistics with the post-reconstruction DR2 BAO measurements, we require a joint covariance between the two probes. 
In the official DR1-only full-shape-plus-BAO analysis \citep{DESI:2024jis,DESI:2024hhd}, this covariance was computed from 1000 \textsc{EZmock} simulations, which effectively capture the relevant cross-correlations.
At the time of writing, the DESI mock catalogues are not yet available.
Notably, the correlation between the pre-reconstructed full-shape power spectrum and the BAO parameters is weak both due to the marginalization over broadband parameters and the added signal-to-noise afforded by reconstruction; for BOSS DR12, this was on the order of $10\%$ at $k > 0.02\hMpc$ \cite{Philcox:2020vvt}.
Compared to DR1, the effective survey volumes in DESI DR2 have increased substantially -- by a factor of 1.8 for QSO and up to 3.1 for ELG at $z_\eff=1.321$ -- as such, we expect the cross-correlation between the DR1 full-shape and the DR2 BAO measurements to be suppressed accordingly. Moreover, we do not expect significant correlations with the bispectrum dataset, since we restrict to $k_{\rm max}^B = 0.08\hMpc$, excising most of the BAO information.

For the above reasons, we neglect any cross-correlations between DR1 full-shape and DR2 post-reconstruction BAO measurements in this paper. This assumption is tested in Appendix \ref{app:cross}, and found to induce only small ($\lesssim0.4\sigma$) shifts in the output cosmological parameters. As a further test, we perform a pure DR1 full-shape-plus-BAO analysis, utilizing an analytic cross-covariance. These results are described in Appendix \ref{app:joint-bao} and are in good agreement with those from the corresponding DESI analysis \citep{DESI:2024hhd} (though now include the bispectrum).

\vskip 6pt
\paragraph{CMB}
To facilitate comparison between probes, we employ the Cosmic Microwave Background (CMB) data from the official \textit{Planck} 2018 release~\cite{Planck:2018nkj}, which includes the high-$\ell$ \textsc{plik} TT, TE, and EE spectra, the low-$\ell$ \textsc{SimAll} EE and low-$\ell$ \textsc{Commander} TT likelihoods. Additionally, we include measurements of the lensing potential auto-spectrum $C_\ell^{\phi\phi}$ from {\it Planck}'s \textsc{npipe} PR4 maps~\cite{Carron:2022eyg} and the Atacama Cosmology Telescope (ACT) Data Release 6 (DR6)~\cite{ACT:2023kun,ACT:2023dou}.\footnote{The likelihood is publicly available at \href{https://github.com/ACTCollaboration/act\_dr6\_lenslike}{https://github.com/ACTCollaboration/act\_dr6\_lenslike}; we use \textsc{actplanck\_baseline} option.}
Hereafter, we will refer to these CMB datasets simply as ``CMB''.

\vskip 6pt
\paragraph{External Priors}
In the analyses which do not include the CMB information, we apply two external priors as in the original EFT analysis of BOSS data \cite{Ivanov:2019pdj}. First, we impose a Gaussian prior on the physical baryon density $\omega_b$ from Big Bang Nucleosynthesis (BBN), given by
\be
\label{bbn}
\omega_b\sim\mathcal{N}(0.02218,0.00055^2)
\ee
Specifically, we adopt results from a recent study~\cite{Schoneberg:2024ifp} that employs the new \textsc{PRyMordial} code~\cite{Burns:2023sgx} to calculate the predictions while marginalizing over uncertainties in the reaction rates.
Secondly, we add a weak Gaussian prior on the spectra index $n_s$, centered at the \textit{Planck} mean value with a standard deviation ten times larger than that of the posterior from the \textit{Planck} baseline analysis~\cite{Planck:2018nkj}:
\be
\label{ns}
n_s\sim\mathcal{N}(0.9649,0.042^2)
\ee
Our priors on $\omega_b$ and $n_s$ match those used in the DESI full-shape analyses~\cite{DESI:2024hhd,DESI:2025ejh}, facilitating direct comparison with the collaboration results. 
When performing analyzing CMB data (either alone or with galaxy clustering), we do not apply the above priors as the parameters are already tightly constrained by the CMB.

\section{Power Spectra}\label{sec: pspec}
\noindent In this work, we estimate the power spectrum and bispectrum using the quasi-optimal estimators discussed in \citep{Philcox:2024rqr} (building on \citep[e.g.,][]{Hamilton:2005kz,Hamilton:2005ma,Hamilton:1999uw,1998ApJ...499..555T,Philcox:2020vbm,Philcox:2021ukg}, as well as \citep{2015arXiv150200635S,2011MNRAS.417....2S,Philcox:2023uwe,Philcox:2023psd,Philcox:2025bvj} for analogous CMB work). Whilst these are formulated somewhat differently to the approaches, the resulting estimators are similar, though feature a number of enhancements such as approximate (and full) `dewindowing', straightforward Gaussian covariance estimation, and arbitrary weighting schemes. Below, we will briefly outline the theoretical basis for the power spectrum estimator and its practical computation, paying particular attention to features not discussed in previous works. Further details can be found in \citep{Philcox:2024rqr}. In \S\ref{subsec: pk-practical} we discuss the practicalities of these estimators applied to DESI data and show the corresponding power spectra in \S\ref{subsec: pk-results}.

\subsection{Formalism}
\noindent Our goal is to estimate some set of power spectrum coefficients, $\{p_\alpha\}$ with $\alpha=1,\ldots,N_{\rm bins}$ (which we will later set equal to bandpowers), from a dataset $d$. We will assume that the dataset is linear in the underlying overdensity field of interest, $\delta_g$, such that
\beq\label{eq: pointing}
    d(\vx) = \int d\vy\,\P(\vx,\vy)\delta_g(\vy)+\text{noise},
\eeq
where $\P$ is a `pointing matrix' and the noise is uncorrelated with $\delta_g$. In the simplest approximation, this is just the mask. A general quadratic estimator for $p_\alpha$ is given by
\beq\label{eq: pk-estimator}
    \hat{p}_\alpha[d] \equiv \sum_{\beta=1}^{N_{\rm bins}}\F^{-1}_{\alpha\beta}\left[p_\beta^{\rm num}[d]-p_\beta^{\rm bias}\right],
\eeq
where $p^{\rm num}$ is a quadratic numerator, depending on the dataset, $d$, $p^{\rm bias}$ is a noise contribution, and $\F$ is a square normalization matrix \citep{Philcox:2024rqr,Philcox:2020vbm,Philcox:2021ukg}. The numerator is given explicitly by
\beq\label{eq: pk-num}
    p^{\rm num}_\alpha[d] &=& \frac{1}{2}\int d\vx\,d\vy\,\frac{\partial\xi(\vx,\vy)}{\partial p_\alpha}[\Si d]^*(\vx)[\Si d](\vy),
\eeq
where $\xi(\vx,\vy)\equiv\av{\delta_g(\vx)\delta_g^*(\vy)}$ is the fiducial two-point function of the overdensity and $\Si(\vx,\vy)$ is an (arbitrary) linear weighting scheme. The data-independent bias term and normalization can be written similarly:
\beq\label{eq: bias-fish}
    p^{\rm bias}_\alpha &=& \frac{1}{2}\int d\vx\,d\vy\,\frac{\partial\xi(\vx,\vy)}{\partial p_\alpha}[\Si\N\Sid](\vy,\vx)\\\nonumber
    \F_{\alpha\beta} &=& \frac{1}{2}\int d\vx\,d\vy\,d\vz\,d\vw\,\frac{\partial\xi(\vx,\vy)}{\partial p_\alpha}[\Si\P](\vy,\vz)\frac{\partial\xi(\vz,\vw)}{\partial p_\beta}[\Si \P]^*(\vx,\vw).
\eeq
where we define the covariance of the data as $\av{dd^\dagger} \equiv \C =  \P\xi\P^\dagger + \N$ for noise term $\N$. 

To compare data and theory it is useful to take the expectation of \eqref{eq: pk-num} with respect to the dataset, $d$. This yields
\beq\label{eq: av-pk-num}
    \mathbb{E}\left[p^{\rm num}_\alpha[d]-p^{\rm bias}_\alpha\right] &=& \frac{1}{2}\int d\vx\,d\vy\,\frac{\partial\xi(\vx,\vy)}{\partial p_\alpha}[\Si\P\xi\P^\dagger\Sid](\vy,\vx)\\\nonumber
    &\approx& \sum_\beta\F_{\alpha\beta}p_\beta,
\eeq
subtracting the shot-noise contribution, and expanding $\xi \approx \sum_{\alpha=1}^{N_{\rm bins}} p_\alpha(\partial\xi/\partial p_\alpha)$ in the second line. This implies that the full estimator is approximately unbiased with $\mathbb{E}\left[\hat{p}_\alpha[d]\right]\approx p_\alpha$, or, in other words, our choice of normalization approximately `unwindows' the statistic \citep[cf.][]{Philcox:2020vbm,Philcox:2021ukg}. In practice, however, it is useful to expand the correlation function in a finer binning, given that the mask-induced variations of the power spectrum within a bandpass can be significant. Defining a set of $N_{\rm fine}$ finely-binned theory band-powers $\{p_\iota^{\rm fine}\}$, we can write the expectation of the full estimator as
\beq
    \mathbb{E}\left[\hat{p}_\alpha[d]\right] &=& \sum_{\iota=1}^{N_{\rm fine}}[\F^{-1}\G]_{\alpha\iota}p_{\iota}^{\rm fine}, \qquad \G_{\alpha\iota} \equiv \frac{1}{2}\int d\vx\,d\vy\,d\vz\,d\vw\,\frac{\partial\xi(\vx,\vy)}{\partial p_\alpha}[\Si\P](\vy,\vz)\frac{\partial\xi(\vz,\vw)}{\partial p^{\rm theory}_\iota}[\Si \P]^*(\vx,\vw)
\eeq
which involves the $N_{\rm bins}\times N_{\rm fine}$ \textit{theory matrix} $\F^{-1}\G$. This is analogous to the \textit{pseudo}-$C_\ell$ approach used in CMB analyses \citep{Alonso:2018jzx} (and similar to that of \citep{Beutler:2021eqq}, which uses a simplified form of $\F$), and facilitates practical application of the residual window function as a matrix multiplication. 

Finally, we can define the covariance of the $\hat{p}_\alpha[d]$ estimator, working in the Gaussian regime (\textit{i.e.}\ ignoring the connected four-point function $\av{dddd}_c$):
\beq\label{eq: cov-pk}
    \mathrm{cov}\left({p}^{\rm num}_\alpha[d],{p}^{\rm num}_\beta[d]\right) &=&  \frac{1}{2}\int d\vx\,d\vy\,d\vz\,d\vw\,\frac{\partial\xi(\vx,\vy)}{\partial p_\alpha}[\Si\C\Sid](\vy,\vz)\frac{\partial\xi(\vz,\vw)}{\partial p_\beta}[\Si\C\Sid](\vw,\vx).
\eeq
In the limit of $\Si\to \P^\dagger\Ci$ (assuming that the inverse covariance exists), the covariance of $\hat{p}_\alpha$ is equal to the inverse normalization, $
\F^{-1}_{\alpha\beta}$: in this case, estimator \eqref{eq: pk-estimator} saturates its Cram\'{e}r-Rao bound and is optimal \citep{Philcox:2024rqr}.

\subsection{Specialization to DESI}
\noindent To practically implement the power spectrum estimator given in \eqref{eq: pk-estimator}, we require a number of survey-specific components including a fiducial form for the two-point function $\xi$ and noise covariance $\N$, a weighting scheme $\Si$, and knowledge of the response function of the data, $\P$ (\textit{i.e.}\ the mask). Below, we discuss each aspect in turn alongside some additional features of interest.

\subsubsection{Two-Point Function} 
\noindent To model the correlation function, we assume a simple form based on the end-point Yamamoto decomposition \citep{2006PASJ...58...93Y}:
\beq\label{eq: 2pf-expansion}
    \xi(\vx,\vy) &=& \frac{1}{2}\int_{\vk}\left[P(\vk;\vx)+P(-\vk;\vy)\right]e^{i\vk\cdot(\vx-\vy)}\\\nonumber
    &=&\frac{1}{2}\int_{\vk}\sum_{\ell=0,2,\ldots}^{\infty} P_\ell(k)\left[\mathcal{L}_\ell(\hk\cdot\hx)+\mathcal{L}_\ell(-\hk\cdot\hy)\right]e^{i\vk\cdot(\vx-\vy)},
\eeq
where $P(\vk;\vd)$ is the power spectrum with anisotropy with respect to $\vd$ and we (symmetrically) expand in Legendre polynomials in the second line.\footnote{For now, we consider even $\ell$: wide-angle effects can source odd $\ell$, as discussed below.} Inserting a set of band-powers $\{p_{\alpha}\}$ via $P_\ell(k;x) \to \sum_{a=1}^{N_k} \Theta_a(k)p_{\alpha}$ (where $\alpha\equiv\{\ell,a\}$), with $\Theta_a(k) = 1$ if $k\equiv|\vk|$ is in bin $a$ and zero else, we have
\beq
    \frac{\partial\xi(\vx,\vy)}{\partial p_\alpha} = \frac{1}{2}\int_{\vk}\Theta_a(k)\left[\mathcal{L}_\ell(\hk\cdot\hx)+(-1)^{\ell}\mathcal{L}_\ell(\hk\cdot\hy)\right]e^{i\vk\cdot(\vx-\vy)}.
\eeq
This can be trivially applied to an input field using Fourier transforms and the spherical harmonic addition theorem \citep[cf.][]{Bianchi:2015oia,Scoccimarro:2015bla}.

\subsubsection{Weighting Scheme}
\noindent As noted above, the optimal power spectrum estimator is achieved by setting $\Si = \mathsf{P}^\dagger \mathsf{C}^{-1}$. In practice, this is expensive to compute since the pixel covariance is high-dimensional. As shown in \citep{Philcox:2024rqr}, if one assumes a constant power spectrum, \textit{i.e.}\ $P(\vk) = P_{\rm FKP}$, the optimal choice becomes simply $\Si = \mathsf{1}$ (noting that we have already incorporated the usual FKP weights). In general, one could use $\Si$ to (linearly) remove contaminated modes in the data -- for example, one could filter out radial (angular) modes, which are most affected by fiber collisions (imaging systematics) \citep[e.g.,][]{Pinol:2016opt,Rosado-Marin:2024xte}. By construction, our estimators are unbiased for all choices of $\Si$, though the variances can differ. For the remainder of this work, we will set $\Si = \mathsf{1}$, following the above considerations.

\subsubsection{Shot-Noise}
\noindent The bias term $p^{\rm bias}_\alpha$ is sourced by Poisson effects in the dataset. As discussed in \citep{Philcox:2024rqr}, the noise covariance is given by $\N(\vx,\vy) = \delta_{\rm D}(\vx-\vy)n_2(\vx)$, where $n_2$ is the doubly-weighted background density, which can be estimated from the data and random catalogs:
\beq\label{eq: doubly-weighted-mask}
    n_2(\vx) &=& (\alpha_2+\alpha^2)\sum_{i=1}^{N_r}w_{r,i}^2\delta_{\rm D}(\vx-\vx_{r,i}), \qquad \alpha_2 = \sum_{i=1}^{N_g}w_{g,i}^2/\sum_{i=1}^{N_r}w_{r,i}^2,
\eeq
implicitly deconvolving the pixel-window, as before. Assuming $\Si=\mathsf{1}$ and simplifying the integrals analytically, bias term \eqref{eq: bias-fish} becomes:
\beq\label{eq: p-bias-ideal}
    p^{\rm bias}_\alpha &=& \frac{1}{2}\int d\vx\,n_2(\vx)\frac{\partial\xi(\vx,\vx)}{\partial p_\alpha} \sim \delta^{\rm K}_{\ell 0}\int_{\vk}\Theta_a(\vk)\left(\sum_{i=1}^{N_g}w_{g,i}^2+\alpha^2\sum_{i=1}^{N_r}w_{r,i}^2\right),
\eeq
which is isotropic and proportional to the bin-volume. Notably, we subtract only the idealized Poisson shot-noise from the data. Scale-dependence and non-Poisson behavior (which are difficult to predict \textit{a priori}) are incorporated in the theory model \S\ref{sec: theory}.

\subsubsection{Integral Constraints}
\noindent In the simplest approximation, the observed dataset is equal to the overdensity field $\delta_g$ multiplied by the background number density, $n$. In this case, the pointing matrix of \eqref{eq: pointing} is simply:
\beq\label{eq: pointing-simple}
    \P(\vx,\vy) = n(\vx)\delta_{\rm D}(\vx-\vy).
\eeq
In practice, the pointing matrix can be a little more subtle. When constructing the random catalog, one usually sets the radial distribution to that of the (suitably binned) galaxy sample. This leads to a `radial integral constraint' -- essentially, $d(\vx)$ contains no radial modes, since they are subtracted out by the random catalog. Following \citep{deMattia:2019vdg}, this can be modeled through the following asymmetric change to the pointing matrix:
\beq\label{eq: pointing-RIC}
    \mathsf{P}(\vx,\vy) = n(\vx)\left[\delta_{\rm D}(\vx-\vy)-\frac{n_{\rm IC}(\vy)\epsilon(x,y)}{\int d\vz\,n_{\rm IC}(\vz)\epsilon(y,z)}\right],
\eeq
where $n_{\rm IC}$ is the mask without FKP weights, and $\epsilon(x,y)=1$ if $x,y$ are in the same radial bin and zero else. This also encodes the global integral constraint, \textit{i.e.}\ that the mean overdensity across the survey is set to zero. This can be isolated by setting $\epsilon=1$.\footnote{To see the effects of the radial integral constraint, we can assume the plane-parallel limit with an infinitely fine radial binning, such that $\epsilon(x,y)\to \delta_{\rm D}(x_\parallel-y_\parallel)$. Applying to an arbitrary field $u(\vx)$,
\beq
    [\mathsf{P}u](\vx) = n(\vx)\left[u(\vx_\perp,x_\parallel)-\frac{\int d\vy_\perp\,n_{\rm IC}(\vy_\perp,x_\parallel)u(\vy_\perp,x_\parallel)}{\int d\vz_\perp\,n_{\rm IC}(\vz_\perp,x_\parallel)}\right];
\eeq
this simply projects out the $x_\parallel$ component, averaged over the survey volume.}

This can be computed by noting that $\epsilon(x,y) = \sum_{i=1}^{N_{z}}\Theta_i(x)\Theta_i(y)$, where $\Theta_i(x)$ is unity if $x$ is in bin $i$ and zero else (across $N_{z}$ redshift bins). Thus:
\beq
    [\mathsf{P}u](\vx) = n(\vx)u(\vx)-n(\vx)\sum_{i=1}^{N_{z}}\Theta_i(x)\frac{\int d\vy\,\Theta_i(y)n_{\rm IC}(\vy)u(\vy)}{\int d\vy\,\Theta_i(y)n_{\rm IC}(\vy)},
\eeq
which simply involves taking the average of $u$ across each redshift bin and combining. Following \citep{DESI:2025fxa}, we use equally spaced $z$-bins with $\delta z = 0.01$ for all samples except QSO, which has $\delta z =0.02$.\footnote{Notably, our treatment of integral constraints affects only the estimator normalizations (since $p^{\rm num}$ does not depend on $\P$. Formally, their presence leads to a slight change in the optimal weighting scheme $\Si$; this is tiny however and will not lead to appreciable change in signal-to-noise.}

\subsubsection{Wide-Angle Effects}
\noindent Due to the large angular footprint of DESI, higher-order terms in the power spectrum expansion of \eqref{eq: 2pf-expansion} can become important for low-redshift samples, particularly if we extend to larger scales than considered in our fiducial analysis. As discussed in \citep{Reimberg:2015jma,Castorina:2017inr,Beutler:2018vpe}, we may incorporate wide-angle effects in the power spectrum by expanding in both Legendre polynomials and in $kx \ll 1$, \textit{i.e.}\ 
\beq
    \xi(\vx,\vy) = \frac{1}{2}\int_{\vk}\sum_{\ell=0,1,\ldots}^\infty\sum_{n=0}^{\infty}P^{(n)}_\ell(k)\left[(kx)^{-n}\mathcal{L}_\ell(\hk\cdot\hx)+(-1)^\ell (ky)^{-n}\mathcal{L}_\ell(\hk\cdot\hy)\right]e^{i\vk\cdot(\vx-\vy)},
\eeq
where $n=0$ is the plane-parallel limit used previously. The resulting band-powers (which can now feature odd $\ell$, provided $\ell+n$ remains even) can be estimated similarly to the above.

The leading source of wide-angle effects is the asymmetry in the `end-point' line-of-sight definition. This usually dominates over the relativistic corrections (except on ultra-large scales not relevant to this work), and can be modeled analytically. As shown in \citep[Appendix B]{Pinon:2024wzd}, the resulting odd-$\ell$ $n=1$ corrections are fully determined by the even-$\ell$ $n=0$ bandpowers, such that the full power spectrum can be written (up to $n=1$)
\beq\label{eq: wide-angle-P}
    P(\vk;\vx) &=& P^{(0)}_0(k) + \mathcal{L}_1(\hk\cdot\hx)\frac{1}{kx}\left[-i\frac{3}{5}\left(3+k\partial_k\right)P_2^{(0)}(k)\right]+\mathcal{L}_2(\hk\cdot\hx)P^{(0)}_2(k)\\\nonumber
    &&\,+\mathcal{L}_3(\hk\cdot\hx)\frac{1}{kx}\left[-i\frac{3}{5}\left(2-k\partial_k\right)P_2^{(0)}(k) -i\frac{10}{9}\left(5+k\partial_k\right)P_{4}^{(0)}(k)\right]+\mathcal{L}_4(\hk\cdot\hx)P_4^{(0)}(k)+\cdots
\eeq
This implies that, instead of measuring the odd-multipoles directly from the data, one can simply measure the even-multipoles with a modified two-point function derivative, $\partial\xi/\partial p_\alpha$. Here, we include this modification in the $\G$ matrix which relates the theoretical and observed power spectra, which allows for our measurements to be wide-angle-corrected at the theory-level (using numerical derivatives to compute $\partial_k$).\footnote{Alternatively, one could utilize \eqref{eq: wide-angle-P} also in the estimator numerators and normalization, essentially debiasing the data rather than biasing the theory. Whilst this is the formally optimal choice, this leads to negligible change in the signal-to-noise and is more difficult to implement since we use wide $k$-space bins in the estimator.} This is analogous to the treatment of \citep{Beutler:2021eqq}, used in the DESI DR1 analyses \citep{DESI:2024aax}, though we avoid explicitly computing odd-$\ell$ correlators.

\subsubsection{Fiber Collisions}
\noindent Due to experimental limitations, DESI cannot measure the redshifts of pairs of galaxies below a certain angular separation (without multiple observations). Being a pairwise systematic, this cannot be mitigated using weights (in contrast to the imaging systematics discussed in \S\ref{sec: data}) -- in the official analyses, this is ameliorated by removing all pairs of points with angular separations $\theta<0.05\degree$ \citep{Pinon:2024wzd} (though many more nuanced approaches have also been suggested \citep[e.g.,][]{Pinol:2016opt,Guo:2011ai,Hahn:2016kiy,Bianchi:2017saf,Bianchi:2019ruq,Burden:2016cba,Bianchi:2024fnl}). Here, we adopt the same approach, which leads to a modification of the power spectrum numerator, normalization, and theory matrix.

In the presence of a $\theta$-cut, the power spectrum numerator transforms as $p_\alpha^{\rm num}\to p_\alpha^{\rm num}-p_\alpha^{\rm fc}$, where
\beq\label{eq: pk-fc}
    p^{\rm fc}_\alpha[d] &=& \frac{1}{2}\int d\vx\,d\vy\,\frac{\partial\xi(\vx,\vy)}{\partial p_\alpha}d(\vx)d(\vy)\Phi(\vx,\vy),
\eeq
setting $\Si = \mathsf{1}$ as above, and defining a function $\Phi(\vx,\vy)$ that is unity if $\vx,\vy$ have an angular separation below $0.05\degree$ (setting $\Phi(\vx,\vx)=0$ to avoid shot-noise). Utilizing the discrete form for the data \eqref{eq: raw-data}, this can be written as a sum over pairs of points:
\beq\label{eq: pk-fc-discrete}
    p^{\rm fc}_\alpha &=& \frac{1}{2}\sum_{ij:\Phi(\vx_i,\vx_j)=1}^{N_g+N_r}w_iw_j\frac{\partial\xi}{\partial p_\alpha}(\vx_i,\vx_j)
\eeq
where we stack both galaxies and randoms, with weights $\{w_{g,i}\}$ and $\{-\alpha w_{r,i}\}$ respectively. Ignoring wide-angle effects, the two-point function takes a simple form (from \ref{eq: 2pf-expansion}):
\beq
    \frac{\partial\xi}{\partial p_\alpha}(\vx_i,\vx_j) = \frac{\left[\mathcal{L}_{\ell}(\hx_{ij}\cdot\hx_i)+(-1)^{\ell}\mathcal{L}_\ell(\hx_{ij}\cdot\hx_j)\right]}{2} i^\ell \int\frac{k^2dk}{2\pi^2}\Theta_a(k)j_\ell(kx_{ij}).
\eeq
with $\vx_{ij} \equiv \vx_i-\vx_j$. Since this sums only over close pairs of galaxies, it can be efficiently computed. 

The change to the normalization can be computed similarly, if one ignores the integral constraint.\footnote{This is justified by noting that the integral constraint is important only on ultra-large scales, whilst the fiber collision effects (which are already fairly small) dominate on smaller scales, due to the angular cut.} Inserting \eqref{eq: pointing-simple} into \eqref{eq: bias-fish} and using the discrete form of the mask \eqref{eq: raw-data}:
\beq\label{eq: fish-fc}
    \F^{\rm fc}_{\alpha\beta} &=& \frac{1}{2}\int d\vx\,d\vy\,\frac{\partial\xi(\vx,\vy)}{\partial p_\alpha}n(\vy)\frac{\partial\xi(\vy,\vx)}{\partial p_\beta}n(\vx)\Phi(\vx,\vy)\\\nonumber
    &=& \frac{\alpha^2}{2}\sum_{ij:\Phi(\vx_i,\vx_j)=1}^{N_r}w_{r,i}w_{r,j}\frac{\partial\xi}{\partial p_\alpha}(\vx_{r,i},\vx_{r,j})\frac{\partial\xi}{\partial p_\beta}(\vx_{r,j},\vx_{r,i})
\eeq
summing only over randoms. The expression for $\G^{\rm fc}_{\alpha\beta}$ is analogous. Practically, these can be efficiently computed by first histogramming all close pairs of galaxies and randoms in bins of $x_{ij}$ (weighted by one or two Legendre polynomials) then assembling the full result after as in \citep{Pinon:2024wzd} (avoiding the need for $\mathcal{O}(N_{r}^2)$ spherical Bessel transforms). Notably, we do not need to perform any rotation on the datavector or window function as done in \citep{DESI:2024jis,Pinon:2024wzd}; this is automatically handled by our modification to the normalization, $\F-\F^{\rm fc}$.

\subsubsection{Covariance}
\noindent As shown in \eqref{eq: cov-pk}, the Gaussian covariance of $p_\alpha^{\rm num}$ takes the same form as the normalization matrix, but with $\Si\P$ replaced with $\Si\C\Sid$.\footnote{In the limit of $\Si \to\P^\dagger\C^{-1}$, the two matrices are equivalent (and the estimator is optimal). In practice, we assume the FKP form for $\Si ( = 1)$, which implies that the covariance and power spectra differ only by a factor $\sim \prod_{i=1}^3(1+\bar{n}P(k_i))/(1+\bar{n}P_{\rm FKP})$, sourced by the scale-dependence of the power spectrum.} To compute this, we require a fiducial pixel covariance, which is defined by
\beq
    \C(\vx,\vy) = \int d\vz\,d\vw\,\P(\vx,\vz)\xi(\vz,\vw)\P^\dagger(\vw,\vy)+n_2(\vx)\delta_{\rm D}(\vx-\vy),
\eeq
where $\P$ is the pointing matrix defined in \eqref{eq: pointing-RIC} and $n_2$ is the doubly-weighted mask of \eqref{eq: doubly-weighted-mask}. The two-point function $\xi$ can be computed using \eqref{eq: 2pf-expansion}, given a fiducial set of power spectrum multipoles $P_\ell(k)$. As discussed in \citep{Philcox:2024rqr} this is more expensive to compute than for $\F$, due to the additional Fourier-transforms required to capture the line-of-sight anisotropy present in $\xi$. In practice, a number of additional sources of error are propagated into the covariance; these are discussed in \S\ref{subsec: pk-practical}.

\subsubsection{Limiting Forms}\label{subsubsec: pk-limit}
\noindent Whilst the above estimators are lengthy and contain a number of non-trivial components, they reduce to standard forms in certain limits. In particular, inserting the two-point function definition into \eqref{eq: pk-num} and simplifying, using $\Si=\mathsf{1}$ and omitting wide-angle effects and fiber collisions, the power spectrum numerator can be written
\beq
    p^{\rm num}_\alpha &\to& \frac{1}{4}\int_{\vk}\Theta_a(k)\left[F^*_\ell[d](\vk)F_0[d](\vk)+(-1)^{\ell}F_{\ell}[d](\vk)F_0^*[d](\vk)\right]\\\nonumber
    F_\ell[d](\vk) &=& \sum_m\frac{4\pi}{2\ell+1}Y^*_{\ell m}(\hk)\int d\vx\,e^{-i\vk\cdot\vx}Y_{\ell m}(\hx)d(\vx).
\eeq
Up to a prefactor, this is equivalent to the form given in \citep{Hand:2017irw} (see also \citep{Scoccimarro:2015bla,Bianchi:2015oia}) and used in the DESI power spectrum analyses \citep{DESI:2024aax}. The same is true for the shot-noise contribution given in \eqref{eq: p-bias-ideal}.\footnote{In practice, we perform Fourier-space integrals on the grid instead of numerically. This allows us to account for discreteness effects.} In expectation, the combined power spectrum numerator yields
\beq
    \mathbb{E}[p^{\rm num}_\alpha[d]-p^{\rm bias}_\alpha] &\to& \frac{1}{2}\int_{\vk\,\vk'}\Theta_a(k)\mathcal{L}_\ell(\hk\cdot\hat{\vd})\left|n(\vk-\vk')\right|^2P(\vk'),
\eeq
working in the plane-parallel limit for simplicity (with global line-of-sight $\hat{\vd}$) and ignoring the integral constraint. This simply convolves the true power spectrum with that of the mask, as expected \citep[e.g.,][]{Beutler:2018vpe,Beutler:2016arn}. 

Under the above assumptions, the normalization matrix can be written
\beq
    \F_{\alpha\beta} &\to& \frac{1}{2}\int_{\vk}\int_{\vk'}\Theta_a(k)\Theta_b(k')\mathcal{L}_\ell(\hk\cdot\hat{\vd})\mathcal{L}_{\ell'}(\hk'\cdot\hat{\vd})|n(\vk-\vk')|^2
\eeq
for $\alpha\equiv\{a,\ell\}, \beta\equiv\{b,\ell'\}$. This can be compared to the analogous form used in traditional `windowed' estimators \citep[e.g.,][]{DESI:2024aax,Hand:2017irw}:
\beq\label{eq: fish-win}
    \F^{\rm win}_{\alpha\beta} &=& \frac{1}{2}\int d\vx\,n^2(\vx)\int_{\vk}\Theta_a(k)\Theta_b(k)\mathcal{L}_\ell(\hk\cdot\hat{\vd})\mathcal{L}_{\ell'}(\hk\cdot\hat{\vd}),
\eeq
which is diagonal and proportional to the bin-volume and the mean squared mask. In practice, one could use either form to normalize the power spectrum, provided that we normalize the theory matrices analogously (\textit{i.e.}\ using $\F^{-1, \rm win}\G$ for the windowed estimators). In this work, we focus on the unwindowed estimators, since they approximately deconvolve the mask (with residual effects by the theory matrix), which allows for more physical $k$-space cuts.\footnote{As noted in \citep{Philcox:2024rqr}, $\F$ can be efficiently computed alongside $\G$, thus the `unwindowed' approach is no more computationally expensive than the usual windowed scheme.} For comparison, we show results obtained using windowed estimators in \S\ref{subsec:window}.

\subsubsection{Summary}
\noindent Our power spectrum estimators account for the following observational and systematic effects (most of which are treated analogously in the official DESI analyses):
\begin{itemize}
    \item \textbf{Anisotropy}: We perform a Legendre decomposition of the power spectrum, adopting the end-point line-of-sight definition and restricting to even $\ell$.
    \item \textbf{Shot-Noise}: We subtract Poisson shot-noise from the power spectra numerator.
    \item \textbf{Integral Constraints}: We account for the radial integral constraint in both the estimator normalization and theory matrix. This automatically includes the global integral constraint.
    \item \textbf{Wide-Angle Effects}: We forward-model the leading-order wide-angle effects by modifying the theory (as in \citep{Beutler:2021eqq}).
    \item \textbf{Fiber Collisions}: We subtract off contributions to the estimator numerator, normalization, and theory matrix, from pairs of points with separations below $0.05\degree$. This is computed by pair-counting. 
    \item \textbf{Covariance}: We compute the Gaussian covariance using \eqref{eq: cov-pk}, which accounts for the mask and shot-noise. This involves a fiducial power spectrum, which we set to the best-fit model obtained from a preliminary analysis of the DESI data.
\end{itemize}

\subsection{Practical Implementation}\label{subsec: pk-practical}
\noindent As discussed in \citep{Philcox:2024rqr}, the power spectrum estimators described above can be efficiently estimated using fast Fourier transforms and (for the bias and normalization terms) Monte Carlo methods. In this work, we use the \polybin implementation of \citep{Philcox:2024rqr,polybin3d}, which has been extended to include wide-angle effects, covariances, and integral constraints. To compute the fiber collision corrections, we additionally develop a fast \textsc{cython} code, which efficiently accumulates all pairs of particles within the required separation. The \polybin code is written in a combination of \textsc{python}, \textsc{cython}, and \textsc{jax} and is heavily optimized to make full use of the available CPU and/or GPU resources. 

Power spectra are measured using the following binning: $\delta k = 0.005\hMpc$ for $k\in[0.005,0.02)\hMpc$, $\delta k = 0.01\hMpc$ for $k\in[0.02,0.25)\hMpc$, $\delta k = 0.02\hMpc$ for $k\in[0.25,0.31)\hMpc$, across $\ell\in\{0,2,4\}$, resulting in a total $N_{\rm bins}=3\times 29 = 87$ bins. This is designed to have good coverage over the primary wavenumber range of interest with additional bins at low- and high-$k$ to capture any mask-induced leakage. For the theory matrix, we adopt a fine binning with $\delta k = 0.001\hMpc$ for $k\in[0.0035,0.4005)$ such that $N_{\rm fine}=3\times 397 = 1191$, with boundaries set by the largest fundamental mode and Nyquist frequency. In practice, we restrict the datavector to $k\in [0.02,0.2)\hMpc$ after assembling the estimator (following \citep{DESI:2024jis}), noting that small-scales are noise-dominated and can be difficult to model perturbatively.

For the largest chunk (QSO-NGC), \polybin requires $\approx 10$ seconds to compute the power spectrum numerator, $p_{\alpha}^{\rm num}[d]$ on a 64-core machine, using the \textsc{mkl\_fft} package to perform Fourier transforms. To assemble the fiber collision term, we perform a pair-count as above, requiring around $1000$ seconds, though this is dominated by the (data-independent) random counts. We compute the bias term, $p_\alpha^{\rm bias}$, by averaging over $50$ Monte Carlo realizations, each of which require $\approx 60$ seconds.\footnote{We use a random subset of $N_r = 10N_g$ randoms for the pair-counting operations to reduce computational costs.} Similarly, the normalization and theory terms (which are computed together) require $2500$ seconds for each of $20$ realizations. The numerator covariance is somewhat more expensive to compute: we require $\approx 8000$ seconds for each of $20$ Monte Carlo iterations for the largest data chunk. This uses a fiducial power spectrum model, which is taken from an initial fit to the DESI data. In each case, the number of Monte Carlo realizations are chosen to ensure convergence to a small fraction of the experimental errors. 

Given the numerator covariance, $\mathbb{C} \equiv \mathrm{cov}(p_\alpha^{\rm num}[d],p_\beta^{\rm num}[d])$, we form the full covariance as follows:
\begin{itemize}
    \item We combine the two galactic caps according to $\mathbb{C}_{\rm tot}=(A_{\rm NGC}^2\mathbb{C}_{\rm NGC}+A_{\rm SGC}^2\mathbb{C}_{\rm SGC})/(A_{\rm NGC}+A_{\rm SGC})^2$ noting that the NGC and SGC datasets are uncorrelated. 
    \item We rescale the covariance according to $\mathbb{C}\to (\F-\F^{\rm fc})\F^{-1}\mathbb{C}\F^{-\dagger}(\F-\F^{\rm fc})^\dagger$ to account for the information loss due to fiber collisions, where $\F^{\rm fc}$ is the contribution to the normalization from close pairs.
    \item We add a correction for the finite number of Monte Carlo simulations used to estimate the noise term, $N_{\rm MC}$. This is computed as the variance across realizations, scaled by $1/N_{\rm MC}$.
    \item We similarly correct for the finite number of simulations used in the theory matrix, computed as the variance of $\G p^{\rm theory}$ across realizations, scaled by $1/N_{\rm MC}$, where $p^{\rm theory}$ is a fiducial theory model.
    \item For the QSO and ELG samples, we marginalize over imaging systematics by adding a term specified by $0.2^2\delta p_\alpha[d]\delta p_\beta[d]$, where $\delta p_\alpha[d] \equiv (p_\alpha^{\rm num}[d]-p_\alpha^{\rm bias}) - \F\F^{-1,\rm nw}(p_\alpha^{\rm num,nw}[d]-p_\alpha^{\rm shot,nw})$. Here, $p_\alpha^{\rm nw}$ are power spectra computed without systematic weights, and the $\F\F^{-1,\rm nw}$ term (which we computed using \eqref{eq: fish-win} for simplicity) accounts for the differences in normalization. This follows the treatment of \citep{DESI:2024jis,DESI:2024rkp}.\footnote{Unlike \citep{DESI:2024jis}, we do not subtract off any systematic bias from the numerator. This correction is small and can only be computed from simulations.}
    \item We form the covariance of $\hat{p}_\alpha$ according to $(\F-\F^{\rm fc})^{-1}\mathbb{C}(\F-\F_{\rm fc})^{-\dagger}$.
\end{itemize}
The resulting matrix is used in the likelihood model, and accounts for the systematics discussed above.\footnote{Note that we do not add an `HOD covariance' such as that suggested by \citep{DESI:2024rkp,DESI:2024jis} since our EFT-based theory model, coupled with the broad priors discussed in \S\ref{sec: theory}, should be able to capture any physical halo realization.} 
We stress that our covariance model does not include any non-Gaussian corrections, \textit{i.e.}\ the connected trispectrum. For the BOSS analyses, these terms were found to negligibly impact parameter inferences \citep{Wadekar:2020hax,Wadekar:2019rdu} (see also \citep{Oddo:2021iwq}), given the comparatively large shot-noise. We expect a similar result for DESI DR1, given that the shot-noise is similar and the non-Gaussian corrections are suppressed at higher redshifts. This approximation could be further tested once a suite of accurate simulations is available.

Our covariance model differs somewhat from that of the official DESI full-shape analysis, which used a sample covariance obtained from a suite of \textsc{ezmock} simulations \citep{DESI:2024jis,Forero-Sanchez:2024bjh}, which includes non-Gaussian contributions. As discussed in \citep{Forero-Sanchez:2024bjh}, this covariance does not match that expected from analytic theory (partly due to fiber collision effects), thus a rescaling factor of $10-40\%$ is applied based on configuration-space results \citep[cf.,][]{Rashkovetskyi:2024eik}. Here, we do not apply such a rescaling since (a) it is specific to the simulation set, (b) our covariance is already calibrated to the actual data, via the input power spectrum model, and (c) any correction should be scale-dependent. Adding the DESI rescaling factor to the analyses of \S\ref{sec: lcdm-results} has only a minor effect, with parameter errorbars increasing by $\lesssim 7\%$. This additionally implies that any biases from the omitted non-Gaussian covariance corrections are small.

\subsection{Results}\label{subsec: pk-results}

\begin{figure}
    \centering
    \includegraphics[width=0.95\linewidth]{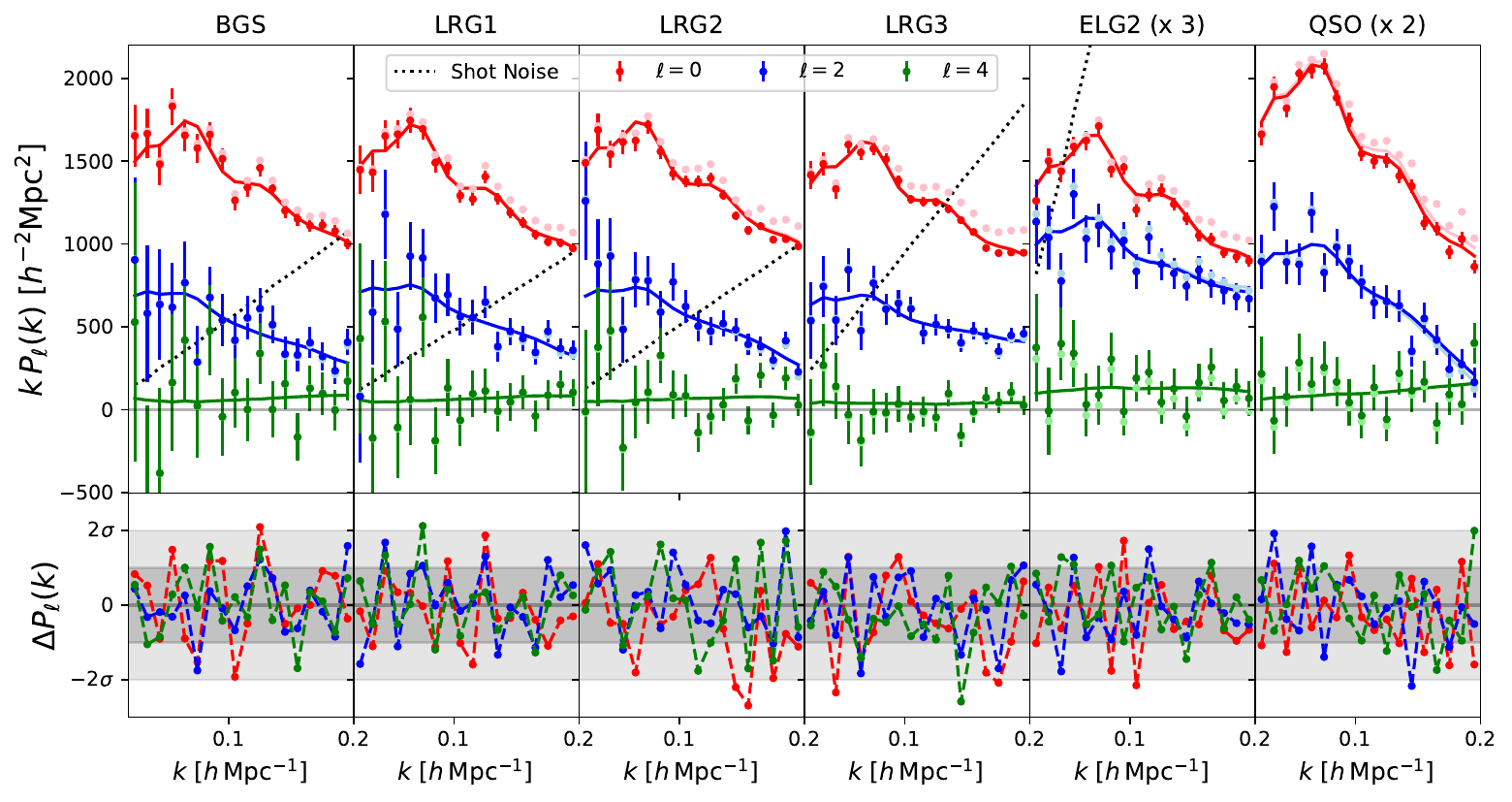}
    \caption{\textbf{DESI $\boldsymbol{P_\ell(k)}$}: Power spectra for the six DESI chunks used in this work. Each panel shows the power spectrum monopole (red), quadrupole (blue), and hexadecapole (green), measured using the \polybin code, as described in \S\ref{sec: pspec}. We include an off-diagonal normalization matrix to approximately deconvolve the window function, which leads to somewhat larger scatter than for conventional windowed estimators, though with no loss of signal-to-noise. For visualization, we rescale the ELG2 and QSO results by a constant factor as shown in the title. The solid lines indicate the best-fit theoretical model, obtained from a joint analysis of the DESI power spectra and bispectra (\S\ref{sec: lcdm-results}). Errorbars indicate the Gaussian errorbars, with corrections for systematic effects and Monte Carlo noise, as shown in Fig.\,\ref{fig: pk-cov} The light points and lines indicate the data without correcting for fiber-collision effects (noting that there is little change to the theory model since we use quasi-unwindowed estimators), and the shot-noise contribution is shown in dashed lines (which is too large to see for QSO). The bottom panels indicate the difference between data and theory, in units of the errorbars.}
    \label{fig: pk-dat}
\end{figure}

\noindent Fig.\,\ref{fig: pk-dat} presents the DESI DR1 power spectra measurements described above. We find strong detections of the monopole ($120-230\sigma$, largest for LRG3) and quadrupole ($20-43\sigma$) in all data chunks, though with only modest detections of the hexadecapole ($0-5\sigma$).\footnote{These are computed as the difference between the best-fit $\chi^2$ with and without each theory contribution.} Due to the DESI observing strategy, the shot-noise contributions to the spectrum are large: the LRG3 and ELG2 samples become shot-noise dominated by $k\approx 0.15\hMpc$ and $0.05\hMpc$ respectively, whilst all the quasar sample is always limited by the number density. In all cases, we find excellent agreement between data and the best-fit theory model (obtained from the $P_\ell+B_0$ dataset, see \S\ref{sec: theory}), achieving a $\chi^2$ of $44.2$, $39.1$, $76.2$, $51.9$, $37.6$ and $47.2$ for the BGS, LRG1, LRG2, LRG3, ELG2 and QSO samples, for 54 data points,\footnote{The number of independent degrees of freedom is somewhat less due to the correlations induced by the mask and the additional contributions to the covariance discussed above.} yielding no evidence for unmodeled systematics. Due to our choice of normalization, neighboring data-points are anticorrelated at $\approx 20\%$; this is in contrast to the $\approx 20\%$ positive correlations found in windowed power spectrum estimators, and leads to a greater apparent scatter in Fig.\,\ref{fig: pk-dat} than in the analogous DESI plots \citep[e.g.,][]{DESI:2024jis}. We caution, however, that the total signal-to-noise is equivalent.

Comparing the light and dark points in Fig.\,\ref{fig: pk-dat}, we find significant variation in the monopole and some in the quadrupole: this is due to the $\theta$-cut introduced to ameliorate fiber-collision systematics. Due to our choice of estimator normalization, this difference is less severe than in DESI \citep[e.g.,][]{DESI:2024aax,DESI:2024jis}; moreover, the theory predictions are only minorly affected by the addition of the $\theta$-cut (since both $\F$ and $\G$ are modified, keeping $\F^{-1}\G$ almost constant). The differences in the numerator, which appear at up to $22\sigma$ for LRG3, indicate that the small-angle correlations are heavily contaminated, as expected.

\begin{figure}
    \centering
    \includegraphics[width=0.95\linewidth]{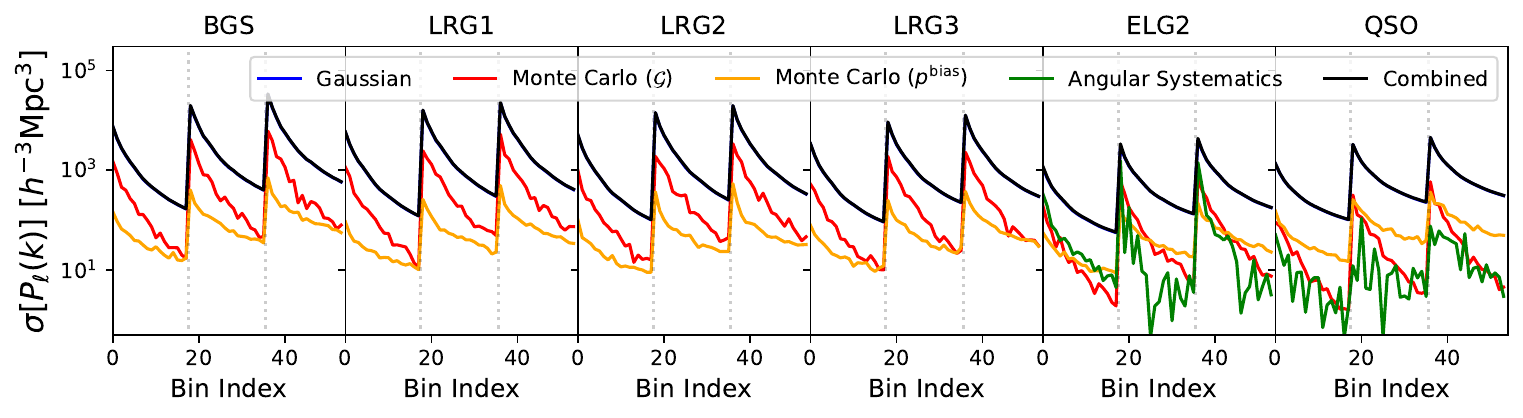}
    \caption{\textbf{$\boldsymbol{P_\ell(k)}$ Errorbars}: Contributions to the power spectrum variances for each data chunk, including the raw Gaussian covariances (blue), corrections for Monte Carlo noise entering the theory matrix (red) and shot-noise (orange) estimators, and, for ELG2 and QSO, angular systematics. The total errorbar is shown in black. For visibility, we stack the three Legendre multipoles into one dimension, as indicated by the vertical dashed lines. In all cases, the errorbar is dominated by the Gaussian contribution, implying that our Monte Carlo estimates of the theory and bias terms are sufficiently converged.}
    \label{fig: pk-cov}
\end{figure}

In Fig.\,\ref{fig: pk-cov} we plot the diagonal of the covariance matrix, splitting according to the various contributions discussed in \S\ref{subsec: pk-practical}. The behavior of the Gaussian covariance is as expected: it is suppressed on small-scales by volume factors and larger by a factor $\sim(2\ell+1)$ for higher-order multipoles, $P_\ell$. We note that the covariance is comparatively large for the low-volume BGS sample, but small for the QSOs due to their large redshift range; this affords a high signal-to-noise QSO detection despite the huge shot-noise (cf.\,Tab.\,\ref{tab: desi-chunks}). To form the \polybin estimators for the power spectrum stochasticity and theory matrix, we utilized Monte Carlo summation: as shown in Fig.\,\ref{fig: pk-cov}, this yields a subdominant error, reducing the signal-to-noise of the data by at most $2\%$; this implies that the summations are appropriately converged. For the theory matrix, the error is dominated by large-scales (due to the smaller number of modes therein), and generally small for the stochasticity term, $p^{\rm bias}$, though larger for QSOs. The angular systematic corrections used by DESI are similarly small and dominated by large-scales \citep{DESI:2024jis}, with at most a $1\%$ correction to the signal-to-noise (for ELG2). Overall, we find that the covariance is dominated by the baseline (Gaussian) contribution, which is an important verification of our pipeline.

\section{Bispectra}\label{sec: bspec}
\noindent Following the discussion of the power spectrum estimators in \S\ref{sec: pspec}, we now proceed to outline our bispectrum estimators and the corresponding observational dataset. These are based on the forms presented in \citep{Philcox:2021ukg,Ivanov:2023qzb,Philcox:2024rqr}, but include additional modifications to account for integral constraints and fiber collisions, as discussed below.

\subsection{Formalism}
\noindent Analogously to \S\ref{sec: pspec}, we wish to estimate a set of bispectrum coefficients, $\{b_\alpha\}$, from the observational dataset $d$. Following \citep{Philcox:2024rqr}, a general estimator takes the form
\beq\label{eq: bk-estimator}
    \hat{b}_\alpha[d] = \sum_{\beta=1}^{N_{\rm bins}}\F_{\alpha\beta}^{-1}\left[b_\beta^{\rm num}[d]-b_\beta^{\rm bias}[d]\right]
\eeq
involving a numerator, a noise term and a normalization matrix, as before. The numerator now involves both a cubic and a linear term: 
\beq\label{eq: b-num}
    b^{\rm num}_\alpha[d]  &=& \frac{1}{6}\int d\vx\,d\vy\,d\vz\,\frac{\partial\zeta(\vx,\vy,\vz)}{\partial b_\alpha}\bigg([\Si d](\vx)[\Si d](\vy)[\Si d](\vz)\\\nonumber
    &&\qquad\qquad\qquad\qquad\qquad\qquad\,-\ [\Si d](\vx)\av{[\Si d](\vy)[\Si d](\vz)} + \text{2 perms.}\bigg)^*.
\eeq
where $\zeta$ is the true three-point function and $\av{\cdots}$ represents an average over some set of simulated realizations of the field with (approximately) the same covariance as the data. As before, this involves the weighting $\Si$; setting this equal to $\Si = \P^{\dagger}\Ci$ results in a minimum-variance estimator. The noise term and normalization read
\beq
    b^{\rm bias}_\alpha[d] &=& \frac{1}{6}\int d\vx\,d\vy\,d\vz\,d\vx'\,d\vy'\,d\vz'\,\frac{\partial\zeta(\vx,\vy,\vz)}{\partial b_\alpha}\left(\Si(\vx,\vx')\Si(\vy,\vy')\Si(\vz,\vz')\N(\vx',\vy',\vz')\right)^*\\\nonumber 
    \F_{\alpha\beta} &=& \frac{1}{6}\int d\vx\,d\vy\,d\vz\,d\vx'\,d\vy'\,d\vz'\frac{\partial\zeta(\vx,\vy,\vz)}{\partial b_\alpha}\left([\Si\mathsf{P}](\vx,\vx')[\Si\mathsf{P}](\vy,\vy')[\Si\mathsf{P}](\vz,\vz')\frac{\partial\zeta(\vx',\vy',\vz')}{\partial b_\beta}\right)^*.
\eeq
where $\N$ is the Poisson contribution to the pixel-space three-point function. Unlike for the power spectrum, we do not compute theory matrices for the bispectrum, since this is computationally prohibitive \citep[cf.][]{Philcox:2024rqr}. Instead, we assume that the unwindowed bispectrum defined by \eqref{eq: bk-estimator} can be directly compared to the theory model (\textit{i.e.}\ that $\zeta= \sum_{\alpha}b_\alpha (\partial\zeta/\partial b_\alpha)$). This is justified by noting that the bispectrum signal-to-noise is much weaker than for the power spectrum, thus any errors caused by residual mask contamination will be small.

As derived in \citep{Philcox:2024rqr}, the Gaussian covariance of the bispectrum numerator is given by
\beq\label{eq: cov-bk}
    \mathrm{cov}(b^{\rm num}_\alpha,b^{\rm num}_\beta)  &=& \frac{1}{6}\int d\vx\,d\vy\,d\vz\,d\vx'\,d\vy'\,d\vz'\,\frac{\partial\zeta(\vx,\vy,\vz)}{\partial b_\alpha}\\\nonumber
    &&\qquad\qquad\qquad\qquad\,\times\,\left([\Si\C\Sid](\vx,\vx')[\Si\C\Sid](\vy,\vy')[\Si\C\Sid](\vz,\vz')\frac{\partial\zeta(\vx',\vy',\vz')}{\partial b_\beta}\right)^*.
\eeq
Analogously to the power spectrum covariance \eqref{eq: cov-pk}, this can be formed by replacing $\Si \mathsf{P}\to\Sid \mathsf{C}\Si$ in the expression for $\F$ (\textit{i.e.}\ assuming an optimal weighting scheme). Whilst this includes all relevant mask induced effects, it ignores non-Gaussian density contractions (such as the bispectrum and trispectrum), which can be important in the squeezed limit \citep[e.g.,][]{Biagetti:2021tua,Salvalaggio:2024vmx}, though we do not use such triangles in the present analysis.

\subsection{Specialization to DESI}\label{subsec: bk-specialization}
\noindent Many of the ingredients entering \eqref{eq: bk-estimator} are common to both the power spectrum and bispectrum estimators. In this section, we discuss the notable differences, including the modeling of the three-point function, the data-dependent shot-noise, and the treatment of fiber-collisions, which is new to this work. All other aspects (namely, the weighting scheme, pointing matrix, and covariance) are treated identically to \S\ref{sec: pspec}. 

\subsubsection{Three-Point Function}
\noindent The three-point function of the underlying data, $\delta_g$, can be expressed in terms of the bispectrum $B(\vk_1,\vk_2,\vk_3)$:
\beq\label{eq: zeta-def}
    \zeta(\vx,\vy,\vz) = \int_{\vk_1\vk_2\vk_3}B(\vk_1,\vk_2,\vk_3)e^{i(\vk_1\cdot\vx+\vk_2\cdot\vy+\vk_3\cdot\vz)}\delD{\vk_1+\vk_2+\vk_3},
\eeq
where the Dirac delta ensures momentum conservation. Restricting to the bispectrum monopole (\textit{i.e.}\ ignoring anisotropies generated by redshift-space distortions), we can expand this in bins:
\beq
    B(\vk_1,\vk_2,\vk_3;\hx_1,\hx_2,\hx_3) \approx \sum_{\alpha=1}^{N_{\rm bins}} \frac{b_\alpha}{\Delta_\alpha}\left[\Theta_{b_1}(k_1)\Theta_{b_2}(k_2)\Theta_{b_3}(k_3)+\text{5 perms.}\right]
\eeq
where $\alpha\equiv\{b_1,b_2,b_3\}$ indexes the bin and the coefficient $\Delta_\alpha$ gives the corresponding degeneracy (six if $b_1=b_2=b_3$, two if $b_1=b_2\neq b_3$ or permutations, one else).\footnote{As discussed in \citep{Ivanov:2023qzb,Philcox:2024rqr}, this can be generalized to anisotropic moments via a Legendre expansion. We defer such analyses to future work.} The three-point functions are thus
\beq\label{eq: zeta-b-deriv}
    \frac{\partial\zeta(\vx,\vy,\vz)}{\partial b_\alpha} = \frac{1}{\Delta_\alpha}\int d\vr\,\int_{\vk_1\vk_2\vk_3}\left[\Theta_{b_1}(k_1)\Theta_{b_2}(k_2)\Theta_{b_3}(k_3)+\text{5 perms.}\right]e^{i\vk_1\cdot(\vx-\vr)}e^{i\vk_2\cdot(\vy-\vr)}e^{i\vk_3\cdot(\vz-\vr)},
\eeq
where we have replaced the Dirac delta by an integral to facilitate efficient computation using Fourier transforms.

\subsubsection{Shot-Noise}
\noindent Stochasticity in the bispectrum is somewhat more complex than in the power spectrum. Following \citep{Philcox:2024rqr}, the Poisson pixel-space noise term can be written as the sum of two contributions:
\beq
    \N(\vx,\vy,\vz) &=& \delta_{\rm D}(\vx-\vy)\delta_{\rm D}(\vx-\vz)\tilde{n}_3(\vx) + \left[\delta_{\rm D}(\vx-\vy)\tilde{n}_2(\vy)n(\vz)\xi(\vy,\vz)+\text{2 perms.}\right],
\eeq
where $\tilde{n}_2$ and $\tilde{n}_3$ are doubly- and triply-weighted masks. This can be estimated from the data as follows (accounting for two-point shot-noise):
\beq
    \widehat{\N}[d](\vx,\vy,\vz) = -2\delta_{\rm D}(\vx-\vy)\delta_{\rm D}(\vx-\vz)n_3(\vx)&& + \left[\delta_{\rm D}(\vx-\vy)d_2(\vy)d(\vz)+\text{2 perms.}\right]
\eeq
defining
\beq
    n_3(\vx) =\left(\alpha_3+\frac{1}{2}\alpha^3+\frac{3}{2}\alpha_2\alpha\right)\sum_{i=1}^{N_r}w_{r,i}^3\delta_{\rm D}(\vx-\vx_{r,i})\quad
    d_2(\vx) = \sum_{i=1}^{N_g} w_{g,i}^2\delta_{\rm D}(\vx-\vx_{g,i}) -\alpha_2\sum_{i=1}^{N_r} w_{r,i}^2\delta_{\rm D}(\vx-\vx_{r,i}),
\eeq
where $\alpha_k \equiv \sum_{i=1}^{N_g}w_{g,i}^k/\sum_{i=1}^{N_r}w_{r,i}^k$. In the limit of $\Si = \mathsf{1}$, we find contributions to the bispectrum scaling as $\bar{n}$ and $\bar{n}^2[P(k_1)+P(k_2)+P(k_3)]$ respectively, ignoring volume factors \citep[cf.,][]{Philcox:2024rqr}. Unlike for the power spectrum, this correction is data-dependent (though one could also estimate it from simulations). As before, we account for departures from the Poisson regime in the theory model.

\subsubsection{Fiber Collisions}\label{subsubsec: bspec-fiber}
\noindent For the power spectrum, we mitigated fiber incompleteness effects by imposing an angular cut on galaxy and random point separations with $\theta<\theta_{\rm cut}$. This could be efficiently achieved by counting all pairs of closely separated particles, which has complexity $\mathcal{O}(Nn_{\rm 2D}\theta_{\rm cut}^2)$, where $n_{\rm 2D}$ is the two-dimensional density of points. Performing an analogous cut for the bispectrum is highly non-trivial, since it would require triplet counts instead of pair-counts, with complexity $\mathcal{O}(N^2n_{\rm 2D}\theta_{\rm cut}^2)$ (noting that only two of the three galaxies need to be close on the sky). Morevoer, the effect of fiber collisions is not localized to a few types of triangle: due to the transformation from angular- to Fourier-space, it affects a wide range of shapes, including squeezed configurations. To ameliorate this, we propose a new estimator for the $\theta$-cut bispectrum, which uses methods analogous to our normalization computation (namely, the Girard-Hutchinson scheme \citep{girard89,hutchinson90} for high-dimensional matrix trace estimation).

Setting $\Si = \mathsf{1}$ and dropping the (small) linear term, the cubic bispectrum numerator involves a sum over three copies of the dataset, $d$. Isolating the terms that include at least one pair of points within a separation $\theta_{\rm cut}$ in \eqref{eq: b-num}, we find:
\beq\label{eq: bk-fc}
    b^{\rm fc}_\alpha[d] &=& \frac{1}{6}\int d\vx\,d\vy\,d\vz\,\frac{\partial\zeta(\vx,\vy,\vz)}{\partial b_\alpha}d(\vx)d(\vy)d(\vz)\\\nonumber
    &&\qquad\qquad\,\times\,\left[\left(\Phi(\vx,\vy)+\text{2 perms.}\right)-\left(\Phi(\vx,\vy)\Phi(\vy,\vz) + \text{2 perms.}\right) + \Phi(\vx,\vy)\Phi(\vy,\vz)\Phi(\vz,\vx)\right]\\\nonumber
\eeq
analogous to \eqref{eq: pk-fc}. This term, which will be subtracted from $b^{\rm num}_\alpha$, encodes three types of fiber collision corrections: (1) a single pair of galaxies within $\theta_{\rm cut}$; (2) two pairs within $\theta_{\rm cut}$; (3) all three galaxies within $\theta_{\rm cut}$. Since the $\Phi(\vx,\vy)$ functions cannot be easily separated in real- or Fourier-space (unlike the $\zeta$ derivative of \ref{eq: zeta-b-deriv}), \eqref{eq: bk-fc} is difficult to evaluate directly, as discussed above.

To decorrelate the integrals, we introduce an additional random field $\epsilon(\vx)\sim \mathcal{N}(0,1)$ \citep[cf.,][]{girard89,hutchinson90}. Explicitly, we can replace
\beq
    \Phi(\vx,\vy)\to \epsilon(\vy)\int d\vw\,\Phi(\vx,\vw)\epsilon(\vw);
\eeq
this gives the same result after averaging over the distribution of $\epsilon$, but explicitly separates the $\vx$ and $\vy$ dependence. For the first term in \eqref{eq: bk-fc}, this gives
\beq
    b^{\rm fc,(1)}_\alpha[d,\epsilon] &=& \frac{1}{2}\int d\vx\,d\vy\,d\vz\,\frac{\partial\zeta(\vx,\vy,\vz)}{\partial b_\alpha}\left(\int d\vw\,d(\vx)\Phi(\vx,\vw)\epsilon(\vw)\right)\left(d(\vy)\epsilon(\vy)\right)d(\vz);
\eeq    
performing a Monte Carlo average over realizations of $\epsilon$ gives the desired result. In practice, it is preferable to evaluate this using a discrete set of $N_g+N_r$ random values, $\{\epsilon_i\}$, rather than a continuous field $\epsilon(\vx)$. This can be achieved by defining the product and convolution of $d$ with $\epsilon$:
\beq\label{eq: de-d-star-e-def}
    [d\epsilon](\vx) &=& \sum_{i=1}^{N_g+N_r} w_i\epsilon_i\delta_{\rm D}(\vx-\vx_i), \qquad [d\ast\epsilon](\vx) = \sum_{i=1}^{N_g+N_r} w_i\delta_{\rm D}(\vx-\vx_i)\sum_{j:\Phi(\vx_i,\vx_j)=1}\epsilon_j,\\\nonumber
\eeq
where we sum over both galaxies and random points as in \eqref{eq: pk-fc-discrete}. The second term can be efficiently computed by iterating over all pairs of close points. With these definitions, the fiber collision term is given by
\beq
    b^{\rm fc,(1)}_\alpha[d,\epsilon] &=& \frac{1}{2}\int d\vx\,d\vy\,d\vz\,\frac{\partial\zeta(\vx,\vy,\vz)}{\partial b_\alpha}[d\ast\epsilon](\vx)[d\epsilon](\vy)d(\vz);
\eeq    
it is straightforward to show that this reduces to the first term in \eqref{eq: bk-fc} when averaging over $\epsilon$, noting that $\av{\epsilon_i\epsilon_j} = \delta^{\rm K}_{ij}$. The (subdominant) second and third terms can be computed similarly, but involve two- and three independent random fields:
\beq
    b^{\rm fc,(2)}_\alpha[d,\epsilon_1,\epsilon_2] &=& -\frac{1}{2}\int d\vx\,d\vy\,d\vz\,\frac{\partial\zeta(\vx,\vy,\vz)}{\partial b_\alpha}[d\ast\epsilon_1](\vx)[d\ast\epsilon_2](\vy)[d\epsilon_1\epsilon_2](\vz)\\\nonumber
    b^{\rm fc,(3)}_\alpha[d,\epsilon_1,\epsilon_2,\epsilon_3] &=& \frac{1}{6}\int d\vx\,d\vy\,d\vz\,\frac{\partial\zeta(\vx,\vy,\vz)}{\partial b_\alpha}[\epsilon_1(d\ast\epsilon_2)](\vx)[\epsilon_2(d\ast\epsilon_3)](\vy)[\epsilon_3(d\ast\epsilon_1)](\vz),
\eeq  
where $[d\epsilon_1\epsilon_2]$ replaces $\epsilon\to\epsilon_1\epsilon_2$ in $[d\epsilon]$ and $[\epsilon_1(d\ast\epsilon_2)]$ replaces $w\to w\epsilon_1,\epsilon\to \epsilon_2$ in $[d\ast\epsilon]$ \eqref{eq: de-d-star-e-def}.

Some comments are in order. Under the above scheme, we stochastically compute the fiber-collision term as a cross-bispectrum between three differently-weighted density fields. Although this involves averaging over set(s) of random fields, it is quick to implement given the efficient nature of the bispectrum estimator; moreover, it typically converges to the desired accuracy within a few Monte Carlo iterations (\textit{i.e.}\ independent $\epsilon_i$ vectors). As discussed in Appendix \ref{app: pk-fc}, a similar scheme can be used to compute the power spectrum fiber collision term. There, we find excellent agreement between the stochastic and explicit fiber collision estimators, which is an important test of the method.

Finally, we note that an analogous method can be used to compute the corrections to the bispectrum Fisher matrix induced by the $\theta$-cut. Ignoring integral constraints (as in the power spectrum), the change to the normalization matrix is given by the sum of
\beq
    \F^{\rm fc,(1)}_{\alpha\beta}[\epsilon] &=& \frac{1}{2}\int d\vx\,d\vy\,d\vz\,\frac{\partial\zeta(\vx,\vy,\vz)}{\partial b_\alpha}[n\ast\epsilon](\vx)[n\epsilon](\vy)n(\vz)\frac{\partial\zeta(\vx,\vy,\vz)}{\partial b_\beta}\\\nonumber
    \F^{\rm fc,(2)}_{\alpha\beta}[\epsilon_1,\epsilon_2] &=& -\frac{1}{2}\int d\vx\,d\vy\,d\vz\,\frac{\partial\zeta(\vx,\vy,\vz)}{\partial b_\alpha}[n\ast\epsilon_1](\vx)[n\ast\epsilon_2](\vy)[n\epsilon_1\epsilon_2](\vz)\frac{\partial\zeta(\vx,\vy,\vz)}{\partial b_\beta}\\\nonumber
    \F^{\rm fc,(3)}_{\alpha\beta}[\epsilon_1,\epsilon_2,\epsilon_3] &=& \frac{1}{6}\int d\vx\,d\vy\,d\vz\,\frac{\partial\zeta(\vx,\vy,\vz)}{\partial b_\alpha}[\epsilon_1(n\ast\epsilon_2)](\vx)[\epsilon_2(n\ast\epsilon_3)](\vy)[\epsilon_3(n\ast\epsilon_1)](\vz)\frac{\partial\zeta(\vx,\vy,\vz)}{\partial b_\beta},
\eeq
where $[n\epsilon]$ and $[n\ast \epsilon]$ follow \eqref{eq: de-d-star-e-def}, but sum only over randoms. This can be computed as for $\F$, simply switching the relevant masks.

\subsubsection{Limiting Forms}
\noindent Inserting \eqref{eq: zeta-b-deriv} into \eqref{eq: b-num} gives the usual expression for the bispectrum numerator in the limit of $\Si = \mathsf{1}$, dropping the linear term:
\beq
    b^{\rm num}_\alpha[d]  &\to& \frac{1}{\Delta_\alpha}\int d\vr\,g_{b_1}[d](\vr)g_{b_2}[d](\vr)g_{b_3}[d](\vr), \qquad g_b[d](\vr) \equiv \int_{\vk}e^{-i\vk\cdot\vr}\Theta_b(k)d(\vk).
\eeq
This can be straightforwardly evaluated using Fourier transforms, and matches the form used in previous analyses \citep[e.g.,][]{Scoccimarro:2015bla,Gil-Marin:2014sta,Gil-Marin:2016wya}. On ultra-large scales, the linear term leads to significantly lower variance: this is given by
\beq
    b^{\rm num,lin}_\alpha[d]  &\to& \frac{1}{\Delta_\alpha}\int d\vr\,g_{b_1}[d](\vr)\av{g_{b_2}[d](\vr)g_{b_3}[d](\vr)}+\text{2 perms.},
\eeq
which can be evaluated using simulations, or, for a uniform sample with known power spectrum, theory. In expectation, the numerator of the bispectrum estimator yields
\beq
    \mathbb{E}[b_\alpha^{\rm num}[d]-b_\alpha^{\rm bias}]&\to& \frac{1}{\Delta_\alpha}\int_{\vk_1+\vk_2+\vk_3=\vec 0}\int_{\vk_1'+\vk_2'+\vk_3'=\vec 0}\Theta_{b_1}(k_1)\Theta_{b_2}(k_2)\Theta_{b_3}(k_3)\\\nonumber
    &&\qquad\qquad\qquad\qquad\qquad\qquad\,\times\,n(\vk_1-\vk_1')n(\vk_2-\vk_2')n(\vk_3-\vk_3')B(\vk_1',\vk_2',\vk_3'),
\eeq
which is a convolution of the true bispectrum with the three-point function of the mask, as expected \citep[e.g.][]{Gil-Marin:2014sta}. The normalization matrix has a similar form:
\beq
    \F_{\alpha\beta}&\to& \frac{1}{\Delta_\alpha\Delta_\beta}\int_{\vk_1+\vk_2+\vk_3=\vec 0}\int_{\vk_1'+\vk_2'+\vk_3'=\vec 0}\Theta_{b_1}(k_1)\Theta_{b_2}(k_2)\Theta_{b_3}(k_3)\left[\Theta_{b_1'}(k_1')\Theta_{b_2'}(k_2')\Theta_{b_3'}(k_3')+\text{5 perms.}\right]\\\nonumber
    &&\qquad\qquad\qquad\qquad\qquad\qquad\,\times\,n(\vk_1-\vk_1')n(\vk_2-\vk_2')n(\vk_3-\vk_3'),
\eeq
(for $\beta \equiv\{b_1',b_2',b_3'\}$), which can be compared to the simplified form found in conventional windowed bispectrum estimators \citep[e.g.,][]{Gil-Marin:2014sta}:
\beq
    \F^{\rm win}_{\alpha\beta}&=& \frac{1}{\Delta_\alpha\Delta_\beta}\int_{\vk_1+\vk_2+\vk_3=\vec 0}\Theta_{b_1}(k_1)\Theta_{b_2}(k_2)\Theta_{b_3}(k_3)\left[\Theta_{b_1'}(k_1)\Theta_{b_2'}(k_2)\Theta_{b_3'}(k_3)+\text{5 perms.}\right]\int d\vx\,n^3(\vx)\\\nonumber
    &=& \frac{\delta_{\alpha\beta}^{\rm K}}{6\Delta_\alpha}\int_{\vk_1+\vk_2+\vk_3=\vec 0}\Theta_{b_1}(k_1)\Theta_{b_2}(k_2)\Theta_{b_3}(k_3)\int d\vx\,n^3(\vx)
\eeq
which is diagonal. As discussed above, we utilize the unwindowed normalization to avoid having to perform expensive window-convolution of the theory predictions \citep[cf.,][]{Pardede:2022udo,wang_window}. Results obtained using the windowed normalization are shown in \S\ref{subsec:window}.

\subsection{Practical Implementation}
\noindent The above estimators can be implemented using Fourier transforms and Monte Carlo methods, and are included in the \polybin package \citep{Philcox:2021ukg,Philcox:2024rqr,polybin3d}. In this work, we adopt a $k$-binning strategy with $\delta k = 0.005\hMpc$ for $k\in[0.005,0.01)\hMpc$, $\delta k=0.01\hMpc$ for $k\in[0.01,0.16)\hMpc$ and $\delta k=0.03\hMpc$ for $k\in[0.16,0.19)\hMpc$, using a total of $N_k = 17$ bins;\footnote{Note that we fix the Nyquist frequency to $k_{\rm Ny}=0.2\hMpc$ for the bispectrum.} this is chosen to ensure good coverage at low-$k$ and to add additional modes at high-$k$ to capture mask-induced leakage.\footnote{We further include triangular configurations where the bin-centers do not satisfy the triangle conditions (but other modes in the bin do). These must be accounted for to prevent biases in the mask-deconvolution.} This corresponds to $N_{\rm bins}=597$, though we restrict to just $N_{\rm bins}=49$ after applying the normalization, dropping modes outside $k\in[0.02,0.08)\hMpc$ to mitigate systematic and modeling uncertainties \citep[e.g.,][]{DESI:2024jis,Ivanov:2021kcd}.\footnote{On these scales, the linear term in the estimator numerator \eqref{eq: b-num} has a minimal impact on the output spectra and could be omitted to expedite computations. We here include it in case of future studies extending to smaller $k_{\rm min}$.}

To compute the linear term in the numerator \eqref{eq: b-num}, we average over $50$ masked Gaussian realizations, which are generated using a fiducial power spectrum model fit to an early version of the data. All other aspects of the estimation follow \S\ref{subsec: pk-practical}, except that we use $N_{\rm MC}=50$ for the bias, normalization, and covariance terms (instead of $N_{\rm MC}=20$ before for the normalization and covariance). For the largest data chunk, computation of the bispectrum numerator requires $260$ seconds on a CPU node; restricting to just the cubic term reduces this by $\sim 100\times$. Each Monte Carlo realization for $b^{\rm bias}$ requires $70$ seconds, whilst those for the normalization and covariance require $930$ and $3000$ seconds respectively. Finally, computing the fiber collision correction required $600$ seconds for each of $20$ numerator realizations, and $1400$ seconds for each of five normalization realizations (which was sufficient to yield convergence), after computing the pairwise convolutions. In total, the full calculation for the largest chunk (almost all of which is data-independent) requires around two node-days.

Finally, we note that the covariance is computed analogously to the power spectrum covariance of \S\ref{subsec: pk-practical}. 
As before, we marginalize over ELG and QSO imaging systematics using unweighted data (though these are highly subdominant for the bispectrum). To account for residual noise in the fiber collision numerator, we add a correction term equal to the variance of the estimate divided by $1/N_{\rm MC}$, with a similar factor involving the variance of $\F^{\rm fc}b_\alpha$ to account for noise in $\F^{\rm fc}$ (adding a factor of $\F^{-1/2}$ to diagonalize). These factors reduce the signal-to-noise by less than $10\%$. As for the power spectrum analysis, we do not include any non-Gaussian contributions to the covariance (sourced by three-, four- and six-point functions); furthermore, we ignore cross-correlations between the power spectra and bispectra. As discussed in \citep{Chan:2016ehg,Barreira:2019icq}, the former assumption is valid given that we restrict to large scales ($k^B_{\rm max}=0.08\hMpc$) and do not include squeezed configurations \citep[cf.][]{Biagetti:2021tua}. Moreover, in a mock analysis adopting similar scale cuts to our work, \citep{Oddo:2021iwq} demonstrate that dropping the cross-covariance between the power spectrum and bispectrum has only a minimal effect on parameter constraints.

\subsection{Results}

\begin{figure}
    \centering
    \includegraphics[width=0.95\linewidth]{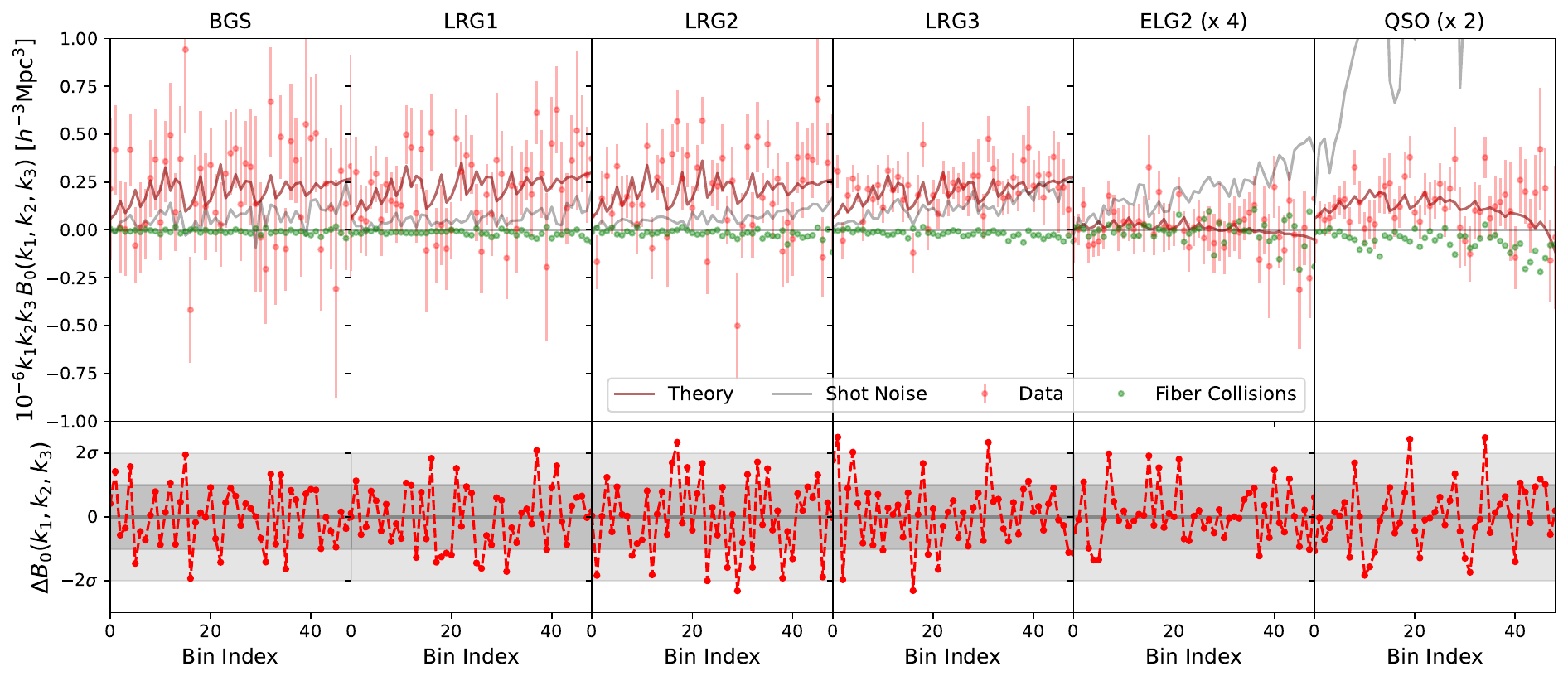}
    \caption{\textbf{DESI $\boldsymbol{B_0(k_1,k_2,k_3)}$}: Bispectra for each of the six DESI data chunks. The red points show the bispectrum monopoles measured using \polybin, with the best-fit theory curve shown as a dark red solid line (obtained from a joint power spectrum and bispectrum analysis). As for the power spectrum, we use (approximately) window-deconvolved estimators, which lead to anticorrelations between neighboring bins and thus considerable scatter in the datapoints. The contributions from shot-noise and the fiber-collision corrections are shown in black and green respectively. In all cases, we compress the data to a single dimension, with largest scales on the left ($k=0.02\hMpc$) and smallest scales on the right ($k=0.08\hMpc$). The bottom panel shows the difference between data and theory, and we find excellent agreement in all cases.}
    \label{fig: bk-dat}
\end{figure}

\noindent Fig.\,\ref{fig: bk-dat} shows the bispectrum measurements for each DESI chunk alongside the best-fit theory curve discussed below. Compared to the power spectrum, we find considerably larger scatter: this is due to the small signal-to-noise of the higher-order statistics on large, quasi-linear, scales. That said, we obtain a strong ($\gtrsim 10\sigma$) detection of the bispectrum in all chunks except for ELG2, reaching $18\sigma$ for LRG2 and $22\sigma$ for LRG3. As for the power spectrum, the stochastic contributions dominate the QSO and ELG2 samples; here, they exhibit a much more complex scale-dependence, as discussed in \S\ref{subsec: bk-specialization}. Due to our choice of normalization, the individual triangles are anticorrelated, which leads to the large scatter shown in the figure; if we were to instead use the windowed normalization ($\F\to\F^{\rm win}$), we would find a much smoother distribution of data-points (as in \S\ref{subsec: pk-results}). Across all scales considered, we find excellent agreement between theory and data, with a $\chi^2$ of $36.1$, $40.0$, $66.8$, $55.9$, $31.4$ and $45.2$ for the BGS, LRG1, LRG2, LRG3, ELG2 and QSO bispectrum samples, respectively, with 49 data points.

\begin{figure}
    \centering
    \includegraphics[width=0.85\linewidth]{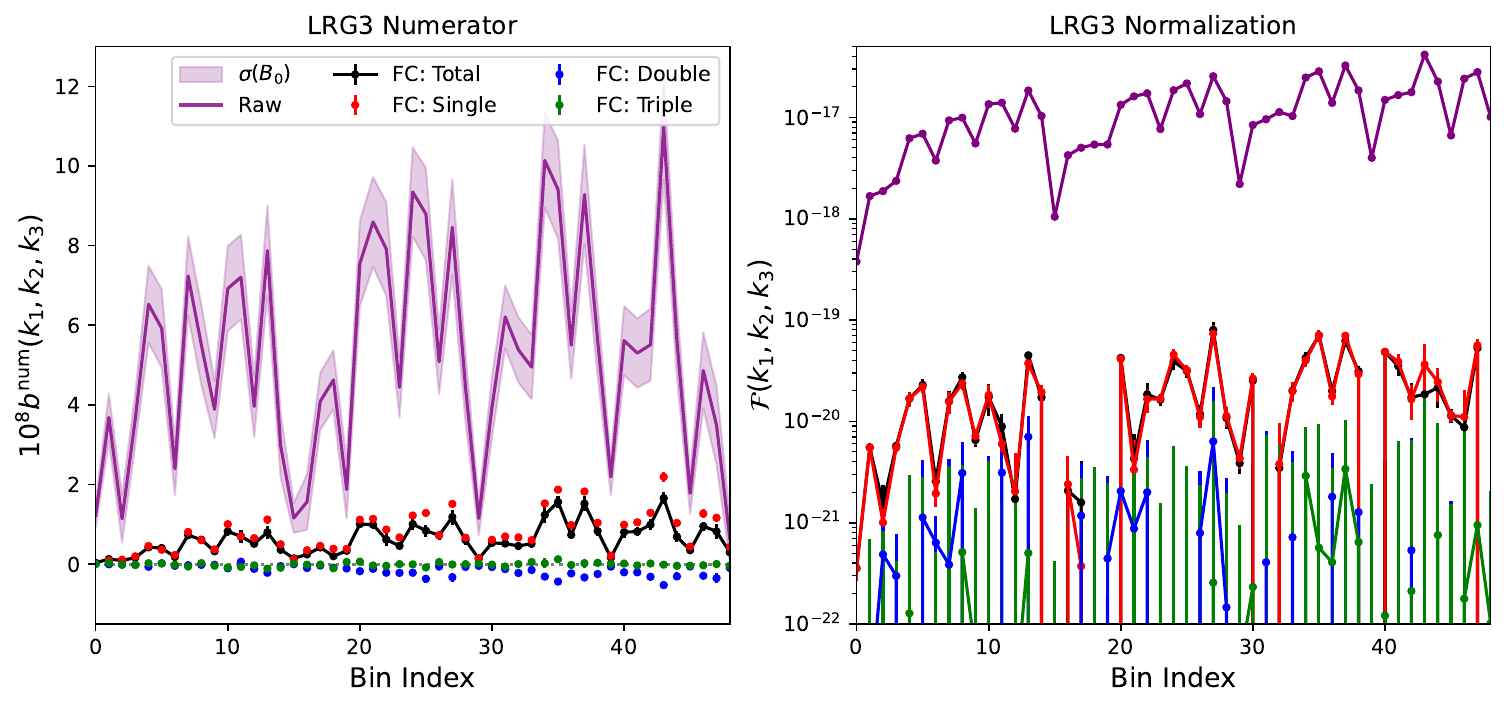}
    \caption{\textbf{Bispectrum Fiber Collisions}: The impact of fiber collisions on the bispectrum numerator (left) and the diagonal of the normalization matrix (right). The purple lines give the fiducial contributions from all triplets of galaxies, with the total fiber-collision contributions are shown in black. This is decomposed into contributions from one (red), two (blue) and three (green) angular cuts, which are computed stochastically, as discussed in \S\ref{subsubsec: bspec-fiber}. We find a clear fiber-collision signal in the numerator at the $(10-20)\%$ level, which is dominated by the single- and double-collision terms. In contrast, the correction to the normalization is small and sourced almost entirely by the single-collision piece.} 
    \label{fig: bk-fc}
\end{figure}

The green points in Fig.\,\ref{fig: bk-dat} show the impact of fiber-collisions, \textit{i.e.}\ the contributions removed using our $\theta$-cut framework \S\ref{subsubsec: bspec-fiber}. Whilst the contribution to each individual bin is small, the net effect can be large, reaching $4.9\sigma$ for QSO (where the effects of angular cuts are most significant due to the larger distance to the sample). To unpack this, we plot the fiber collision contributions to the bispectrum numerator and normalization ($b^{\rm fc}$ and $\F^{\rm fc}$) in Fig.\,\ref{fig: bk-fc}. Analogously to the power spectrum case, the effects are present on all scales, but are larger at higher-$k$ and for equilateral triangles. Removing contributions from $\theta<\theta_{\rm cut}$ changes the numerator by around $10-20\%$ but the normalization only by $\sim 1\%$; this mismatch indicates that there is excess signal in the observational dataset at low-$\theta$, as expected. Splitting according to the three terms described in \S\ref{subsubsec: bspec-fiber}, we find that numerator contributions are dominated by terms with both one and two factors of $\Theta$ (\textit{i.e.}\ explicit restrictions on both one and two sides of the triangle, whilst the normalization is sensitive only to the first contribution (which is the cheapest to compute). 

\begin{figure}
    \centering
    \includegraphics[width=0.95\linewidth]{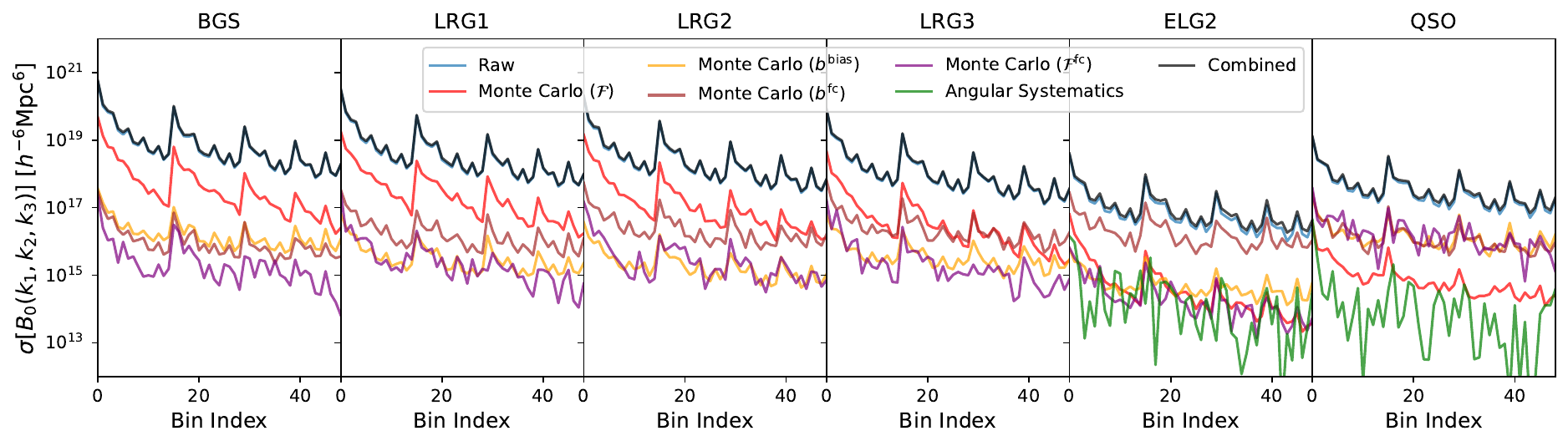}
    \caption{\textbf{$\boldsymbol{B_0(k_1,k_2,k_3)}$ Errorbars}: Contributions to the diagonal of the bispectrum covariance matrices, restricting to $k\in[0.02,0.08)\hMpc$ in all cases. We include the Gaussian piece (blue), Monte Carlo noise in the normalization (red), noise term (orange), fiber collision numerator (brown), and fiber collision normalization (purple), as well as imaging systematics corrections (green). All triangular configurations are stacked into one-dimension following Fig.\,\ref{fig: bk-dat}. As for the power spectra (Fig.\,\ref{fig: pk-cov}), the covariance is strongly dominated by the Gaussian contribution, with subdominant Monte Carlo errors and negligible contribution from angular systematics.}
    \label{fig: bk-cov}
\end{figure}

Finally, we plot the contributions to the bispectrum covariance in Fig.\,\ref{fig: bk-cov}. Despite the various Monte Carlo summations used in the estimator (to compute the normalization, stochasticity, fiber collision numerator, and fiber collision normalization), we observe that the output covariance is dominated by the masked Gaussian contribution as in \S\ref{subsec: pk-results}. The most significant contributions arise from the normalization term, or, for the high-redshift samples, the fiber collision terms; however, they affect the total signal-to-noise only by $<10\%$, or $<2\%$ for the BGS and LRG samples. This implies that the number of Monte Carlo realizations used above is conservative. Furthermore, we find negligible impact from the angular systematic covariance, implying that the systematic weights have minimal impact on the large-scale bispectrum signal. As above, we caution that our covariance matrix does not take non-Gaussian effects into account; whilst this is likely a valid assumption for $k<0.08\hMpc$ \citep{Oddo:2021iwq,Chan:2016ehg,Barreira:2019icq}, it is likely to break down on smaller scales and for highly squeezed triangles \citep[e.g.,][]{Biagetti:2021tua}.

\section{Theory and Analysis}\label{sec: theory}
\noindent We now discuss the theoretical model used to fit the above power spectrum and bispectrum datasets. Our treatment closely follows previous works \citep[e.g.,][]{Philcox:2021kcw,Ivanov:2021kcd} and is similar to the perturbative models used for the power spectrum in \citep{DESI:2024jis,Maus:2024dzi,Noriega:2024eyu,Lai:2024bpl}. We utilize the EFT, as implemented in the \textsc{class-pt} code~\cite{Chudaykin:2020aoj,classpt}, which is an extension of the \textsc{class} Boltzmann code \citep{Lesgourgues:2011re,Blas:2011rf}).\footnote{Various other perturbation theory codes exist, including \textsc{pybird}, \textsc{velocileptors}, \textsc{folps}, \textsc{pbj} and \textsc{class one-loop} \citep{DAmico:2020kxu,Chen:2020fxs,Noriega:2022nhf,Moretti:2023drg,Linde:2024uzr}. A comparison between several codes is presented in \citep{Maus:2024sbb}.} For consistency, the power spectrum (bispectrum) is computed up to one-loop (tree-level) order within perturbation theory, though one could push to smaller scales by adding one-loop corrections to the bispectrum \citep[e.g.,][]{Philcox:2022frc,DAmico:2022ukl,DAmico:2022osl,Bakx:2025pop}. 
Our model includes all the necessary ingredients to accurately describe galaxy clustering on large scales, including the leading-order non-linear corrections, galaxy bias, infrared resummation, ultraviolet counterterms and stochasticity. Below, we provide a brief summary of the model; further information can be found in~\cite{Ivanov:2019pdj,Philcox:2021kcw,Ivanov:2021kcd,Chudaykin:2024wlw}.

\subsection{Power Spectrum}
\noindent Schematically, our model for the power spectrum multipoles can be written as
\be
\label{Ptheory}
P_\ell(k) = P_\ell^{\rm tree}(k) + P_{\ell}^{\rm 1-loop}(k) + P_{\ell}^{\rm ct}(k) + P_\ell^{\rm stoch}(k),
\ee
where the four terms denote the standard linear theory (Kaiser) prediction, the deterministic one-loop perturbation theory correction, the counterterms, and the stochastic contributions. The counterterm model includes both the leading-order terms and the next-to-leading order $k^4$ contribution, the latter used to address fingers-of-God effects (\cite{Jackson:2008yv}) as explained in~\cite{Chudaykin:2020aoj,Ivanov:2024xgb,Chudaykin:2024wlw}. As in previous works, each term is resummed to address the impact of long-wavelength displacements; this leads to a suppression of the wiggly part of the spectrum \citep[e.g.,][]{Ivanov:2018gjr,Blas:2016sfa,Senatore:2017pbn}. We also account for the effect of the fiducial cosmology used to convert angles and redshifts into three-dimensional coordinates (commonly known as Alcock-Paczynski distortions~\cite{Alcock:1979mp}) by a redefinition of the wavenumber and angles, following \citep[e.g.,][]{Philcox:2021kcw}. Power spectra are evaluated at the effective redshift
\beq
    z^{(2)}_{\rm eff} = \frac{\int_{z_{\rm min}}^{z_{\rm max}} z\,n^2(z)dV(z)}{\int_{z_{\rm min}}^{z_{\rm max}} n^2(z)dV(z)},
\eeq
where $n(z)$ is the radial distribution of galaxies, including both FKP and systematics weights. This is computed by histogramming the random catalogs in fine redshift bins, weighted by their comoving volume, with numerical values given in Tab.\,\ref{tab: desi-chunks}. 

\subsection{Bispectrum}
\noindent The bispectrum model takes the following form
\beq\label{eq: bispectrum-model}
	B(\vk_1,\vk_2) = B^{\rm tree}(\vk_1,\vk_2) + B^{\rm ct}(\vk_1,\vk_2) + B^{\rm stoch}(\vk_1,\vk_2),
\eeq
where the three terms are the tree-level contribution, counterterms and stochasticity, respectively. 
While the counterterms, strictly speaking, contribute at one-loop order, we include the $k^2$-term in the tree-level model to address fingers-of-God effects, as discussed in \citep{Ivanov:2021kcd}. To include the effect of long-wavelength displacements, the linear matter power spectrum is replaced by its resummed version. Model \eqref{eq: bispectrum-model} can be compared to data by averaging over Fourier-space modes and binning in $k_1,k_2,k_3$ to obtain the bin-averaged bispectrum monopole. We further account for the coordinate distortion effect by a redefinition of wavenumbers and angles \citep[e.g.,][]{Philcox:2021kcw}. 
Similar to \citep{Ivanov:2021kcd}, we adopt the integral approximation by replacing the sum over discrete Fourier modes with a continuous integral over momentum, itself estimated via Gauss-Legendre quadrature. We found that the discreteness effects associated to the finite resolution of the Fourier grid are of order of $\sim1\%$ and can be safely neglected.\footnote{Previous works found larger effects leading to the introduction of discreteness weights \cite{Ivanov:2021kcd} in the continuum mode approximation. Here, the bias is small since (a) we explicitly restrict the continuum mode integration to configurations obeying the triangle conditions when performing the integral, and (b) we use a larger $\kmin=0.02\hMpc$ than in previous works.
} Bispectra are evaluated at the following effective redshift:
\beq
    z^{(3)}_{\rm eff} = \frac{\int_{z_{\rm min}}^{z_{\rm max}} z\,\tilde{n}^3(z)dV(z)}{\int_{z_{\rm min}}^{z_{\rm max}} \tilde{n}^3(z)dV(z)},
\eeq
where $\tilde{n}(z) \equiv n(z)w_B(z)$ is the radial distribution including additional bispectrum weights, $w_B(z)\propto n^{-1/3}(z)$ following \citep{Chen:2024vuf}. By construction, this is equal to the power spectrum effective redshift, which ensures that both statistics depend on the same set of bias parameters.

\subsection{EFT Parameters}
\label{subsec:params}

\noindent Our model for the power spectrum and bispectrum is specified by the following nuisance parameters (cf.\,Tab.\,\ref{tab:priors}),
\beq\label{eq: nuisance-params}
	\{\color{blue}{b_1}\color{black},\color{blue}{b_2}\color{black},\color{blue}{b_{\mathcal{G}_2}}\color{black}, b_{\Gamma_3}\} \times \{c_{0}, c_{2}, c_{4}, \tilde c, \color{red}c_1\color{black}\} \times \{\color{blue}P_{\rm shot}\color{black}, a_0, a_2, \color{red}B_{\rm shot}\color{black}, \color{red}A_{\rm shot}\color{black}\}
\eeq
where the first set describes galaxy bias (linear, quadratic, tidal, and third-order biases respectively), the second gives the counterterms for the power spectrum monopole, quadrupole, and hexadecapole, fingers-of-God effect and bispectrum, while the final set accounts for stochasticity. Parameters in blue enter both the power spectrum and bispectrum, whilst those in black (red) enter just the power spectrum (bispectrum). We introduce separate set of nuisance parameters for each DESI data chunk, leading to 84 free parameters, though all but 18 (six copies of $b_1,b_2,b_{\G_2}$) are analytically marginalized in the likelihoods \citep{Philcox:2020zyp}.
Tab.\,\ref{tab:priors} also details our priors, whose ranges
are motivated by EFT naturalness arguments, as well as measurements from hydrodynamical and $N$-body simulations~\cite{Ivanov:2024xgb,Ivanov:2024dgv}. 

At the power spectrum level, the parameters $c_\ell$ in \eqref{eq: nuisance-params} correspond to the leading-order counterterms for multipoles $\ell=0,2,4$, while $\tilde c$ represents a next-to-leading order $k^4$ redshift-space contribution, which is used to model fingers-of-God effects~\cite{Chudaykin:2020aoj}. In addition, our tree-level bispectrum model includes a $k^2$ fingers-of-God-term, parameterized by the nuisance parameter $c_1$~\cite{Ivanov:2021kcd}. These terms arise by noting that the scale controlling stochastic velocities, $\sigma_{\rm FoG}$, is significantly larger than that of dark matter and halo non-linearities, $k_{\rm NL}^{-1}$ for many realistic galaxy populations. To ensure that our model remains under perturbative control up to $k_{\rm max}$, we can expand the stochastic terms to one-higher-order than the standard mode-coupling contributions (which avoids power counting difficulties due to the scale hierarchy); this results in a set of derivative counterterms $\sim k^4P_{11}$ and $k^2P_{11}^2$ for the power spectrum and bispectrum respectively. Given the statistical precision of the DESI data, we require only a single counterterm in each case, which additionally ensures that the one-loop stochastic terms (e.g., $a_2$) match those obtained from field-level models \citep[e.g.,][]{Ivanov:2024dgv}.

Our stochasticity model for the power spectrum is given by~\cite{Ivanov:2021kcd}
\begin{align}
\label{stoch}
P_{\rm stoch}(k,\mu)&=\frac{1}{\bar n}\left[1+P_{\rm shot} + a_0 \left(\frac{k}{k_{\rm NL}}\right)^2 + a_2\mu^2 \left(\frac{k}{k_{\rm NL}}\right)^2 \right]\,,
\end{align}
where $k_{\rm NL}=0.45~\hMpc$, $\mu\equiv ({\bf k\cdot\hat z})/k$ and $\bar n$ is the background galaxy density. Here, $\Pshot$ captures the residual constant shot-noise, while $a_0$, $a_2$ describe the scale-dependent stochastic corrections at one-loop order. The latter quantities arise as a Taylor expansion in $k^2$ and $k_\parallel^2\equiv (\vk\cdot\hz)^2$ (and from renormalization of contact operators); as such, there is no component scaling as $k^2\mu^4$. We do not include the $k^4\mu^4$ contribution discussed in \citep{Maus:2024dzi,DESI:2024jis} since it enters only at two-loop order in perturbation theory. 

The stochastic contribution to the bispectrum is specified by 
\begin{align}
B_{\rm stoch}(\vk_1,\vk_2,\vk_3)&=\frac{1}{\bar n}b_1^2P^{\rm tree}(k_1)\left[\Bshot+\beta\mu_1^2(\Pshot+\Bshot)+\beta^2\mu_1^4\Pshot\right] +\text{cycl.} + \frac{1}{\bar n^2}A_{\rm shot}
\end{align}
where $\beta=f/b_1$. This involves both the power spectrum parameter $\Pshot$ and two additional stochastic parameters, $\Bshot$ and $\Ashot$, which account for deviations from the Poissonian noise. 
We note that the Poisson shot-noise has been subtracted from the data at the level of estimators; as such, the Poisson limit is reproduced with $\Pshot$, $\Bshot$, $\Ashot\to 0$.
Our stochasticity model is general; it does not impose any assumptions about correlations between the bispectrum and power spectrum of the stochastic overdensity component, as in the Poissonian case.

In the official DESI analyses \citep{DESI:2024jis,DESI:2024hhd}, the parameters $b_{\Gamma_3}$, $a_0$ and $\tilde{c}$ are set to zero. Within EFT, however, these parameters can take non-zero values, with the first two the power spectrum model at the same order as the usual quadratic biases $b_{\mathcal{G}_2}$ and $b_2$, and the third motivated by scale hierarchies (as discussed above).\footnote{It is particularly important to include $b_{\Gamma_3}$ when adding the bispectrum, since this breaks the bias parameter degeneracies arising in the power spectrum.} Moreover, these parameters have been detected for dark matter halos and simulated galaxies~\cite{Ivanov:2024hgq,Ivanov:2024xgb,Ivanov:2024dgv,Sullivan:2025eei}, and correspond to important physical effects, such as halo exclusion. In this work, we vary the above parameters with broad Gaussian priors (noting that $a_n$ are scaled by $k_{\rm NL}^{-2}\bar{n}^{-1}$, cf.\,\ref{stoch}). A detailed comparison between our EFT model and variants that fix $b_{\Gamma_3}$, $a_0$, or $\tilde{c}$ is presented in \S\ref{sec:EFTmodel}.

\subsection{Analysis Procedure}
\label{sec:analysis}

\noindent Table~\ref{tab:priors} summarizes the cosmological and EFT nuisance parameters and the priors applied to them.
\begin{table}[t] 
    \centering
    \begin{tabular}{|lllll|}
    \hline
    data / model & parameter & default & prior & units \\  
    \hline
    \textbf{DESI }
    & $H_0$ &---& $\mathcal{U}[20, 100]$ & $[\kmsMpc]$\\
    & $\omega_b$ &---& $\mathcal{N}(0.02218, 0.00055^2)$ &---\\
    & $n_s$ &---& $\mathcal{N}(0.9649, 0.042^2)$ &---\\  
    & $\omega_c$ &---& $\mathcal{U}[0.001, 0.99]$ &---\\
    & $\ln(10^{10} A_\mathrm{s})$ &---& $\mathcal{U}[1.61, 3.91]$ &---\\    
    \hline    
    \textbf{CMB} 
    & $\tau$ & 0.0544 & $\mathcal{U}[0.01, 0.8]$ &---\\
    & $\omega_b$ &---& $\mathcal{U}[0.005, 0.1]$ &---\\ 
    & $n_s$ &---& $\mathcal{U}[0.8, 1.2]$ &---\\      
    \hline
    \textbf{nuisance} 
    & $b_1 \sigma_8(z)$ &  & $\mathcal{U}[0, 3]$ &---\\  (sampled) 
    & $b_2 \sigma_8^2(z)$ &  & $\mathcal{N}[0, 5^2]$ &---\\    
     & $b_{\mathcal{G}_2} \sigma_8^2(z)$ &  & $\mathcal{N}[0, 5^2]$ &---\\ 
    \hline 
    \textbf{nuisance} & $A\, b_{\Gamma_3}$ &  & $\mathcal{N}\left(\frac{23}{42}(b_1-1),1^2\right)$ &---\\
    (analytically & $A\, c_0$ &  & $\mathcal{N}(0,30^2)$ & $[\Mpc/h]^2$\\
    marginalized) & $A\, c_2$ &  & $\mathcal{N}(30,30^2)$ & $[\Mpc/h]^2$\\
    & $A\, c_4$ &  & $\mathcal{N}(0,30^2)$ & $[\Mpc/h]^2$\\
    & $A^2\, \tilde{c}$ &  & $\mathcal{N}(400,400^2)$ & $[\Mpc/h]^4$\\
    & $A\, c_1$ &  & $ \mathcal{N}(0,5^2)$ & $[\Mpc/h]^2$\\
    & $P_{\rm shot}$ &  & $\mathcal{N}(0,1^2)$ &---\\
    & $a_{0}$ &  & $\mathcal{N}(0,1^2)$ &---\\
    & $a_{2}$ &  & $\mathcal{N}(0,1^2)$ &---\\
    & $A\, B_{\rm shot}$ &  & $\mathcal{N}(0,1^2)$ &---\\
    & $A_{\rm shot}$ &  & $\mathcal{N}(0,1^2)$ &---\\
   \hline 
    \end{tabular}
    \caption{\textbf{Model Parameters}: 
    Parameters and priors used in our analyses. Here, $\mathcal{U}$ refers to a uniform prior in the given range, whilst $\mathcal{N}(\mu,\sigma^2)$ denotes a Gaussian distribution with mean $\mu$ and variance $\sigma^2$. Bias parameters $b_1\sigma_8$, $b_2\sigma_8^2$, $b_{\mathcal{G}_2}\sigma_8^2$ are directly sampled in the MCMC chains, whilst the nuisance parameters that appear quadratically in the likelihood are marginalized over analytically; see text for details. We use $A\equiv\sigma_8^2(z)/\sigma_{8,{\rm ref}}^2(z)$, where $\sigma_{8,{\rm ref}}^2(z)$ is the late-time fluctuation amplitude at the \textit{Planck} 2018 fiducial cosmology. We note that the BBN and $n_s$ priors are added by default in the DESI data analysis, but dropped when we combine with the CMB.
    }
    \label{tab:priors}
\end{table}
For the DESI-only analyses, we vary five cosmological parameters: the Hubble constant $H_0$, the physical densities of cold dark matter and baryons $\omega_c\equiv\Omega_ch^2$ and $\omega_b\equiv\Omega_bh^2$, and the amplitude and spectral index of the primordial density fluctuations, $A_s$ and $n_s$. When using the CMB likelihoods, we additionally include the optical depth to reionization, $\tau$.
Additionally, we have a separate set of EFT nuisance parameters for each of the six DESI data chunks, which are required to model the power spectrum and bispectrum of the DESI DR1 data. 
Some of these parameters are degenerate with the cosmological parameters potentially leading to projection effects. To mitigate these effects, we impose physically motivated priors on the nuisance parameters as described below.

We directly sample the parameters $b_1\sigma_8(z)$, $b_2\sigma_8^2(z)$, $b_{\mathcal{G}_2}\sigma_8^2(z)$ (with $\sigma_8(z)$ evaluated at the effective redshift of the corresponding sample). This somewhat differs from several previous works \citep[e.g.,][]{Philcox:2021kcw,Ivanov:2021kcd} but more closely corresponds to the parameter combinations the data can constrain \citep[e.g.][]{Ivanov:2019pdj,Ivanov:2019hqk,Maus:2024dzi,Tsedrik:2025hmj}.
We adopt Gaussian uninformative priors on combinations involving the quadratic bias parameters $b_2$ and $b_{\mathcal{G}_2}$.\footnote{In power-spectrum-only analyses some bias parameters can take large, nonphysical values; however, they return to agreement with the prediction of galaxy/halo formation models once the higher-order statistics is included (for details see \S\ref{sec:bias}). For consistency, we recommend using the complete theoretical model, which includes the one-loop power spectrum and the tree-level bispectrum.}
Additionally, we rescale the parameters that enter quadratically into the likelihood with the late-time fluctuation amplitude based on how they appear in the theoretical model, and marginalize over those combinations analytically.
This is a more natural choice as the resulting combinations are directly constrained by observation, which effectively reduces degeneracies between the cosmological and nuisance parameters (see \citep{Tsedrik:2025hmj} for a recent discussion). 
This parameterization mitigates the projection effects would favor low values of $\sigma_8\equiv\sigma_8(z=0)$ if the results were to be interpreted
naively~\cite{DESI:2024jis}.
We validate our choice of the priors on mock data in \S\ref{sec: consistency}.

We perform Markov Chain Monte Carlo (MCMC) analyzes to sample from the posterior distributions adopting the Metropolis-Hastings algorithm, as implemented in the \textsc{Montepython} code~\cite{Audren:2012wb,Brinckmann:2018cvx}, with plots and marginalized constraints are obtained using the \textsc{getdist} package \citep{Lewis:2019xzd}.\footnote{\href{https://getdist.readthedocs.io}{https://getdist.readthedocs.io}} As discussed above, we limit our analyses to the $\Lambda$CDM cosmological model in this work, assuming a spatially flat Universe with two massless and one massive neutrino species with $M_\nu=0.06\eV$.

\section{\texorpdfstring{$\Lambda$CDM Constraints}{LCDM Constraints}}
\label{sec: lcdm-results}

\noindent In this section, we present the main cosmological results of our work: $\ld$ parameter constraints obtained from the DESI DR1 full-shape and DESI DR2 BAO measurements, as well as their combination with CMB data. These utilize the redshift-space power spectrum and bispectrum monopole data shown in Figs.~\ref{fig: pk-dat} and~\ref{fig: bk-dat}, respectively. Figs.~\ref{fig: pk-cov} and~\ref{fig: bk-cov} display the diagonal elements of the covariance matrices for the power spectrum and bispectrum.
Fig.~\ref{fig: bk-fc} illustrates the impact of fiber collisions for the LRG3 bispectrum.

\subsection{Cosmological Results}
\label{sec:results}
\begin{figure}[!t]
	\centering
	\includegraphics[width=0.95\textwidth]{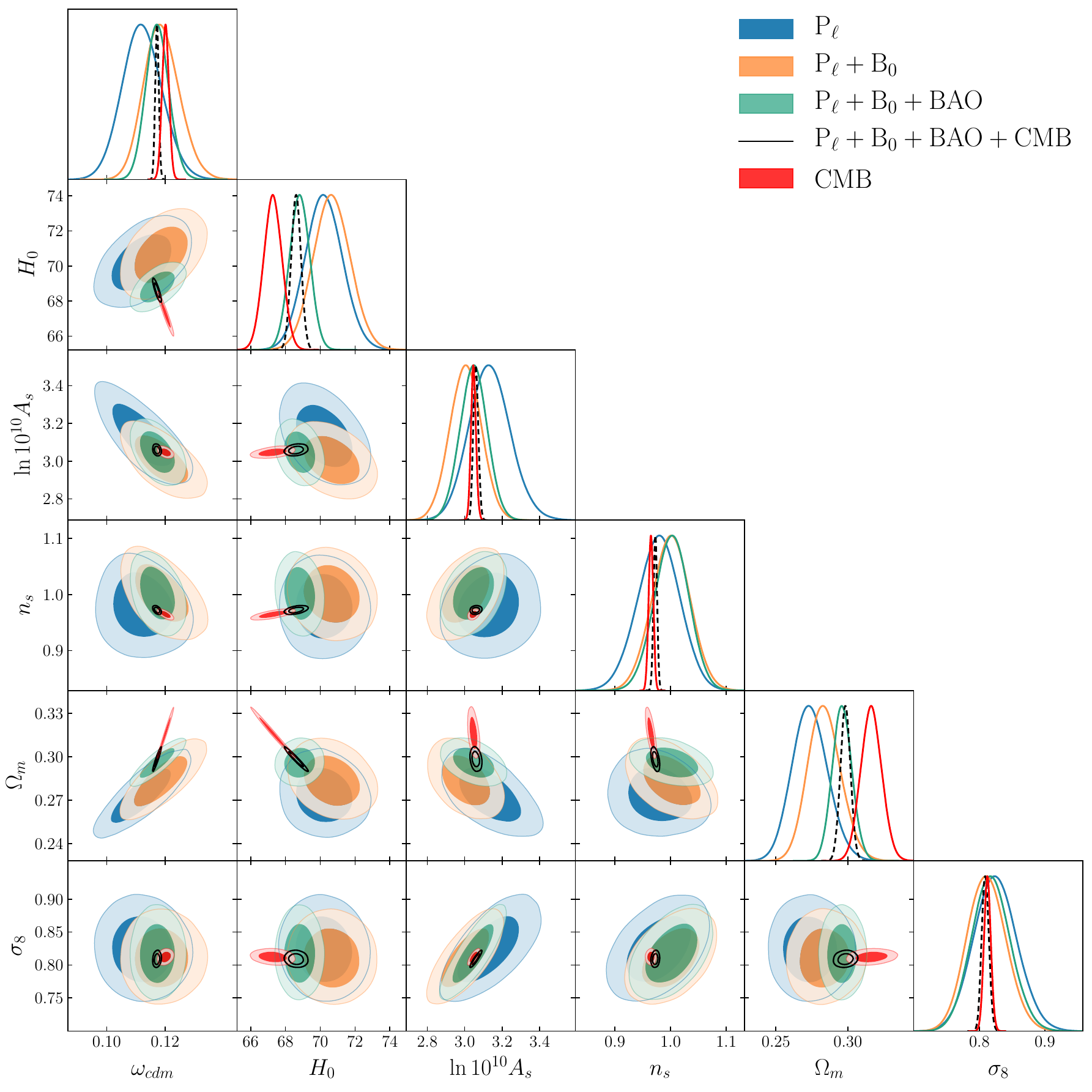}
	\caption{\textbf{Main Parameter Constraints}: Cosmological parameter constraints in the $\Lambda$CDM analyses including our new measurements of the DESI DR1 power spectrum ($P_\ell$) and the large-scale bispectrum monopole ($B_0$), the official DESI DR2 BAO measurements ($\rm BAO$) and the \textit{Planck} cosmic microwave background (CMB) data. For the analyses without the CMB, a BBN prior on the baryon density as well as a weak Gaussian prior on the spectral index $n_s$ were adopted. Tab.\,\ref{tab:main2} shows the corresponding parameter constraints for each analysis shown here. We report no tension between the DESI and CMB datasets, with the main parameters consistent within $2\sigma$.
    }
    \label{fig:main2}
\end{figure}

\noindent Fig.~\ref{fig:main2} shows the main cosmological results of this work, 
with one-dimensional marginalized constraints given in Tab.~\ref{tab:main2}.
Additional parameter constraints, including all bias parameters sampled directly in the MCMC chains, are presented in Tab.~\ref{tab:full}. From Fig.~\ref{fig:main2}, we may draw two immediate conclusions. Firstly, as more datasets are included into the analysis, the posterior means are generally consistent, while the error-bars reduce, indicating that the various datasets are statistically compatible. Secondly, we find consistency between the CMB and LSS datasets, with the main results from the DESI power spectrum, bispectrum and BAO data matching those from the CMB data at $95\%$ confidence level. Below, we consider each analysis in turn.

\begin{table}[!t]
    \centering
  \begin{tabular}{|c|cccc|ccc|} \hline
    \textbf{Dataset} 
    & $\omega_{\rm cdm}$
    & $H_0$ 
    & ${\ln(10^{10}A_s)}$ 
    & $n_s$ 
    & $\Omega_m$ 
    & $\sigma_8$
    & $S_8$
    \\
    \hline
    $P_\ell(k)$ 
    & $\enspace 0.1122_{-0.0067}^{+0.0068}\enspace$ 
    & $\enspace 70.22_{-1.06}^{+1.06}\enspace$ 
    & $\enspace 3.135_{-0.114}^{+0.104}\enspace$ 
    & $0.978_{-0.038}^{+0.038}$ 
    & $\enspace 0.274_{-0.013}^{+0.012}\enspace$ 
    & $\enspace 0.825_{-0.033}^{+0.033}\enspace$ 
    & $\enspace 0.787_{-0.036}^{+0.036}\enspace$ 
    \\
    $P_\ell(k)+B_0(k)$ 
    & $0.1189_{-0.0065}^{+0.0055}$ 
    & $70.67_{-1.05}^{+1.05}$ 
    & $3.005_{-0.084}^{+0.083}$ 
    & $1.001_{-0.034}^{+0.034}$ 
    & $0.284_{-0.012}^{+0.010}$ 
    & $0.811_{-0.031}^{+0.028}$ 
    & $0.789_{-0.035}^{+0.032}$ 
    \\
    $P_\ell(k)+B_0(k)+{\rm BAO}$ 
    & $0.1174_{-0.0039}^{+0.0039}$ 
    & $68.82_{-0.58}^{+0.58}$ 
    & $3.047_{-0.072}^{+0.072}$ 
    & $1.002_{-0.031}^{+0.031}$ 
    & $0.296_{-0.007}^{+0.007}$ 
    & $0.818_{-0.029}^{+0.029}$ 
    & $0.813_{-0.031}^{+0.031}$ 
    \\
    $P_\ell(k)\!+\!B_0(k)\!+\!{\rm BAO}\!+\!{\rm CMB}$ 
    & $0.1172_{-0.0006}^{+0.0006}$ 
    & $68.61_{-0.28}^{+0.28}$ 
    & $3.059_{-0.014}^{+0.012}$ 
    & $0.973_{-0.0033}^{+0.0033}$ 
    & $0.298_{-0.003}^{+0.003}$ 
    & $0.809_{-0.005}^{+0.005}$ 
    & $0.807_{-0.007}^{+0.007}$ 
    \\
    $\,\, {\rm CMB}\,\,$ 
    & $0.1202_{-0.0012}^{+0.0012}$
    & $67.28_{-0.53}^{+0.53}$
    & $3.047_{-0.013}^{+0.013}$ %
    & $0.965_{-0.0041}^{+0.0041}$ 
    & $0.316_{-0.007}^{+0.007}$
    & $0.812_{-0.005}^{+0.005}$
    & $0.834_{-0.012}^{+0.012}$
    \\
  \hline
    \end{tabular}
    \caption{\textbf{Main Parameter Constraints}: Mean and 68\% confidence intervals on $\ld$ cosmological parameters from the main analyses of this work. This is summarized in Tab.\,\ref{tab:main} and includes results from the DESI DR1 power spectrum ($P_\ell$) and bispectrum ($B_0$), DR2 BAO, and the \textit{Planck} primary and lensing anisotropies (CMB). The two-dimensional constraints are shown in Fig.\,\ref{fig:main2}. Additional results obtained in a joint analysis with DR1 BAO are presented in Tab.\,\ref{tab:main2jointbao}\,\&\,Fig.\,\ref{fig:main2jointbao}.
    }
\label{tab:main2}
\end{table}

First, we consider the results obtained using the DESI DR1 power spectrum alone. The $P_\ell$-only analysis provides a $4\%$ constraint on $\sigma_8$, equal to $0.825\pm0.033$, which is in excellent agreement with that inferred from the CMB data (see below for comparison to the official DESI constraints).
In contrast, the posterior means of $\Omega_m$ and $H_0$ deviate more strongly from those obtained in the CMB analysis, at $2.9\sigma$ and $2.5\sigma$, respectively. As we will see shortly, these shifts are driven by unconstrained degeneracy directions at the level of the power spectrum data and disappear when combined with the additional measurements.

The inclusion of the bispectrum tightens the $\omega_{cdm}$ and $\Omega_m$ posteriors by $11\%$ and $8\%$, respectively; this gain is sourced by the information contained within the broadband shape of the bispectrum \cite{Ivanov:2021kcd}.
The central value of $\Omega_m$ is shifted higher by $0.8\sigma$, reducing the difference with the CMB-based value to the $2.4\sigma$ level.
The bispectrum sharpens the $\sigma_8$ constraint by $10\%$ (in accordance with the fractional improvements found for BOSS~\cite{Philcox:2021kcw}).
This improvement is driven by the broadband shape of the bispectrum which helps to break parameter degeneracies, yielding a more accurate estimate of the late-time fluctuation amplitude.
In contrast, the $H_0$ constraint remains largely unchanged -- this occurs because the bispectrum is measured over a limited wavenumber range (recalling that we set $\kmax=0.08\hMpc$) which lacks the BAO information.
The inclusion of the bispectrum has a much larger impact on the constraints on bias parameters, as detailed in \S\ref{sec:bias}.

The addition of the DESI DR2 BAO data provides a significant improvement in the precision of parameter measurements. 
The $H_0$ error-bar reduces by a factor of two; this extra constraining power comes from the high precision transverse distance measurements.
Moreover, the $\omega_{cdm}$ and $\Omega_m$ posteriors shrink by approximately $40\%$. 
Inclusion of the BAO leads to notable shifts in the posteriors: the mean value of $H_0$ decreases by $1.7\sigma$, while $\Omega_m$ increases by $1.1\sigma$.
These shifts bring the parameter constraints into agreement with those from the CMB analysis at the $2\sigma$ level.
This underlines the importance of the post-reconstruction BAO data, which effectively break parameter degeneracies present at the level of the full-shape data.
Note that in this work, we have adopted the {\it Planck} 2018 likelihood; comparing our DESI results to the best-fits from the newer \texttt{HiLLiPoP}+\texttt{LoLLiPoP}  likelihoods~\cite{Tristram:2020wbi,Tristram:2023haj}, based on the \textit{Planck} PR4 maps, we find agreement within $1.8\sigma$ for all parameters.

Finally, we combine our DESI power spectrum, bispectrum, and BAO likelihood with the \textit{Planck} data on the CMB anisotropies (supplemented with ACT lensing). Whilst the CMB contours remain much smaller than those from the galaxy clustering datasets alone, it is clear from Fig.\,\ref{fig:main2} that the addition of the DESI data leads to considerable narrowing and shifts of the posterior densities. This occurs due to the breaking of some degeneracies, especially the geometric degeneracy present in the primary CMB. The joint constraints on $H_0$ and $\omega_{\rm cdm}$ and $S_8$ are significantly enhanced compared to the CMB-only results; the constraint on $\omega_{\rm cdm}$, for example, tightens by a factor of two. These improvements are much larger than those found with the previous galaxy survey (BOSS), which reported a 30\% improvement in $\omega_{\rm cdm}$ when combining the CMB and LSS data~\cite{Ivanov:2019hqk,Philcox:2020vvt}.

\subsection{Comparison to Previous Analyses}
\noindent Next, we compare our results to those from the official DESI DR1 full-shape-only analysis of the power spectrum monopole and quadrupole~\cite[Tab.\,10]{DESI:2024jis}, namely $\Omega_m=0.284^{+0.10}_{-0.11}$, $H_0=70.0\pm1.0$ and $\sigma_8=0.839\pm0.034$. The nominal difference in the means of these three parameters between the official analysis and our $P_\ell$ analysis (including the hexadecapole) is $(0.8\sigma,0.2\sigma,0.4\sigma)$, expressed in units of the standard deviation. 
To explore the effect of including the hexadecapole and to more closely reproduce the DESI setup, we repeat our analysis using only the monopole and quadrupole, adopting the same theoretical model as before (for details see \S\ref{sec:EFTmodel}).
We find very similar cosmological constraints with parameter differences remaining below $0.6\sigma$ compared to the official DESI analysis~\cite{DESI:2024jis}. Considering that the two analyses rely on different analysis pipelines, prior choices, theory codes, power spectrum estimators, covariance estimators, binning choices, and some systematics corrections,
our results are generally in good agreement. 
This test further demonstrates that our results are robust with respect to the inclusion of the hexadecapole in full agreement with the findings of~\cite{Lai:2024bpl}. 

Notably, our power spectrum analysis features slightly larger error-bars on cosmological parameters compared to those of \cite{DESI:2024jis}, with differences up to $15\%$.
This can be attributed to the different treatment of systematic effects, our theoretical model for the covariance matrix, and our conservative bias parameter priors.
When the cubic bias parameter $b_{\Gamma_3}$ is fixed to the coevolution dark matter relation and the stochastic counterterm $a_0$ is set to zero (as in DESI), our errorbars (excluding the hexadecapole) are consistent with those of DESI \cite{DESI:2024jis} to within $10\%$ (for further discussion see \S\ref{sec:EFTmodel}).

We also compare our parameter constraints from the $P_\ell + B_0 + {\rm BAO}$ analysis with those from the official DESI DR1 full-shape-plus-BAO analysis~\citep{DESI:2024hhd}, namely $\Omega_m=0.2962\pm0.0095$, $H_0=68.56\pm0.75$ and $\sigma_8=0.842\pm0.034$.
We find that the mean value of $\Omega_m$ is in near-perfect agreement, while the differences for $H_0$ and $\sigma_8$ are $0.3\sigma$ and $0.7\sigma$, respectively.
Given that our analysis incorporates the newer DESI DR2 BAO measurements and includes the bispectrum, this level of agreement with the official DESI DR1 results is very good. When performing a joint analysis with the DR1 BAO using an analytic cross-covariance, we find even closer agreement, as discussed in Appendix \ref{app:joint-bao}.
Fig.~\ref{fig:comp} summarizes our results and those from the official DESI DR1 full-shape-only~\cite{DESI:2024jis} and full-shape-plus-BAO analysis~\citep{DESI:2024hhd}.

\begin{figure}[!t]
    \hspace{2em}
    \includegraphics[width=0.42\textwidth]{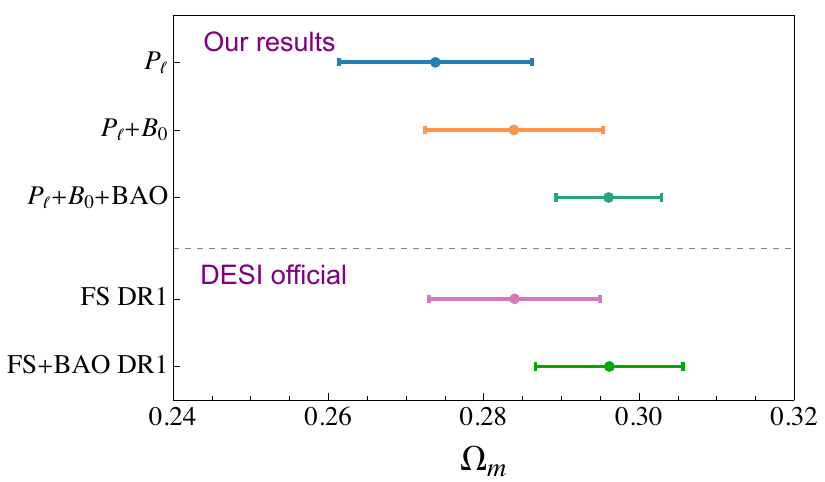}
    \includegraphics[width=0.332\textwidth]{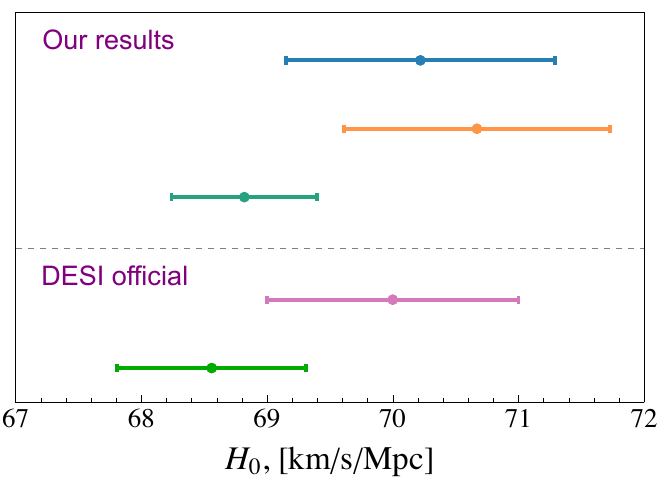}
    \includegraphics[width=0.45\textwidth]{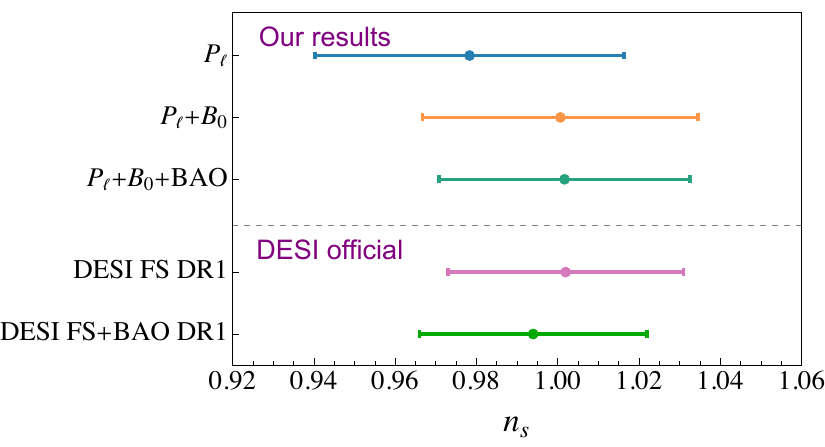}
    \includegraphics[width=0.33\textwidth]{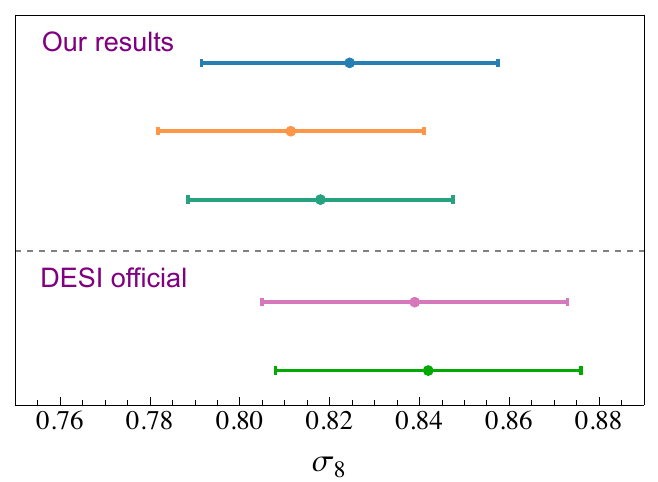}
	\caption{\textbf{Comparison with the DESI analyses}: Constraints on the main cosmological parameters from the our analyses of the $P_\ell$, $P_\ell+B_0$ and $P_\ell+B_0+{\rm BAO}$ data, as well as from the official DESI DR1 full-shape-only~\cite{DESI:2024jis} and (DR1 only) full-shape-plus-BAO~\citep{DESI:2024hhd} analyses.
    Error-bars show the 68\% confidence intervals. Our results are broadly in agreement, though slight differences arise due to the differences in the covariance matrix and bias parameter priors.
    }
    \label{fig:comp}
\end{figure}

It is also interesting to compare our results with the previous analysis of the BOSS DR12 power spectrum multipoles and bispectrum monopole~\cite{Philcox:2021kcw}. 
We find a very good agreement for all parameters with the exception of $\sigma_8$, which is higher in our analysis and closely aligns with the {\it Planck} value. This is likely a combination of (a) our use of rescaled EFT priors, which minimize projection effects (as discussed in \S\ref{sec:analysis}), and (b) differences in the dataset itself, as evidenced through the differing $\sigma_8$ results obtained from analyses of BOSS and DESI lensing cross-correlations \citep[e.g.,][]{Maus:2025rvz,Chen:2024vuf}, frequentest
tests~\cite{Philcox:2021kcw,Chen:2021wdi},
and the stability of 
low $\sigma_8$ results
with respect to physically consistent choices of 
priors~\cite{Ivanov:2024xgb}.

\subsection{Galaxy Bias Relations}
\label{sec:bias}

\noindent Galaxy bias is a key characteristic of perturbation theory models. 
The quadratic and tidal biases, $b_2$ and $b_{\mathcal{G}_2}$, enter the power spectrum in a degenerate manner, making it difficult to measure these parameters from the power spectrum data alone.
The tree-level bispectrum introduces additional shape dependencies, allowing these parameters to be directly constrained from the data. 
The bispectrum information can also be used to test the popular bias relations for dark matter halos~\cite{Desjacques:2016bnm,Lazeyras:2015lgp} and to assess consistency with simulation-based measurements~\cite[e.g.,][]{Ivanov:2024dgv,Ivanov:2024xgb,Zhang:2024thl}.

\begin{figure}[!t]
	\centering
	\includegraphics[width=0.95\textwidth]{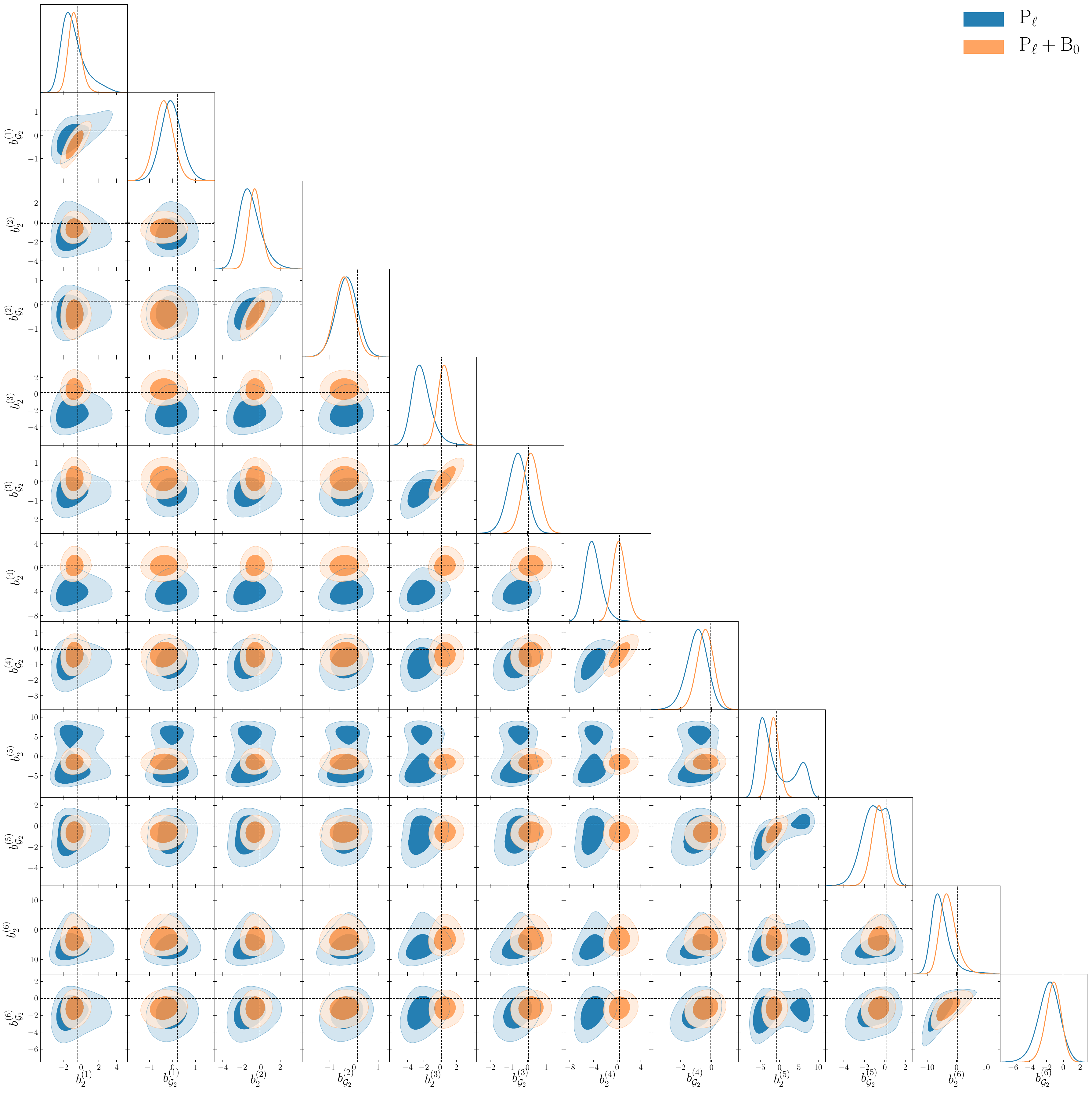}
	\caption{\textbf{Bias Parameters}: Quadratic bias parameters from the analyses of the $P_\ell$ (blue) and the $P_\ell+B_0$ (orange). Vertical and horizontal lines show the predictions for dark matter halos extracted from the   \textsc{Quijote} simulations from~\cite{Ivanov:2024xgb}.
    The corresponding parameter constraints are shown in Tab.~\ref{tab:full}. The bispectrum data significantly sharpens constraints on $b_2$ and $b_{{\mathcal G}_2}$ for all DESI data chunks.
    The $P_\ell+B_0+{\rm BAO}$ analysis yields nearly identical constraints to that from $P_\ell+B_0$ data and is thus omitted from this figure. Upper scripts
    (1)-(6) correspond to BGS, LRG1, LRG2, LRG3, ELG2, and QSO samples, respectively.
    }
    \label{fig:bias}
\end{figure}

Fig.~\ref{fig:bias} shows the constraints on the quadratic galaxy bias parameters for the six DESI data chunks,
with one-dimensional marginalized constraints provided in Tab.~\ref{tab:full}.
The dashed lines represent the direct measurements of $b_2$ and $b_{\mathcal{G}_2}$ for dark matter halos, obtained from the high-resolution \textsc{Quijote} simulations~\cite{Villaescusa-Navarro:2019bje}.
Specifically, we plot the cubic fits to the halo $b_2(b_1)$ and $b_{\mathcal{G}_2}(b_1)$ measurements from~\cite{Ivanov:2024xgb};
\begin{align}
 \label{quij1}
& b_2(b_1) = -0.016  b_1^3 +  0.39 b_1^2+  0.013 b_1 -1.50  \\
 \label{quij2}
& b_{\mathcal{G}_2}(b_1)=0.016 b_1^3 -0.50 b_1^2
+1.40 b_1 -0.80 
\end{align}
where $b_1$ is evaluated at its marginalized mean value obtained from the $P_\ell + B_0$ analysis for each sample.
In the absence of the bispectrum, we find highly non-Gaussian posteriors, with much of the $b_2$ posteriors below zero. 
This behavior is in tension with the simulation-based halo bias relation given above for almost all data chunks.
When bispectrum is included, the posterior widths shrink dramatically; for the ELG2 sample the constraints on $b_2$ and $b_{{\mathcal G}_2}$ are improved by $70\%$ and $40\%$, respectively.
The inclusion of the bispectrum also shifts the posterior means closer to the \textsc{Quijote} halo values (described below).

\begin{figure}[!t]
	\centering
	\includegraphics[width=0.48\textwidth]{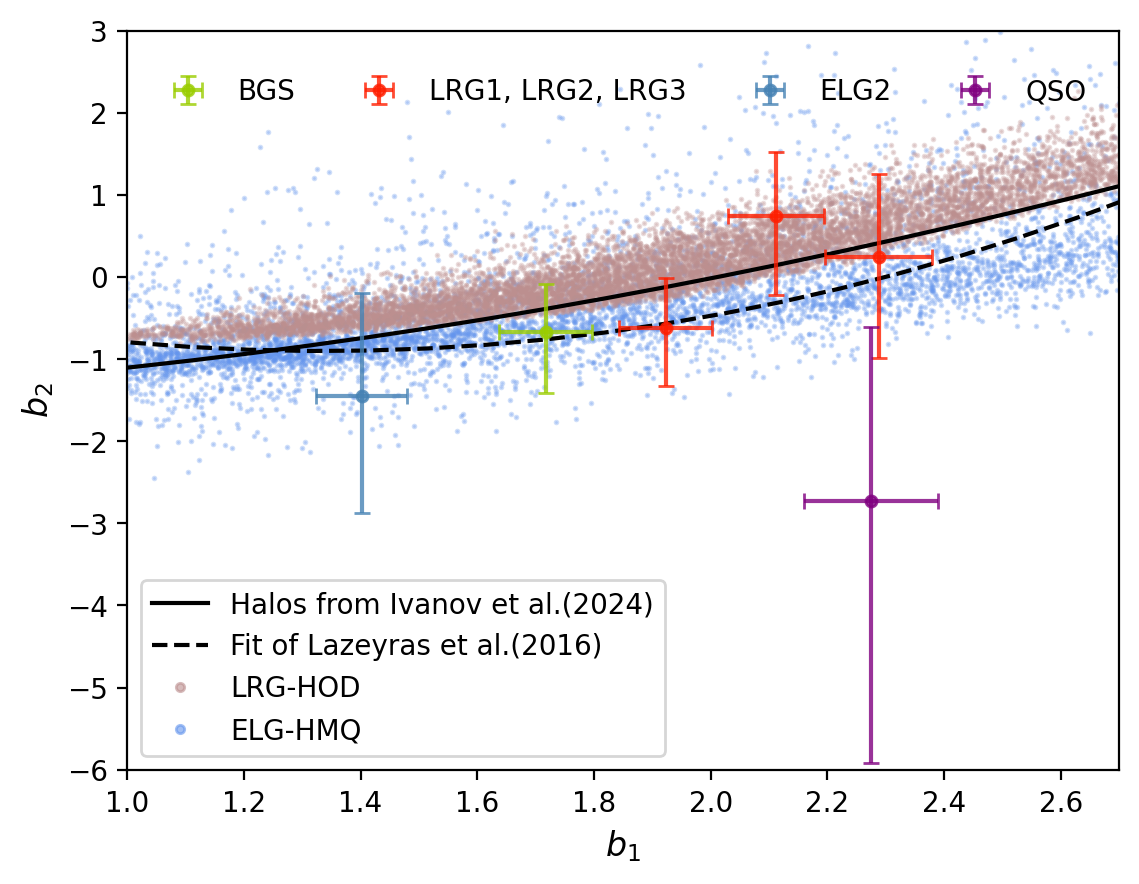}
    \includegraphics[width=0.48\textwidth]{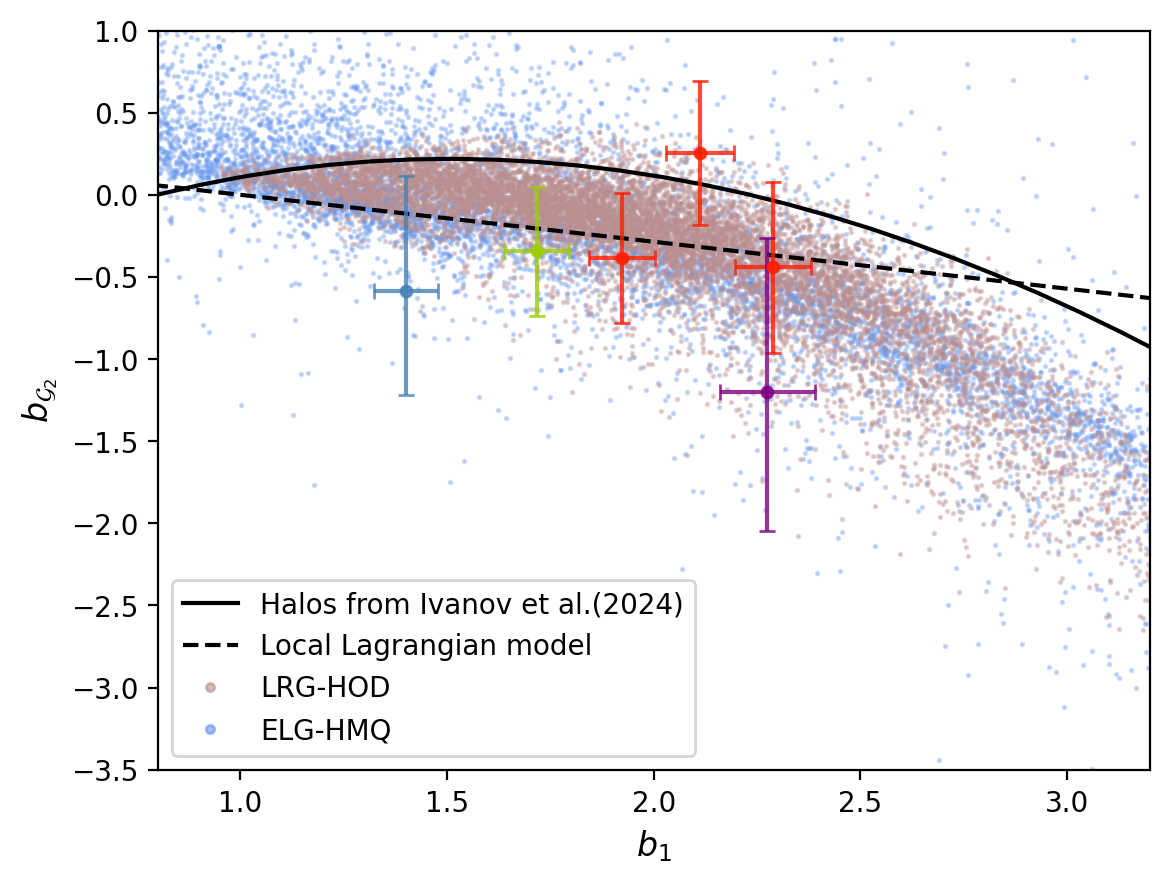}
	\caption{\textbf{Bias Relations}: The linear, quadratic and tidal bias parameters extracted from an analysis of the $P_\ell+B_0$ data for the six DESI data chunks: BGS (green), three LRGs (red), ELG2 (blue) and QSO (purple). The black curves represent
    results for dark matter halos:
    the solid curve
    depicts 
    the fit to the direct measurements from \textsc{Quijote} halo catalogs~\cite{Ivanov:2024xgb}, whilst the dashed curves show the peak-background split 
    prediction for $b_2$ from~\cite{Lazeyras:2015lgp} and the local Lagrangian bias 
    model prediction for  $b_{{\mathcal G}_2}$, respectively~\cite{Desjacques:2016bnm} The bias parameters measured from the LRG-HOD and ELG-HMQ mock catalogs~\cite{Ivanov:2024dgv} are shown as discrete data points.
    }
    \label{fig:bias2}
\end{figure}

Fig.~\ref{fig:bias2} presents the measured quadratic and tidal biases from the $P_\ell+B_0+{\rm BAO}$ analysis as a function of the linear bias, $b_1$, and compares them with commonly used halo bias relations and direct simulation-based measurements.
In particular, we show the popular predictions for $b_2(b_1)$ based on the background-split argument~\cite{Lazeyras:2015lgp}, and for $b_{\mathcal{G}_2}(b_1)$ within the local Lagrangian model~\cite{Desjacques:2016bnm}. 
Also shown are direct measurements from the \textsc{Quijote} halo catalogs~\cite{Ivanov:2024xgb}, as well as from realistic galaxy mocks~\cite{Ivanov:2024dgv,Ivanov:2024xgb}, constructed using the Halo Occupation Distribution (HOD) and High Mass Quenched (HMQ) models for LRGs and ELGs tracers, respectively.

For $b_2(b_1)$, our measurements are consistent with both the predictions for dark matter halos and the results obtained from the HOD/HMQ mock catalogs.
The $b_2$ values from the LRG-HOD mocks lie systematically above the halo curve from the \textsc{Quijote} simulation, while those from ELG-HMQ mocks tend to lie below it; however these differences are too small to be detected in the DESI data.
Future data releases will increases sensitivity to the non-linear bias parameters which helps to identify deviations from the dark matter halo bias relation.

For $b_{\mathcal{G}_2}(b_1)$, the situation is different. While the DESI measurements remain broadly consistent with dark matter halo predictions, all points -- except for LRG2 -- fall systematically below the halo curve.\footnote{Notably, the Local Lagrangian model tends to underestimate $b_{\mathcal{G}_2}$ relative to the \textsc{Quijote} results, suggesting that the Local Lagrangian model provides less accurate predictions.}
The LRGs and ELG2 measurements show better agreement with the results derived from the HOD/HMQ-based galaxy catalogs.
Our results provide
an important 
evidence for the departure
of the bias parameter of 
galaxies away from the 
dark matter halos. Such departures are 
expected in theory,
and can be easily 
described in the context
of the halo model~\cite{Akitsu:2024lyt,Ivanov:2025qie}.  
Hence, our results
clearly suggest against the practice
of applying the halo bias relations
to galaxies.

\section{Consistency Tests}
\label{sec: consistency}

\noindent The analysis pipeline associated with the \textsc{class-pt} code has been validated using several high-fidelity simulation suites~\cite{Chudaykin:2020ghx,Chudaykin:2020hbf,Ivanov:2021zmi,Chudaykin:2022nru,Philcox:2021kcw,Chudaykin:2024wlw}
and two masked data challenges~\cite{Nishimichi:2020tvu,Beyond-2pt:2024mqz}.\footnote{The robustness of the `unwindowed' estimators used in this work was demonstrated on the \textsc{nseries}
and \textsc{beyond-2pt}
catalogues which implement the effects of the survey geometry and selection functions; for details see Refs.~\cite{Chudaykin:2020ghx,Philcox:2021kcw,Beyond-2pt:2024mqz}. We further stress that the windowed and unwindowed estimators differ only by an invertible matrix multiplication jointly applied to the data, theory, and covariance.}
There are two main sources of concern: (1) a theory-driven systematic bias arising from truncating the perturbation theory at a given order; (2) prior volume projection effects from the marginalization over nuisance parameters in the presence of non-Gaussian posterior correlations.\footnote{There are also the prior weight effects, which arise from a restrictive choice of priors when the location and width of the prior on some parameter pulls the posterior away from the value preferred by the data. We adopt conservative priors on all nuisance parameters, such that this effect is negligible in our analyses.}
These are very different in nature: the error arising from the inaccuracy of the theoretical model is a genuine systematic bias, whilst shifts caused by the prior volume effects vanish when using more informative priors or larger data volumes. 
For the scale-cuts adopted in this work, the true theory-systematic bias was previously found to be small, with the largest systematic shift in $\sigma_8$ being $\sim1\%$ for LRGs~\citep{Ivanov:2021kcd,Philcox:2021kcw}.
No theory systematic bias was detected in EFT analyses of ELGs up to $k_{\rm max}=0.25~\hMpc$~\citep{Ivanov:2021zmi} and quasars up to $k_{\rm max}=0.3~\hMpc$~\citep{Chudaykin:2022nru}.\footnote{The theory-systematic bias in principle depends on the galaxy selection, redshift, number density etc, but these parameters from the simulations used in~\citep{Ivanov:2021kcd,Philcox:2021kcw,Ivanov:2021zmi,Chudaykin:2022nru} are quite similar to the DESI samples.} 
This constitutes a small and negligible fraction of the statistical uncertainty in the actual data analysis, indicating that the EFT model for the power spectrum and bispectrum described above is accurate and the associated scale cuts are appropriate.

Whilst the shifts induced by prior volume effects do not correspond to a physical bias (appearing just due to Bayes' theorem), they can complicate interpretation of the Bayesian marginalized posteriors. 
If the conservative priors are used, the marginalization shifts are more significant for current spectroscopic surveys for which the prior weight of the posterior may be more significant than the likelihood contribution. A multi-chunk analysis of the four simulation samples~\cite{Chudaykin:2024wlw} used to reproduce the actual BOSS data analysis, showed a $\sim 2\sigma$ marginalization shift in $\sigma_8$ with respect to the ground truth when adopting the EFT pipeline used in \cite{Philcox:2021kcw}.
This occurs since the magnitude of projection effects increases with the number of nuisance parameters (or equivalently, with the number of data chunks). 
In this work, we adopt EFT priors
similar to the ones used in \cite{Philcox:2021kcw}
in terms of the allowed range of the EFT parameters. 
However, we impose our priors in 
a different form, specifically multiplying the bias 
terms by some powers of $\sigma_8$ in order to mitigate the projection effects, as detailed in \S\ref{sec:analysis}. 

In the following section, we validate the updated analysis pipeline and quantify the impact of projection effects for the different analyses. We further present additional tests of our pipeline, assessing the impact of the hexadecapole moment and the effect of fixing certain EFT parameters. We additionally compare results obtained using unwindowed and conventional power spectrum and bispectrum estimators.

\subsection{Parameter Projection Effects}
\label{sec:proj}

\noindent To test the so-called
parameter projection effects (see \citep[e.g.,][]{Ivanov:2019pdj,Chudaykin:2020ghx,Philcox:2021kcw,Chudaykin:2024wlw,Paradiso:2024yqh,Tsedrik:2025hmj} for the original discussions
in the context of the EFT-based full-shape analysis and some recent work) we generate a noiseless synthetic data vector using the \textsc{class-pt} pipeline. 
As a first step, we compute the best-fit model from the true $P_\ell+B_0$ data analysis (cf.\,\S\ref{sec: lcdm-results}), fitting 
the cosmological parameters $\{\omega_c$, $H_0$, $\ln(10^{10} A_\mathrm{s})\}$ along with six copies of ($b_1\sigma_8$, $b_2\sigma_8^2$, $b_{\mathcal{G}_2}\sigma_8^2$), corresponding to each DESI data chunk included in the full-shape analysis, while fixing $\omega_b=0.02218$ and $n_s=0.9649$.
\footnote{Since they enter the likelihood quadratically, the nuisance parameters that are analytically marginalized can be recovered from the chains {\it a posteriori}.}
This model, which includes the power spectrum and bispectrum statistics at six different redshifts, simulating the DESI samples, is then used as mock data, assuming the same covariance as in previous sections.
Additionally, we generate a synthetic noiseless BAO DESI DR2 post-reconstructed data vector for the BGS, LRG1, LRG2, LRG3+ELG1, ELG2, QSO samples, as well as for the Lyman-$\alpha$ forest auto-correlation and cross-correlation with the QSOs, replicating the BAO measurements described in \S\ref{sec:data_extra}. Secondly, we fit the generated mock data with the same theoretical model used in the actual main analysis.
Since the mock data vector is generated with the same theory pipeline used in the analysis, the true input parameter vector will exactly minimize the mock data likelihood. Thus, any shifts in the posterior means away from the truth encode prior volume effects. This provides a clean test of marginalization-induced biases.

\begin{figure}[!t]
	\centering
	\includegraphics[width=0.95\textwidth]{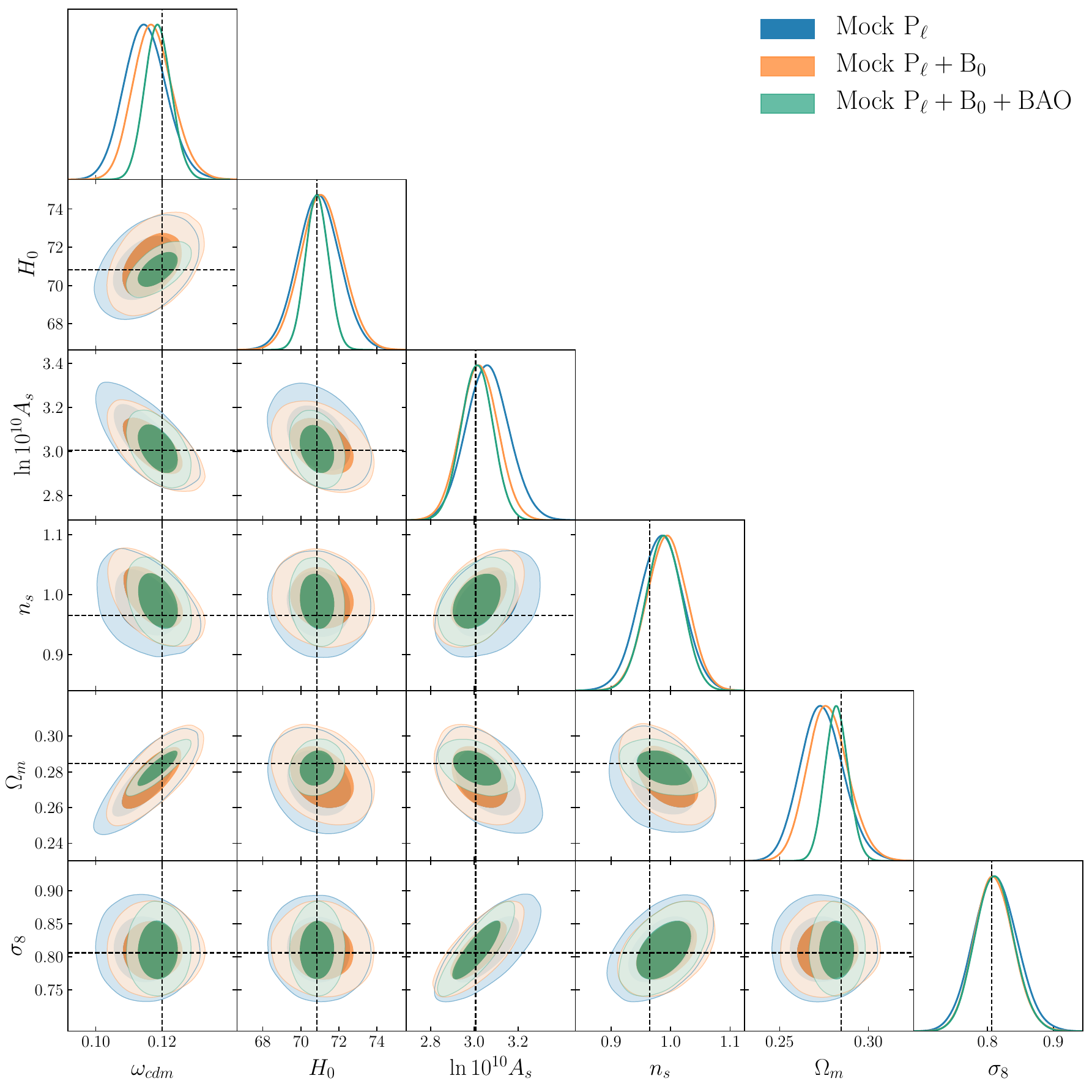}
	\caption{\textbf{Projection Effects}: Cosmological parameter constraints estimated from the synthetic mock data computed with our pipeline. The results are shown for three data combinations: $P_\ell$ (blue), $P_\ell+B_0$ (orange) and $P_\ell+B_0+{\rm BAO}$ (green), emulating our three DESI analyses. The corresponding shifts in the posterior means from the input parameters (shown in dashed lines) are presented in Tab.~\ref{tab:synth}. We find good recovery of the input cosmological parameters indicating that prior volume effects are well controlled.
     }
    \label{fig:synth}
\end{figure}
\begin{table}[!t]
    \centering
  \begin{tabular}{|c|cccccc|} \hline
    \textbf{Mock data} 
    & $\omega_{\rm cdm}$
    & $H_0$ 
    & ${\ln(10^{10}A_s)}$ 
    & $n_s$
    & $\Omega_m$ 
    & $\sigma_8$
    \\
    \hline
    $P_\ell$
    & $-0.77\sigma$ %
    & $0.09\sigma$ %
    & $0.56\sigma$ %
    & $0.52\sigma$ %
    & $-0.87\sigma$ %
    & $0.18\sigma$ %
    \\
    $P_\ell+B_0$
    & $-0.44\sigma$
    & $0.24\sigma$ 
    & $0.19\sigma$ 
    & $0.82\sigma$ 
    & $-0.63\sigma$ 
    & $0.14\sigma$ 
    \\
    $P_\ell+B_0+{\rm BAO}$
    & $-0.28\sigma$
    & $0.04\sigma$ 
    & $0.07\sigma$ 
    & $0.77\sigma$ 
    & $-0.37\sigma$ 
    & $0.16\sigma$
    \\
  \hline
    \end{tabular}
    \caption{\textbf{Projection Effects}: Differences between the true and posterior parameter values obtained from an analysis of synthetic mock data. We show results for three data combinations, expressed in units of the standard deviation from the corresponding analysis. Whilst we find some $\lesssim 1\sigma$ shifts, these reduce when the data vector is enlarged and are small for key cosmological parameters such as $\sigma_8$. We caution that $n_s$ cannot be meaningfully constrained from the DESI data alone.
    }
\label{tab:synth}
\end{table} 

Fig.~\ref{fig:synth} shows the one- and two-dimensional marginalized posterior distributions for the cosmological parameters by analyzing different combinations of the noiseless mock data, with the impacts of projection effects shown in Tab.\,\ref{tab:synth}. As more mock datasets are included into the analysis, the magnitude of the shifts generally decreases, matching our expectation. In the $P_\ell$ analysis, all the parameters are recovered within a $68\%$ confidence level (CL) with the largest discrepancies observed for $\omega_{cdm}$ and $\Omega_m$. Our results are consistent with those of the simulation-based DESI analyses~\cite{DESI:2024jis}.\footnote{The DESI collaboration quantifies the prior volume effects using the 25 Abacus DESI DR1 complete mocks. 
While this approach provides reliable estimates of prior volume effects, it can, strictly speaking, be sensitive to statistical noise in the simulations and theory-driven systematics associated to the theoretical pipeline.
In contrast, tests based on the noiseless synthetic mock data vector -- generated with the same theory pipeline -- are unaffected by these sources of uncertainties and thus represent a cleaner estimate of prior volume effects.}
The addition of the bispectrum further reduces parameter shifts to $\lesssim0.4\sigma$ for all parameters, except for $n_s$, which demonstrates a larger marginalization bias at the level of $0.82\sigma$.
This bias can be attributed to the fact that the DESI full-shape data poorly constrain the spectral slope which increases degeneracies with other parameters and leads to a larger marginalization shift. As such, the interpretation of the results for $n_s$ should be done with caution; this situation is expected to improve with the inclusion of additional data as in~\cite{Chen:2024vuf}, or with the use of simulation-based priors~\cite{Cabass:2024wob,Ivanov:2024hgq,Ivanov:2024xgb,Akitsu:2024lyt,Chen:2025jnr,Ivanov:2025qie} (see also \cite{DESI:2025wzd}).
When adding the mock BAO dataset, the marginalization shifts are less than $\approx0.3\sigma$ for all parameters except $n_s$. Importantly, the shifts in $\sigma_8$ are small; this is due to the rescaled priors defined in \S\ref{sec:analysis}.
All in all, results demonstrate that our analysis pipeline is robust and can be applied to DESI data to yield accurate 
and easily interpretable parameter constraints.\footnote{Ref.~\citep{DESI:2024jis} found that the power spectrum hexadecapole sourced significant parameter projection effects; we find a different behavior in this work since we do not include the (two-loop-order) $k^4\mu^4$ stochasticity term.}

We do not quantify the magnitude of projection effects in the presence of the CMB data for the following two reasons. First, generating reliable CMB mock data is challenging due to the complexity of modeling foregrounds and instrument noise across multiple frequencies. Second, the parameter shifts caused by marginalization effects are significantly suppressed once CMB data are included, becoming a negligible fraction of the statistical uncertainty~\cite{DESI:2024hhd}.

\subsection{Effects of the Hexadecapole and Fixing EFT Parameters}
\label{sec:EFTmodel}

\noindent Here, we present a series of robustness tests of our theoretical pipeline. We re-analyze the full-shape DESI data without the hexadecopole moment and assess the impact of adopting more restrictive models for the EFT bias parameters.
The results of these tests are summarized in Tab.~\ref{tab:test}.
\begin{table}[!t]
    \centering
    \renewcommand{\arraystretch}{1.5}
  \begin{tabular}{|c|ccc|ccc|} \hline
    \textbf{Dataset} 
    & $\omega_{\rm cdm}$
    & $H_0$ 
    & ${\ln(10^{10}A_s)}$ 
    & $\Omega_m$ 
    & $\sigma_8$
    & $S_8$
    \\
    \hline
    $P_0(k),P_2(k)$ 
    & $\enspace 0.1133_{-0.0072}^{+0.0066}\enspace$ %
    & $\enspace 70.23_{-1.07}^{+1.07}\enspace$ %
    & $\enspace 3.118_{-0.117}^{+0.107}\enspace$ %
    & $\enspace 0.276_{-0.014}^{+0.012}\enspace$ %
    & $\enspace 0.824_{-0.034}^{+0.034}\enspace$ %
    & $\enspace 0.790_{-0.037}^{+0.037}\enspace$ %
    \\
    $P_0(k),P_2(k);\,b_{\Gamma_3}^{\rm coev},\,a_0=0$ 
    & $0.1156_{-0.0068}^{+0.0060}$ %
    & $70.47_{-1.05}^{+1.05}$ %
    & $3.082_{-0.102}^{+0.095}$ %
    & $0.279_{-0.013}^{+0.011}$ %
    & $0.825_{-0.033}^{+0.033}$ %
    & $0.796_{-0.036}^{+0.036}$ %
    \\
    $P_0(k),P_2(k);\,b_{\Gamma_3}^{\rm coev},\,a_0=0,$  
    & \multirow{2}{*}{$0.1162_{-0.0069}^{+0.0061}$} %
    & \multirow{2}{*}{$70.48_{-1.05}^{+1.05}$} %
    & \multirow{2}{*}{$3.060_{-0.105}^{+0.096}$} %
    & \multirow{2}{*}{$0.280_{-0.013}^{+0.012}$} %
    & \multirow{2}{*}{$0.818_{-0.033}^{+0.033}$} %
    & \multirow{2}{*}{$0.790_{-0.037}^{+0.034}$} %
    \\
    $\tilde c=400\,[\Mpc/h]^4$ 
    &  %
    &  %
    &  %
    &  %
    &  %
    &  %
    \\
    $P_0(k),P_2(k);\,b_{\Gamma_3}^{\rm coev},\,a_0,\tilde c=0$
    & $0.1160_{-0.0068}^{+0.0059}$ 
    & $70.51_{-1.05}^{+1.05}$ 
    & $3.107_{-0.096}^{+0.096}$ 
    & $0.279_{-0.013}^{+0.011}$ 
    & $0.838_{-0.033}^{+0.033}$ 
    & $0.809_{-0.037}^{+0.037}$ 
    \\\hline
    DESI DR1 $P_0(k),P_2(k)$
    & $-$ 
    & $70.0\pm1.0$ 
    & $3.10\pm0.1$ 
    & $0.284^{+0.10}_{-0.11}$ 
    & $0.839\pm0.034$ 
    & $-$ 
    \\
  \hline
    \end{tabular}
    \caption{\textbf{Robustness Tests}: Mean and 68\% confidence intervals on $\ld$ cosmological parameters from analyses of the DESI DR1 power spectrum monopole and quadrupole (excluding the hexadecopole), using four different prescriptions for the EFT bias parameters: (1) the baseline model used in the main analysis (cf.\,Tab.\,\ref{tab:priors}); (2) a model with $b_{\Gamma_3}$ fixed to the coevolution dark matter prediction and $a_0 = 0$; (3) the same as model (2), but additionally fixing $\tilde c=400\,[\Mpc/h]^4$; (4) the same as model (2), but with $\tilde c = 0$, analogous to the official DESI analysis. The priors are applied to all DESI data chunks. We also quote the results from the official DESI DR1 full-shape-only analysis of the power spectrum monopole and quadrupole~\cite[Tab.\,10]{DESI:2024jis}.
    }
\label{tab:test}
\end{table}

The inclusion of the $\ell=4$ power spectrum moment shifts the means of the cosmological parameters by less than $0.2\sigma$.
This slightly reduces the error-bars with the largest effect observed for $\Omega_m$, where the uncertainty decreases by $5\%$.
These findings demonstrate the robustness of the main cosmological results with respect to the inclusion of the hexadecapole moment, consistent with the conclusions of~\cite{Lai:2024bpl}.

Next, we assess the robustness of the simplified EFT frameworks used by the DESI collaboration. When fixing the cubic bias parameter, $b_{\Gamma_3}$ and the stochastic counterterm $a_0$, the shifts in posterior means remain below $0.3\sigma$, and the corresponding statistical uncertainties are reduced by up to $10\%$.
We emphasize that, although this simplified EFT model works well for the power spectrum, it can induce larger shifts in the posteriors when adding the bispectrum, which sources tighter constraints on the non-linear bias parameters (for details see \S\ref{sec:bias}).
When further fixing the next-to-leading order counterterm parameter $\tilde c$ (describing fingers-of-God effects), the means of parameter constraints shift by less $0.2\sigma$, while the corresponding standard deviations remain essentially intact. In this test, $\tilde c$ was fixed to a non-zero value of $400\,[\Mpc/h]^4$ across all DESI data chunks. This choice is motivated by our baseline analysis (cf.\,\S\ref{sec: lcdm-results}), which shows evidence for a non-zero $\tilde c$ for all DESI data chunks except for ELG2 and QSO.
This indicates enhanced velocity dispersion induced by the
strong fingers-of-God effect, which breaks the naive EFT counting and thus necessitates the inclusion of the next-to-leading order redshift-space contribution.
To test it, we set $\tilde c=0$ and find larger deviations in the posterior means: the inferred value of $\sigma_8$ changes by $0.4\sigma$ compared to the analysis with free $\tilde c$, and by $0.6\sigma$ relative to the analysis with fixed value $\tilde c=400\,[\Mpc/h]^4$.
These results show the importance of including the next-to-leading order counterterm parameter and the non-trivial impact of non-linear redshift space distortions; fixing $\tilde{c}$ to zero introduces moderate shifts in the parameter recovery. 

\subsection{Comparing Windowed and Unwindowed Estimators}
\label{subsec:window}

\noindent Finally, we validate the unwindowed estimators used in this work. To this end, we compute the standard `windowed' estimates of the DESI power spectrum and bispectrum using \polybin, alongside the associated theory matrices (encoding the window convolution) and covariances. Unlike in the fiducial analysis, we require a theory matrix for the bispectrum (as well as the power spectrum), which applies the window to the theoretical model.

If we do not impose scale cuts, the windowed and unwindowed bispectrum pipelines are equivalent, since they differ only by a matrix multiplication (cf.\,\S\ref{subsubsec: pk-limit}).\footnote{Formally, this requires the same binning for the theory matrix and the data such that the theory matrix is square. Given the small signal-to-noise of the bispectrum, this is usually a good approximation, provided the $k$-bins are fairly thin.} In practice, however, there are considerable differences in their application. For the windowed estimator, we must compute the theoretical bispectrum across a wide range of scales before applying the theory matrix, whilst for the unwindowed estimator, we require the theory only in the scale-range of interest (since we have already approximately deconvolved the mask). This results in considerably slower sampling in the windowed case due to the fast increase in bispectrum dimensionality with $k_{\rm max}$ and the need to apply a matrix multiplication at each step in the MCMC chain. Moreover, one requires a model for the bispectrum beyond $k\approx 0.1~\hMpc$, which should properly include higher-order terms outside the regime of validity. The unwindowed estimator requires stochastic computation of a normalization matrix; this is equivalent to the window matrix however (which is difficult to compute analytically \citep[cf.,][]{Pardede:2022udo}), and is itself computationally inexpensive compared to the Gaussian covariance.

\begin{figure}
    \centering
    \includegraphics[width=0.48\linewidth]{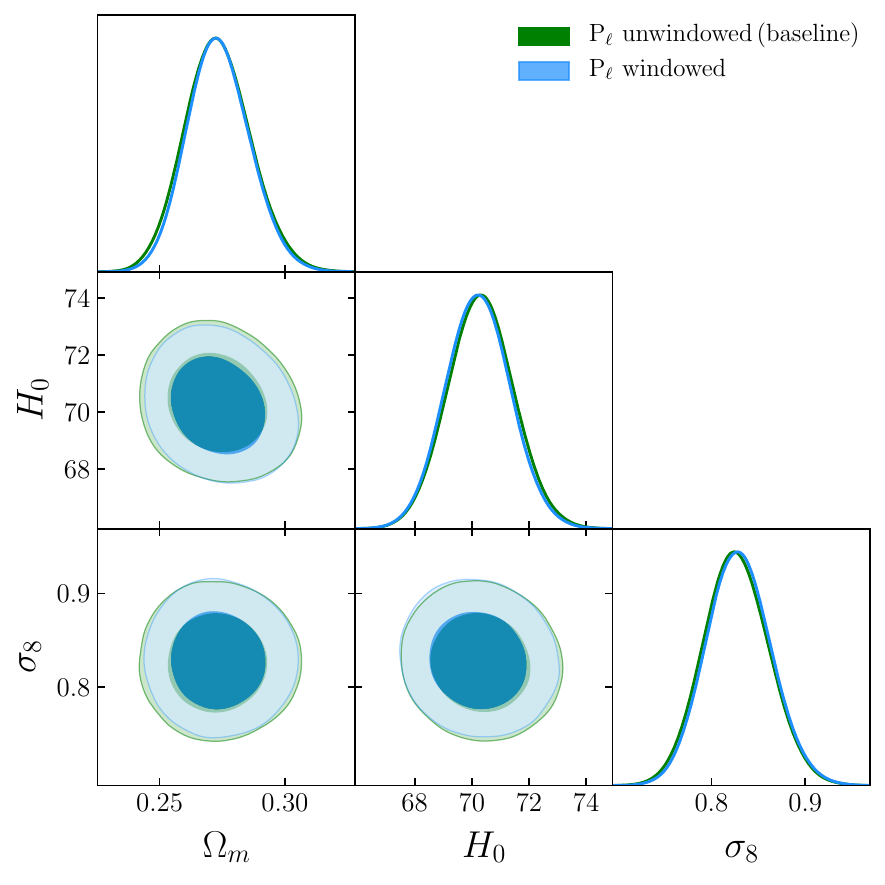}
    \includegraphics[width=0.48\linewidth]{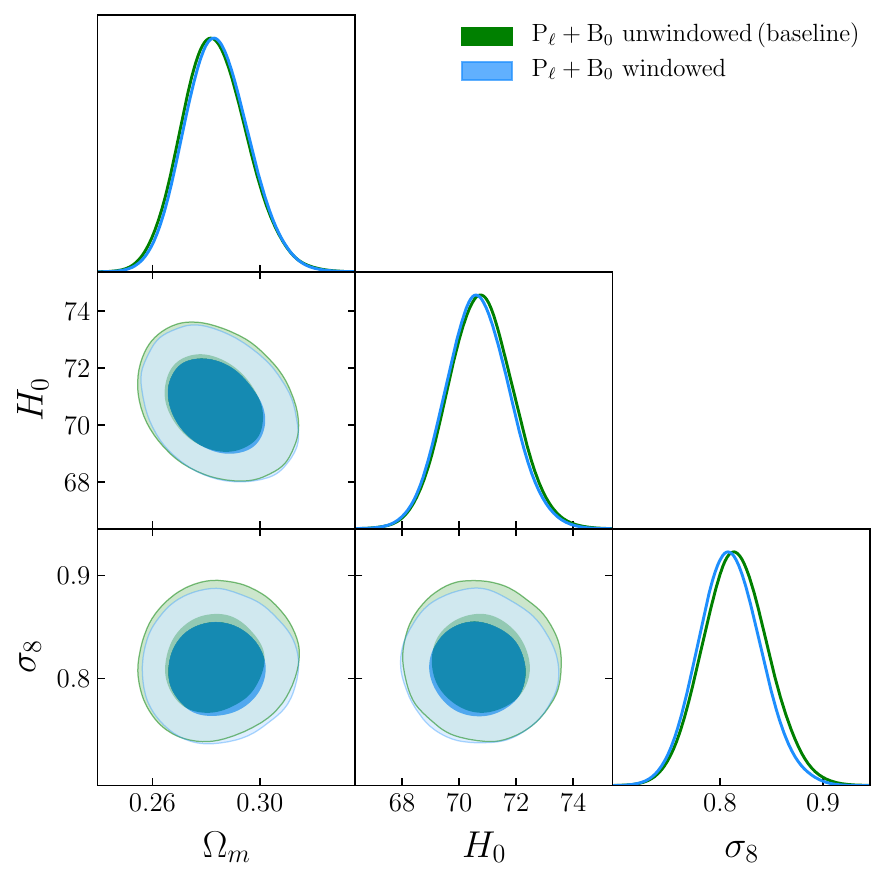}
    \caption{\textbf{Validation of the windowed approach}:
    Comparison of cosmological posteriors obtained from a DESI analysis of the power spectrum ({\it left panel}) and the power spectrum and bispectrum ({\it right panel}) using the unwindowed (\textit{i.e.}\ window-deconvolved) estimators of this work and the more common `windowed' estimators.
    The results show a near-perfect agreement, validating our approach.
    }
    \label{fig:window}
\end{figure}

Fig.~\ref{fig:window} compares the cosmological constraints obtained using the unwindowed and windowed DESI full-shape power spectra and bispectra. In both the power spectrum only and power spectrum and bispectrum analyses, we find nearly identical constraints from windowed and unwindowed estimators; this provides an excellent confirmation of our pipeline, and motivates our use of (fast) unwindowed estimators. Due to the similarity in the posteriors, we do not provide the marginalized credible intervals for the windowed scenario in this work.

\section{Conclusions}\label{sec: conclusions}

\noindent In this work, we have begun a through reanalysis of the full-shape galaxy clustering dataset from the first year of DESI data \citep[e.g.,][]{DESI:2024aax,DESI:2024jis,DESI:2024hhd}. Starting from the public data release \citep{DESI:2025fxa} — which provides only the processed galaxy and random catalogs — we have computed the redshift-space power spectrum multipoles and the bispectrum monopole using quasi-optimal estimators implemented in the \polybin code. These are closely related to the standard `windowed' estimators, but feature approximate window deconvolution, analogous to the pseudo-$C_\ell$ estimators used in the CMB community. Our measurements and binning matrices account for a number of systematic effects in the power spectrum, following the DESI analyses \citep{DESI:2024aax,DESI:2024jis,DESI:2024hhd}, including survey geometry, wide-angle effects, stochasticity, integral constraints, imaging systematics, and fiber collisions. We additionally account for these effects in the bispectrum, including via a novel stochastic estimator for fiber-collisions which removes close pairs without resorting to brute-force summation over triplets of particles. Finally, we have computed mask- and sample-dependent theoretical covariance matrices for each data-chunk of interest, incorporating the above systematic effects.

By combining the robust data with an accurate theoretical model, we can obtain tight constraints on the underlying cosmological parameters. After validating that our model is insensitive to projection effects, we placed tight constraints on $\ld$ parameters from DESI DR1 alone, with marginalized bounds on $\Omega_m$, $H_0$ and $\sigma_8$ that are consistent with \textit{Planck} CMB measurements, particularly when the bispectrum is included. Notably, the addition of the bispectrum monopole improves constraints on $\sigma_8$ and $\Omega_m$ by about $10\%$ and shifts $\Omega_m$ by approximately $1\sigma$ toward the CMB-preferred value (though see the covariance caveats below). In combination with the BAO constraints from DESI DR2 \citep{DESI:2025zpo,DESI:2025zgx}, the central values shift, preferring lower values of $H_0$ and higher values of $\Omega_m$, in closer agreement with \textit{Planck}. These shifts highlight the importance of including multiple statistics, allowing for breaking of parameter degeneracies. In combination with the CMB, we particularly tight constraints, with $\omega_{\rm cdm}$ tightening by a factor of two.

Whilst our pipeline accounts for a number of systematic effects (many of which are based on the official DESI treatment \citep{DESI:2024jis}), there remain several sources of uncertainty. In particular, we assume Gaussian covariances for all statistics, incorporating the mask but ignoring higher-order correlations (such as the cross-covariance between the power spectrum and bispectrum). Additionally, we have neglected the correlation between the DR2 BAO and DR1 full-shape data, motivated by the significantly higher resolution of the former. Whilst we expect these approximations to be valid on the scales probed herein (given previous studies \citep[e.g.,][]{Wadekar:2020hax,Oddo:2021iwq,Barreira:2021ukk,Philcox:2020vvt} and the tests of Appendix \ref{app:cross}), they could be further validated using numerical simulations, though a full study would require a suite of mocks that were (a) sufficiently high resolution to capture the necessary gravitational non-Gaussianities, (b) sufficiently numerous to obtain precise covariance estimates, and (c) correctly accounted for systematic effects such as fiber collisions (without resorting to simple rescaling factors \citep[cf.][]{Forero-Sanchez:2024bjh}).

A natural next step is to apply our pipeline to cosmological models beyond $\ld$. This is of particular importance given the potential hints of new physics in DESI BAO data, such as time-varying dark energy, non-zero spatial curvature, and novel physical components \citep[e.g.,][]{DESI:2025zgx,DESI:2025ejh,Chen:2025mlf,Green:2024xbb,Philcox:2025faf}. Oftentimes, parameter degeneracies can limit such analyses; in these scenarios, we expect to benefit strongly from the inclusion of the bispectrum in our analyses. Furthermore, the three-point function is an excellent probe of primordial non-Gaussianity; tight constraints on the corresponding $f_{\rm NL}$ parameters can be obtained by repeating previous analyses with the high-resolution DESI data \citep[e.g.,][]{Cabass:2022wjy,Cabass:2022ymb,Cabass:2024wob,DAmico:2022gki}.

Our analysis may also be extended by folding in additional statistics. Whilst we have already incorporated the reconstructed power spectrum, through joint analyses with the (correlated) DR1 and (approximately uncorrelated) DR2 BAO \citep[cf.,][]{Philcox:2020vvt,DESI:2024hhd}, there are several other sources of information such as 
the bispectrum multipoles and the transverse power spectrum (real space proxy $Q_0$); these could source somewhat tighter constraints on $\sigma_8$ \citep{Ivanov:2023qzb,Ivanov:2021fbu}. Finally, it will be advantageous to consider cross-correlations with other datasets, such as weak gravitational lensing \citep[e.g.,][]{Chen:2024vuf}.

\vskip 8pt
\acknowledgments
{\small
\begingroup
\hypersetup{hidelinks}
\noindent 
We thank Stephen Chen, Hector Gil-Marin, Jamie Sullivan, and Martin White for fruitful discussions. AC acknowledges funding from the Swiss National Science Foundation. We are additionally grateful to the anonymous referee for insightful feedback. The work of MMI was performed in part at Aspen Center for Physics, which is supported by National Science Foundation grant PHY-2210452. OHEP is a Junior Fellow of the Simons Society of Fellows and thanks the fauna of \href{https://www.flickr.com/photos/198816819@N07/54659520816/in/dateposted-public/}{Sllovakia} for emotional support. 
The computations in this work were run at facilities supported by the Scientific Computing Core at the Flatiron Institute, a division of the Simons Foundation, as well as at the Helios cluster at the Institute for Advanced Study, Princeton.
\endgroup
\vskip 4pt
\noindent This research used data obtained with the Dark Energy Spectroscopic Instrument (DESI). DESI construction and operations is managed by the Lawrence Berkeley National Laboratory. This material is based upon work supported by the U.S. Department of Energy, Office of Science, Office of High-Energy Physics, under Contract No. DE–AC02–05CH11231, and by the National Energy Research Scientific Computing Center, a DOE Office of Science User Facility under the same contract. Additional support for DESI was provided by the U.S. National Science Foundation (NSF), Division of Astronomical Sciences under Contract No. AST-0950945 to the NSF’s National Optical-Infrared Astronomy Research Laboratory; the Science and Technology Facilities Council of the United Kingdom; the Gordon and Betty Moore Foundation; the Heising-Simons Foundation; the French Alternative Energies and Atomic Energy Commission (CEA); the National Council of Humanities, Science and Technology of Mexico (CONAHCYT); the Ministry of Science and Innovation of Spain (MICINN), and by the DESI Member Institutions: \url{www.desi.lbl.gov/collaborating-institutions}. The DESI collaboration is honored to be permitted to conduct scientific research on I’oligam Du’ag (Kitt Peak), a mountain with particular significance to the Tohono O’odham Nation. Any opinions, findings, and conclusions or recommendations expressed in this material are those of the authors and do not necessarily reflect the views of the U.S. National Science Foundation, the U.S. Department of Energy, or any of the listed funding agencies.
}

\appendix

\section{Stochastic Fiber Collision Power Spectra}\label{app: pk-fc}
\noindent In this appendix, we discuss a stochastic method to estimate the fiber collision correction to the power spectrum, analogous to that used for the bispectrum in \S\ref{subsec: bk-specialization}. Starting from \eqref{eq: pk-fc}, we can introduce a unit random field $\epsilon$ with $\av{\epsilon(\vy)\epsilon(\vz)}_\epsilon =\delta_{\rm D}(\vy-\vz)$ to separate the two data fields. Explicitly, we define 
\beq\label{eq: pk-fc-stoch1}
    p^{\rm fc}_\alpha[d,\epsilon] &=& \frac{1}{2}\int d\vx\,d\vy\,\frac{\partial\xi(\vx,\vy)}{\partial p_\alpha}\left[d(\vx)\Phi(\vx,\vz)\epsilon(\vz)\right]\left[d(\vy)\epsilon(\vy)\right],
\eeq
which is equal to $p_\alpha^{\rm fc}[d]$ upon averaging over realizations of $\epsilon$. In the notation of \eqref{eq: de-d-star-e-def}, this can be written as
\beq\label{eq: pk-fc-stoch}
    p^{\rm fc}_\alpha[d,\epsilon] &=& \frac{1}{2}\int d\vx\,d\vy\,\frac{\partial\xi(\vx,\vy)}{\partial p_\alpha}\left[d\ast\epsilon\right](\vx)\left[d\epsilon\right](\vy),
\eeq
which is simply the cross-power-spectrum between the transformed fields $d\ast\epsilon$ and $d\epsilon$ (which can be obtained from the discrete catalogs). Ignoring the integral constraint, the change to the normalization matrix \eqref{eq: fish-fc} can be expressed similarly:
\beq
    \F^{\rm fc}_{\alpha\beta}[\epsilon] = \frac{1}{2}\int d\vx\,d\vy\,\frac{\partial\xi(\vx,\vy)}{\partial p_\alpha}[n\ast\epsilon](\vx)\frac{\partial\xi(\vy,\vx)}{\partial p_\beta}[n\epsilon](\vy),
\eeq
where $\av{\F^{\rm fc}[\epsilon]}_\epsilon=\F^{\rm fc}$, as before. This can be computed analogously to the usual normalization matrix, but with the two masks replaced by $n\ast\epsilon$ and $n\epsilon$.

\begin{figure}
    \centering
    \includegraphics[width=0.9\linewidth]{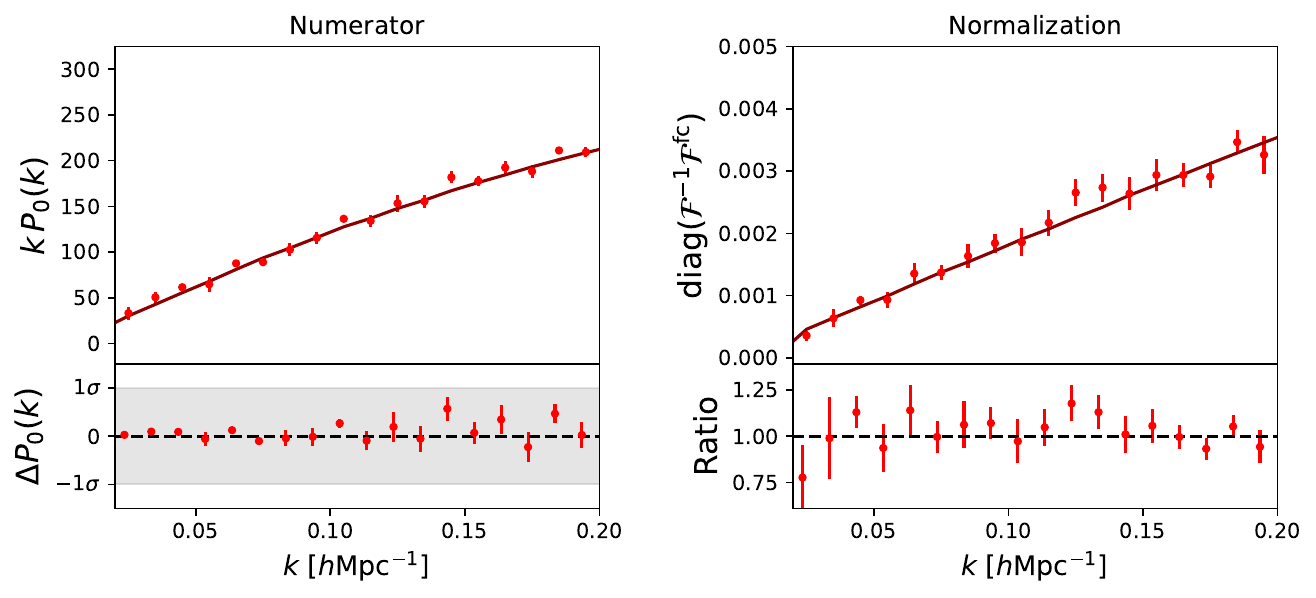}
    \caption{\textbf{Fiber Collision Estimator}: Comparison of stochastic (points) and deterministic (lines) algorithms for computing the impact of fiber collisions on the power spectrum monopole of the NGC LRG3 chunk. The methods compute the change to the power spectrum numerator (left) and normalization (right) induced by removing pairs of points with $\theta<\theta_{\rm cut}$. The bottom panels show the error in the stochastic method with respect to the errorbars (left) or the ratio (right). In all cases, we find excellent agreement between the deterministic pair-counting solution. This validates the stochastic method, which is used in the bispectrum analyses of this work.}
    \label{fig: stoch-pk}
\end{figure}

To test this approach, we compute the fiber collision corrections to the power spectrum numerator and normalization both stochastically (using the $\epsilon$ approach above) and deterministically (pair-counting, as in DESI \citep{Pinon:2024wzd}), focusing on LRG3-NGC chunk, which shows the largest fiber collision signal (\S\ref{subsec: pk-results}). The stochastic contributions are obtained across ten random realizations for the numerator and normalization, and are computed analogously to the bispectrum contributions discussed in \S\ref{subsec: bk-specialization}. Since our goal is to validate the bispectrum methodology, we focus on the $\ell=0$ mode, since we do not include anisotropic bispectra. Both terms requires $\approx 30$ node-minutes per iteration; this could be significantly reduced with a faster algorithm for computing $n\ast \epsilon$, which dominates the runtime. 

A comparison between the two methods is shown in Fig.\,\ref{fig: stoch-pk}. For both the numerator and the normalization (which encodes the window function), we find excellent agreement between the deterministic pair-counting approach and our stochastic method, despite the small number of Monte Carlo realizations. For all $k$-bins considered, the Monte Carlo errors in the numerator are a small fraction of the experimental errors, indicating convergence; furthermore, the ratio of the normalization is consistent with unity. As discussed above, the change to the normalization is small, reaching just $0.4\%$ by $k = 0.2\hMpc$. The close agreement between methods implies that our stochastic approach is both precise and accurate, and validates its application to the bispectrum analyses, for which there is no easily-applicable deterministic analog.

\section{Full Parameter Constraints}
\label{app:tables}

\noindent In this Appendix, we provide the full marginalized constraints on all cosmological and nuisance parameters sampled within the Markov Chain analysis. Parameters entering the likelihood quadratically are marginalized over analytically, and are thus excluded. The corresponding results for the four main analyses are displayed in Tab.~\ref{tab:full}. 
\begin{table}[!t]
    \centering
  \begin{tabular}{|c|cccc|} \hline
    \multirow{2}{*}{\diagbox[width=2.0cm]{\small Param.}{\small Dataset}}
    & \multirow{2}{*}{$\enspace P_\ell\enspace$}
    & \multirow{2}{*}{$\enspace P_\ell+B_0\enspace $} 
    & \multirow{2}{*}{$\enspace P_\ell+B_0+{\rm BAO}\enspace $}
    & $\enspace P_\ell+B_0+{\rm BAO}\enspace$
    \\
    & 
    &  
    & 
    & $+\,{\rm CMB}$
    \\
    \hline
    $\omega_{\rm cdm}$
    & $\enspace 0.1122_{-0.0067}^{+0.0068} \enspace$ %
& $\enspace 0.1189_{-0.0065}^{+0.0055} \enspace$ %
& $\enspace 0.1174_{-0.0039}^{+0.0039}\enspace$ %
& $\enspace 0.1172_{-0.0006}^{+0.0006}\enspace$ %
    \\
    $H_0$ 
    & $70.22_{-1.06}^{+1.06}$ %
& $70.67_{-1.05}^{+1.05}$ %
& $68.82_{-0.58}^{+0.58}$ %
& $68.61_{-0.28}^{+0.28}$ %
    \\
    ${\ln(10^{10}A_s)}$ 
    & $3.135_{-0.114}^{+0.104}$ %
& $3.005_{-0.084}^{+0.083}$ %
& $3.047_{-0.072}^{+0.072}$ %
& $3.059_{-0.014}^{+0.012}$ %
    \\
    $n_s$ 
    & $0.978_{-0.038}^{+0.038}$ %
& $1.001_{-0.034}^{+0.034}$ %
& $1.002_{-0.031}^{+0.031}$ %
& $0.973_{-0.003}^{+0.003}$ %
    \\\hline
    $b_1^{(1)}$ 
    & $1.603_{-0.103}^{+0.115}$ %
& $1.729_{-0.081}^{+0.081}$ %
& $1.717_{-0.079}^{+0.079}$ %
& $1.690_{-0.051}^{+0.051}$ %
    \\
    $b_2^{(1)}$ 
    & $-0.830_{-1.617}^{+0.730}$ %
& $-0.706_{-0.737}^{+0.592}$ %
& $-0.673_{-0.742}^{+0.583}$ %
& $-0.808_{-0.718}^{+0.572}$ %
    \\
    $b_{\mathcal{G}_2}^{(1)}$ 
    & $-0.057_{-0.467}^{+0.425}$ %
& $-0.380_{-0.400}^{+0.401}$ %
& $-0.345_{-0.395}^{+0.394}$ %
& $-0.411_{-0.395}^{+0.389}$ %
    \\
    $b_1^{(2)}$ 
    & $1.791_{-0.108}^{+0.120}$ %
& $1.932_{-0.085}^{+0.078}$ %
& $1.923_{-0.080}^{+0.080}$ %
& $1.892_{-0.044}^{+0.044}$ %
    \\
    $b_2^{(2)}$ 
    & $-1.110_{-1.332}^{+0.858}$ %
& $-0.596_{-0.718}^{+0.616}$ %
& $-0.617_{-0.705}^{+0.610}$ %
& $-0.796_{-0.690}^{+0.581}$ %
    \\
    $b_{\mathcal{G}_2}^{(2)}$ 
    & $-0.307_{-0.442}^{+0.443}$ %
& $-0.406_{-0.402}^{+0.404}$ %
& $-0.387_{-0.396}^{+0.395}$ %
& $-0.471_{-0.391}^{+0.390}$ %
    \\
    $b_1^{(3)}$ 
    & $1.929_{-0.122}^{+0.134}$ %
& $2.138_{-0.086}^{+0.085}$ %
& $2.112_{-0.083}^{+0.082}$ %
& $2.079_{-0.041}^{+0.041}$ %
    \\
    $b_2^{(3)}$ 
    & $-2.239_{-1.338}^{+0.918}$ %
& $0.631_{-0.943}^{+0.775}$ %
& $0.739_{-0.957}^{+0.787}$ %
& $0.544_{-0.912}^{+0.765}$ %
    \\
    $b_{\mathcal{G}_2}^{(3)}$ 
    & $-0.557_{-0.481}^{+0.527}$ %
& $0.183_{-0.440}^{+0.441}$ %
& $0.253_{-0.439}^{+0.439}$ %
& $0.172_{-0.441}^{+0.439}$ %
    \\
    $b_1^{(4)}$ 
    & $2.015_{-0.139}^{+0.157}$ %
& $2.298_{-0.094}^{+0.094}$ %
& $2.288_{-0.092}^{+0.092}$ %
& $2.251_{-0.042}^{+0.042}$ %
    \\
    $b_2^{(4)}$ 
    & $-4.040_{-1.575}^{+1.160}$ %
& $0.432_{-1.285}^{+1.040}$ %
& $0.242_{-1.233}^{+1.006}$ %
& $-0.219_{-1.093}^{+0.936}$ %
    \\
    $b_{\mathcal{G}_2}^{(4)}$ 
    & $-0.941_{-0.602}^{+0.710}$ %
& $-0.394_{-0.533}^{+0.535}$ %
& $-0.441_{-0.520}^{+0.520}$ %
& $-0.633_{-0.488}^{+0.491}$ %
    \\
    $b_1^{(5)}$ 
    & $1.261_{-0.114}^{+0.154}$ %
& $1.421_{-0.079}^{+0.079}$ %
& $1.402_{-0.078}^{+0.077}$ %
& $1.366_{-0.044}^{+0.052}$ %
    \\
    $b_2^{(5)}$ 
    & $-0.854_{-5.626}^{+7.927}$ %
& $-1.420_{-1.440}^{+1.264}$ %
& $-1.450_{-1.429}^{+1.258}$ %
& $-1.857_{-1.381}^{+1.205}$ %
    \\
    $b_{\mathcal{G}_2}^{(5)}$ 
    & $-0.912_{-0.992}^{+1.448}$ %
& $-0.605_{-0.671}^{+0.675}$ %
& $-0.587_{-0.632}^{+0.700}$ %
& $-0.778_{-0.659}^{+0.660}$ %
    \\
    $b_1^{(6)}$ 
    & $2.057_{-0.157}^{+0.197}$ %
& $2.298_{-0.117}^{+0.117}$ %
& $2.275_{-0.115}^{+0.115}$ %
& $2.213_{-0.061}^{+0.074}$ %
    \\
    $b_2^{(6)}$ 
    & $-5.209_{-3.707}^{+1.889}$ %
& $-2.685_{-3.206}^{+2.150}$ %
& $-2.728_{-3.195}^{+2.114}$ %
& $-3.948_{-2.730}^{+1.830}$ %
    \\
    $b_{\mathcal{G}_2}^{(6)}$ 
    & $-1.737_{-1.192}^{+1.374}$ %
& $-1.197_{-0.850}^{+0.962}$ %
& $-1.199_{-0.848}^{+0.935}$ %
& $-1.620_{-0.826}^{+0.816}$ %
    \\
    \hline
    $\Omega_m$ 
    & $0.274_{-0.013}^{+0.012}$ %
& $0.284_{-0.012}^{+0.010}$ %
& $0.296_{-0.007}^{+0.007}$ %
& $0.298_{-0.003}^{+0.003}$ %
    \\
    $\sigma_8$ 
    & $0.825_{-0.033}^{+0.033}$ %
& $0.811_{-0.031}^{+0.028}$ %
& $0.818_{-0.029}^{+0.029}$ %
& $0.809_{-0.005}^{+0.005}$ %
    \\
    $S_8$ 
    & $0.787_{-0.036}^{+0.036}$ %
& $0.789_{-0.035}^{+0.032}$ %
& $0.813_{-0.031}^{+0.031}$ %
& $0.807_{-0.007}^{+0.007}$ %
    \\
  \hline
    \end{tabular}
    \caption{\textbf{All Parameter Constraints}: Full parameter constraints from the $\Lambda$CDM analysis of the DESI data including the pre-reconstruction power spectrum and bispectrum data ($P_\ell+B_0$), post-reconstruction BAO DR2 measurements, as well as their combination with the CMB data. We give the mean values and 68\% confidence intervals for cosmological and nuisance parameters in all cases, with derived parameters shown in the final section.
    The inclusion of the bispectrum sharpens constraints on the quadratic bias parameters by up to $70\%$.
    }
\label{tab:full}
\end{table}

\section{Analysis including the Cross-Correlation of DR1 Full-Shape and DR2 BAO}
\label{app:cross}

\noindent In this appendix, we present a joint analysis of the DESI DR1 full-shape and DR2 BAO measurements, approximately accounting for the cross-correlation between the pre-reconstruction full-shape power spectrum and the Alcock Paczynski parameters extracted from post-reconstruction spectra (taken from \citep{DESI:2024uvr}). This test validates the main analysis of this work, wherein we neglect the correlation between the DR1 full-shape and DR2 BAO measurements. Notably, a significant fraction of the DR2 BAO information comes from Lyman-$\alpha$ data \citep{DESI:2025zpo}, which is independent from the galaxy samples considered herein.  

To model the dimensionless cross-correlation between the pre-reconstruction power spectrum and the BAO parameters, we use 2048 public \textsc{Patchy} mock catalogs, generated for SDSS BOSS analyses \citep{Kitaura:2015uqa}. 
The \textsc{Patchy} mocks use an approximate gravity solver, enabling a fast generation of a large suite of galaxy catalogs and providing a reliable estimate of the cross-correlation. 
This is expected to be largely insensitive to survey-specific galaxy clustering and observational systematics; therefore, we consider it a reasonable estimate of the cross-correlation for the DESI data, allowing us to test the assumption of independence.

We model the dimensionful cross-covariance between the galaxy power spectrum multipoles, $P^{\rm pre}_\ell(k)$, and the Alcock Pacznyski parameters, $\alpha_i$ with $i\in\{\parallel,\perp\}$, as follows
\be
\textrm{cov}[P^{\rm pre}_\ell(k),\alpha_i]=\textrm{corr}[P^{\rm pre}_\ell(k),\alpha_i]\sqrt{\textrm{var}[P^{\rm pre}_\ell(k)]\textrm{var}[\alpha_i]}\times\frac{V_{\rm DR1}}{V_{\rm DR2}}.
\ee
The cross-correlation coefficient $\textrm{corr}[P_\ell(k)],\alpha_i]$ is estimated from the \textsc{Patchy} mock catalogs corresponding to the CMASS NGC sample, spanning $0.43 < z < 0.7$ (with effective redshift $z_{\rm eff} = 0.56$). We adopt the CMASS NGC footprint because it provides the highest signal-to-noise ratio for the BAO parameter measurements~\cite{Philcox:2020vvt}. Above, $\textrm{var}[P^{\rm pre}_\ell(k)]$ denotes the variance of the pre-reconstructed DESI DR1 power spectrum, estimated from the Gaussian covariance for $P^{\rm pre}_\ell$ defined in \S\ref{sec: pspec}, while $\textrm{var}[\alpha_i]$ represents the variance of the $\alpha_i$ parameters corresponding to the DESI BAO DR2 data, taken from \citep{DESI:2024uvr}.
In addition, we include a volume factor to account for the partial overlap between the DR1 and DR2 samples, which reduces their covariance. Specifically, we adopt the following factors, calculated using the effective volumes of the DR1 and DR2 samples: $V_{\rm DR1}/V_{\rm DR2}=0.45$, $0.53$, $0.53$, $0.34$, $0.33$, and $0.56$ for the BGS, LRG1, LRG2, LRG3, ELG2, and QSO samples, respectively \citep{DESI:2024uvr,DESI:2025zgx}.~\footnote{The volume factor does not fully capture survey geometry effects and provides only an approximate estimate of the correlation between the two overlapping samples, which is appropriate for our study.}  
For the BGS sample, only the angle-averaged distance measurement is provided, so we estimate the cross-correlation using the isotropic BAO parameter $\alpha_{\rm iso}\equiv\alpha_{\parallel}^{2/3}\alpha_{\perp}^{1/3}$, setting $\mathrm{var}[\alpha_{\parallel}]=\mathrm{var}[\alpha_{\perp}]=9/5\,\mathrm{var}[\alpha_{\rm iso}]$ for simplicity.
This assumption has no impact on our results: the parameter constraints remain almost unchanged when the cross-covariance for the BGS sample is neglected.
Finally, we ignore the cross-covariance with the bispectrum since this statistic is limited to very large scales, $\kmax^{B_0}=0.08\hMpc$, which do not contain significant BAO information \cite{Philcox:2021kcw}. 

\begin{figure}[!t]
	\centering
	\includegraphics[width=0.6\textwidth]{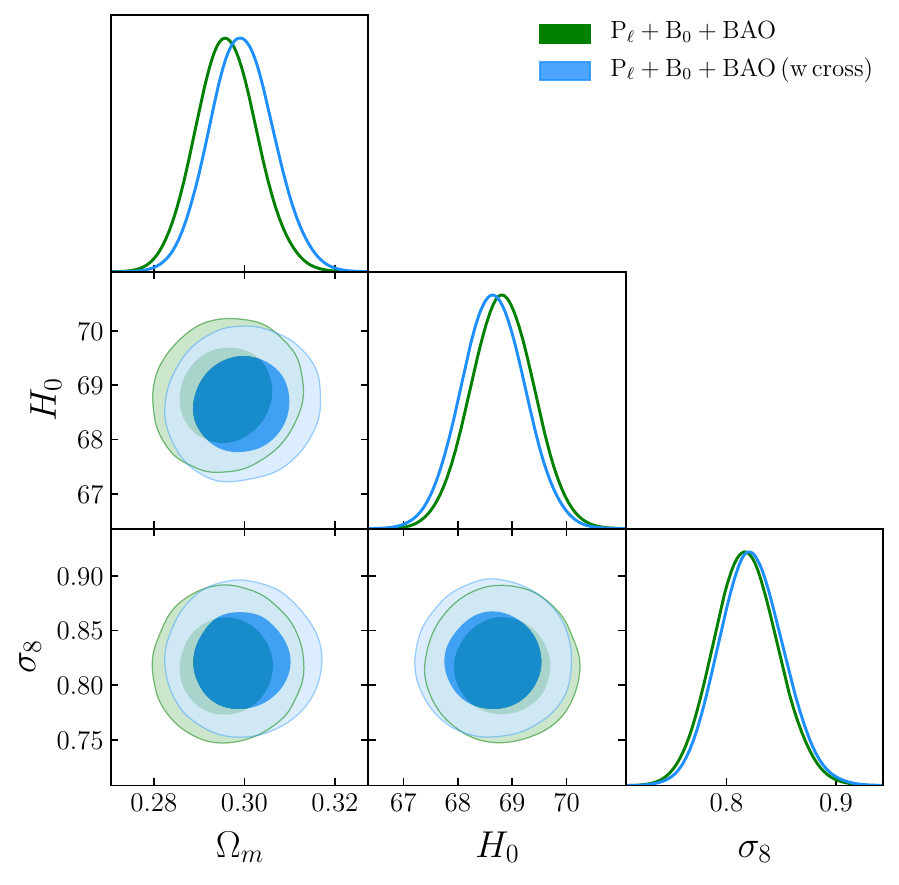}
	\caption{\textbf{Validation of the DR2 BAO Analysis}: Comparison of cosmological constraints obtained from the DESI DR1 power spectrum and bispectrum monopole and the DESI DR2 BAO measurements, when neglecting the cross-correlation between the full-shape and BAO measurements (green), and when using an approximate correlation across all six data chunks (blue). 
    This is evaluated using the cross-correlation estimated from 2048 Patchy mocks for the CMASS NGC sample, rescaled to the DESI observables, as described in the text. The full parameter constraints are listed in Tab.\,\ref{tab:cross}. We find only small shifts in practice, validating the independence assumption used in the main analyses of this work.
    }
    \label{fig:cross}
\end{figure}

Fig.~\ref{fig:cross} compares the cosmological constraints from the $P_\ell(k)+B_0(k)+{\rm BAO}$ analysis both setting the cross-correlation to zero and approximating it using the algorithm described above.
The full parameter constraints are listed in Tab.~\ref{tab:cross}.
\begin{table}[!t]
    \centering
  \begin{tabular}{|c|ccc|cc|} \hline
    \textbf{Dataset} 
    & $\omega_{\rm cdm}$
    & $H_0$ 
    & ${\ln(10^{10}A_s)}$ 
    & $\Omega_m$ 
    & $\sigma_8$
    \\
    \hline
    $P_\ell(k)+B_0(k)+{\rm BAO}$ 
    & $\enspace 0.1174_{-0.0039}^{+0.0039}\enspace$ 
    & $\enspace 68.82_{-0.58}^{+0.58}\enspace$ 
    & $\enspace 3.047_{-0.072}^{+0.072}\enspace$ 
    & $\enspace 0.296_{-0.007}^{+0.007}\enspace$ 
    & $\enspace 0.818_{-0.029}^{+0.029}\enspace$ 
    \\
    $P_\ell(k)+B_0(k)+{\rm BAO}$ (w cross)
    & $0.1183_{-0.0040}^{+0.0040}$ 
& $68.65_{-0.58}^{+0.58}$ 
& $3.052_{-0.072}^{+0.072}$ 
& $0.299_{-0.007}^{+0.007}$ 
& $0.823_{-0.029}^{+0.029}$ 
    \\
  \hline
    \end{tabular}
    \caption{\textbf{Validation of the DR2 BAO Analysis}: Mean and 68\% confidence intervals on $\ld$ cosmological parameters from the main analysis of this work, $P_\ell(k)+B_0(k)+{\rm BAO}$, and that when modeling the cross-covariance between the full-shape pre-reconstruction power spectrum and post-reconstruction Alcock Paczynski measurements.}
\label{tab:cross}
\end{table}
We find that including the cross-covariance induces small shifts in the cosmological parameters. In particular, $\Omega_m$ increases by $\approx 0.4\sigma$, while variations in the other parameters remain below $0.3\sigma$. Overall, the impact of the cross-covariance in the joint analysis of the DESI DR1 full-shape and DR2 BAO data is small, confirming the validity of the main analysis presented in this work. As a further test, we present a joint analysis using an analytic cross-covariance (and restricting to the DR1 BAO dataset) in the next section.

\section{Joint Analysis with DR1 BAO}\label{app:joint-bao}

\noindent In this appendix, we present a joint analysis of the DESI DR1 full-shape and BAO data. This differs from the main text, wherein we utilize DR2 BAO, neglecting its cross-correlation with DR1 full-shape. Whilst we expect weaker cosmological parameter constraints when using DR1 compared to DR2, our results are somewhat more robust, since we account for the cross-correlation between the two datasets.

\subsection{Methodology}
\noindent To perform a joint analysis, we adopt a framework similar to \citep{Philcox:2020vvt} (see also \citep{Chen:2021wdi,DESI:2024hhd}). First, we compute the post-reconstruction power spectrum for each of the six data-chunks listed in Tab.\,\ref{tab: desi-chunks} (omitting ELG1 and Ly-$\alpha$, for consistency with the pre-reconstruction analysis). These are computed using \polybin, as in \S\ref{sec: pspec}, with reconstruction performed utilizing the DESI \textsc{pyrecon} code \citep{DESI:2024aax}.\footnote{Available at \href{https://github.com/cosmodesi/pyrecon}{https://github.com/cosmodesi/pyrecon}.} Following DESI, we adopt the ``IterativeFFTReconstruction`` method, in the `RecSym` framework. Furthermore, we assume the same window function and normalization matrix for the pre- and post-reconstruction power spectra \citep[cf.,][]{DESI:2024aax}. Given that the BAO feature is visible up to large-$k$, we fix $k_{\rm min}^{\rm BAO}=0.02\hMpc$ and $k_{\rm max}^{\rm BAO}=0.29\hMpc$, adopting the same $\Delta k = 0.01\hMpc$ bins as before, but restrict to $\ell_{\rm max}^{\rm BAO}=2$.\footnote{Initial testing demonstrated that our results are not sensitive to the bin-width.} Unlike the official DESI analyses, we include a correction for fiber-collisions in the post-reconstruction spectra -- this is achieved using the stochastic algorithm outlined in Appendix \ref{app: pk-fc} (applied before the reconstruction shifts), allowing the $\theta_{\rm cut}$ to be imposed at the level of the observed, rather than transformed, field.

Next, we build a joint covariance matrix for the pre- and post-reconstruction power spectrum. For the reconstructed spectra, we use the same approach as \S\ref{sec: pspec}, computing a theoretical Gaussian covariance using an input smooth power spectrum model obtained by fitting an early version of the post-reconstruction data, $P^{\rm post}_{\ell}(k)$. A similar approach can be used to form the cross-covariance, schematically replacing $\mathbb{C}[P^{\rm post},P^{\rm post}]\to \tfrac{1}{2}\left(\mathbb{C}[P^{\rm pre},P^{\rm post}]+\mathbb{C}[P^{\rm cross},P^{\rm cross}]\right)$, where $\mathbb{C}$ is some covariance matrix function, and $P_\ell^{\rm cross}(k)$ is a smooth model for the cross-spectrum of pre- and post-reconstruction density fields, fitted from data. Notably, this approach differs from \citep{Philcox:2020vvt,DESI:2024hhd}, which computed the cross-covariance of $P_\ell(k)$ with the Alcock-Paczynski parameters $\alpha_\parallel,\alpha_\perp$, themselves inferred from the post-reconstruction spectra. Since $\alpha_\parallel, \alpha_\perp$ fully encapsulate the BAO signal, the two approaches yield equivalent constraints (assuming Gaussianity); however, the method used herein does not require simulations.\footnote{Computing the analytic covariance of $P^{\rm pre}_\ell(k)$ and $\alpha_\parallel,\alpha_\perp$ is challenging since the latter parameters are obtained from $P^{\rm post}_\ell(k)$ non-linearly.} The above method provides an accurate model for the joint covariance of the pre- and post-reconstruction DR1 spectra (which are almost 100\% correlated at small $k$); however, it cannot be straightforwardly applied to the DR2 data, since the underlying spectra are not publicly available.

Finally, we construct a Gaussian likelihood for the two sets of power spectra, utilizing the joint covariance. For $P^{\rm pre}_\ell(k)$ our theoretical model matches \S\ref{sec: theory}, whilst for $P^{\rm post}_\ell(k)$, we adopt the model described in \citep{DESI:2024aax}, where the BAO is parametrized by Alcock Paczynski parameters, biases, damping scales, and spline polynomials (with analytically marginalized coefficients). Unlike the fiducial DESI analyses, we do not attempt to infer $\alpha_\parallel,\alpha_\perp$ explicitly, instead sampling the underlying cosmological parameters. When combining with the bispectrum dataset, we ignore the cross-covariance with the post-reconstruction power spectra, noting that we restrict to very large scales.

\subsection{Results}
\begin{figure}[!t]
	\centering
	\includegraphics[width=0.65\textwidth]{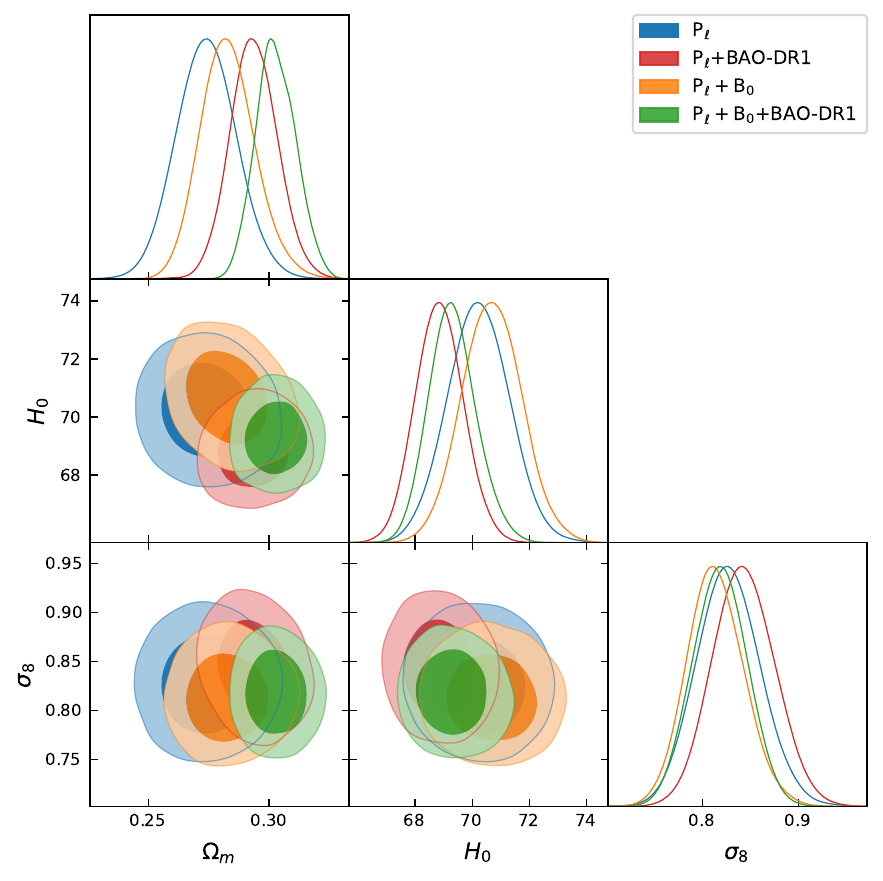}
	\caption{\textbf{Joint Analysis with DR1 BAO}: $\Lambda$CDM constraints from the DESI DR1 dataset, including both full-shape and BAO information. The DR1 BAO information is incorporated via a joint likelihood featuring an analytic cross-covariance between pre- and post-reconstruction power spectra. One-dimensional constraints are shown in Tab.\,\ref{tab:main2jointbao}. The joint analyses find similar constraints to those obtained using DR2 BAO (Fig.\,\ref{fig:main}), though with weaker bounds on $H_0$. As before, we find that the bispectrum significantly enhances parameter constraints.}
    \label{fig:main2jointbao}
\end{figure}

\noindent The results of the DR1 analysis are shown in Tab.\,\ref{tab:main2jointbao}\,\&\,Fig.\,\ref{fig:main2jointbao}. First, we perform a DR1 BAO-only analysis, which constrains the $\ld$ expansion history. In combination with BBN (as before), we find $H_0 = 68.7_{-1.0}^{-0.9}\hun$ and $\Omega_m = 0.293^{0.022}_{0.021}$. Since we do not include data from the ELG1 region nor the Lyman-$\alpha$ forest, the constraints are somewhat weaker than those of the official DESI DR1 analysis \citep{DESI:2024mwx} ($\Omega_m = 0.295\pm0.015, H_0 = 68.5\pm0.8\hun$). If we excise these datapoints from the official DESI DR1 BAO likelihood, we find $H_0 = 68.8_{-1.1}^{+1.2}\hun$ and $\Omega_m = 0.298_{-0.023}^{+0.025}$, in excellent agreement with our results. This is a non-trivial test of both our pipeline and that of \citep{DESI:2024mwx}, noting that our constraints are obtained using independent power spectrum estimators, alternative covariance choices, different modeling of fiber collisions (here via a $\theta_{\rm cut}$ in pre-reconstruction-space), and without prior compression to Alcock-Paczynski parameters.

In Fig.\,\ref{fig:main2jointbao}, we show the impact of the DR1 BAO information on the full-shape $\ld$ constraints. The results are similar to those involving the DR2 BAO (Fig.\,\ref{fig:main}); we find a shift to larger $\Omega_m$ and smaller $H_0$, with a fair increase in the $\Omega_m-H_0$ figure-of-merit. Though not independent, the results using DR1 and DR2 BAO are fully consistent, and achieve similar errors on $\sigma_8$, though the DR2 bounds on $\Omega_m$ and $H_0$ are tighter by a factor of $\approx 40\%$, due to the strong constraints on expansion history. As in the main text, the bispectrum adds significant constraining power, additionally with a shift to somewhat larger $\Omega_m$. Finally, we can compare our joint power spectrum and DR1 BAO analyses to those of \citep{DESI:2024hhd}. As shown in Tab.\,\ref{tab:main2jointbao}, the agreement is excellent (up to the Lyman-$\alpha$ information included in the latter), with no deviations above $0.4\sigma$.

\begin{table}[!t]
    \centering
  \begin{tabular}{|c|ccc|cc|} \hline
    \textbf{Dataset} 
    & $\omega_{\rm cdm}$
    & $H_0$ 
    & ${\ln(10^{10}A_s)}$ 
    & $\Omega_m$ 
    & $\sigma_8$
    \\
    \hline
    BAO-DR1 & $0.1155_{-0.0120}^{+0.0135}$ & $68.74_{-0.94}^{+1.01}$ & -- & $0.2930_{-0.0212}^{+0.0225}$ & -- \\
    $P_\ell(k)+$BAO-DR1  & $0.1166_{-0.0055}^{+0.0057}$ & $68.86_{-0.82}^{+0.84}$ & $3.150_{-0.097}^{+0.099}$ &  $0.2937_{-0.0094}^{+0.0100}$ & $0.842_{-0.032}^{+0.033}$\\
    $P_\ell(k)+B_0(k)+$BAO-DR1  & $0.1224_{-0.0048}^{+0.0055}$ & $69.27_{-0.80}^{+0.85}$ & $3.021_{-0.080}^{+0.075}$
    & $0.3023_{-0.0079}^{+0.0087}$ & $0.819_{-0.028}^{+0.028}$\\\hline
    Official$^*$ BAO-DR1 & $0.1180_{-0.0136}^{+0.0157}$ &  $68.75_{-1.08}^{+1.17}$ & -- &  $0.2981_{-0.0231}^{+0.0254}$ & --\\
    Official $P_\ell(k)+$BAO-DR1 & -- & $68.56\pm 0.75$ & -- & $0.2962\pm 0.0095$ & $0.842\pm 0.034$
    \\
    \hline
    \end{tabular}
    \caption{\textbf{Joint Analysis with DR1 BAO}: Mean and 68\% confidence intervals on $\ld$ cosmological parameters from analyses incorporating the DESI DR1 post-reconstruction BAO information. All results are obtained using \polybin power spectra, using analytic covariance matrices, which include the cross-covariance between pre- and post-reconstructed power spectra. The fourth row gives the official BAO-only results from DESI DR1, omitting the ELG1 and Lyman-$\alpha$ data-chunks, which are not included in our analysis. The final row gives the baseline constraints from \citep{DESI:2024hhd}, including all tracers. We find excellent consistency with the official results, and note that the addition of the bispectrum monopole significantly improves parameter constraints.}    
\label{tab:main2jointbao}
\end{table}

\bibliographystyle{apsrev4-2}
\bibliography{refs}

\end{document}